\documentclass[twocolumn]{aastex63}

\usepackage{natbib}

\received{October 2020}
\revised{February 2021}
\accepted{Accepted for publication in Icarus, march 2021}
\submitjournal{Icarus}

\shorttitle{Tidal degassing of the early Moon}
\shortauthors{Charnoz et al.}
\graphicspath{{./}{figures/}}

\begin{document}

\title{Tidal pull of the Earth strips the proto-Moon of its volatiles}

\correspondingauthor{Sébastien Charnoz}
\email{charnoz@ipgp.fr}

\author{Sébastien Charnoz}
\affiliation{Université de Paris, Institut de Physique du Globe de Paris, CNRS, F-75005 Paris,France}

\author{Paolo A. Sossi}
\affiliation{Institute of Geochemistry and Petrology, ETH Zürich, CH-8092, Zürich, Switzerland}

\author{Yueh-Ning Lee}
\affiliation{Department of Earth Sciences, National Taiwan Normal University, 88, Sec. 4, Ting-Chou Road, Taipei City 11677, Taiwan}

\author{Julien Siebert}
\affiliation{Université de Paris, Institut de Physique du Globe de Paris, CNRS, F-75005 Paris,France}

\author{Ryuki Hyodo}
\affiliation{ISAS, JAXA, Sagamihara, Japan}

\author{Laetitia Allibert}
\affiliation{Université de Paris, Institut de Physique du Globe de Paris, CNRS, F-75005 Paris,France}

\author{Francesco C. Pignatale}
\affiliation{Université de Paris, Institut de Physique du Globe de Paris, CNRS, F-75005 Paris,France}

\author{Maylis Landeau}
\affiliation{Université de Paris, Institut de Physique du Globe de Paris, CNRS, F-75005 Paris,France}

\author{Apurva V. Oza}
\affiliation{Physikalisches Institut, Universität Bern, Bern, Switzerland}

\author{Frédéric Moynier}
\affiliation{Université de Paris, Institut de Physique du Globe de Paris, CNRS, F-75005 Paris,France}



\begin{abstract}

Prevailing models for the formation of the Moon invoke a giant impact between a planetary embryo and the proto-Earth \citep{Canup_2004, Cuk_Stewart_2012}. Despite similarities in the isotopic and chemical abundances of refractory elements compared to Earth's mantle, the Moon is depleted in volatiles \citep{Wolf_Anders_1980}. Current models favour devolatilisation via incomplete condensation of the proto-Moon in an Earth-Moon debris-disk
\citep{Charnoz_Michaut_2015,Canup_2015,Lock_2018}. However the physics of this protolunar disk is poorly understood and thermal escape of gas is inhibited by the Earth's strong gravitational field \citep{Nakajima_Stevenson_2014}. Here we investigate a simple process, wherein the Earth's tidal pull promotes intense hydrodynamic escape from the liquid surface of a molten proto-Moon assembling at 3-6 Earth radii. Such tidally-driven atmospheric escape persisting for less than 1 Kyr at temperatures  $\sim 1600-1700$ K reproduces the measured lunar depletion in K and Na, assuming the escape starts just above the liquid surface. These results are also in accord with timescales for the rapid solidification of a plagioclase lid at the surface of a lunar magma ocean \citep{Elkins-Tanton_ET_AL_2011}. We find that hydrodynamic escape, both in an adiabatic or isothermal regime, with or without condensation, induces advective transport of gas away from the lunar surface, causing a decrease in the partial pressures of gas species ($P_s$) with respect to their equilibrium values ($P_{sat}$). The observed enrichment in heavy stable isotopes of Zn and K \citep{Paniello_2012, Wang_Jacobsen_2016} constrain $P_s/P_{sat}$ \textgreater 0.99, favouring a scenario in which volatile loss occurred at low hydrodynamic wind velocities ($<1\%$ of the sound velocity) and thus low temperatures. We conclude that tidally-driven atmospheric escape is an unavoidable consequence of the Moon's assembly under the gravitational influence of the Earth, and provides new pathways toward understanding lunar formation.

\end{abstract}



\section{Introduction}

While it is widely accepted that the Moon formed after a giant impact, the  mechanisms by which it acquired its current physical state and composition remain debated  \citep{Canup_2004, Cuk_Stewart_2012, Nakajima_Stevenson_2014, Lock_2018, Cameron_1991}. Lunar mare basalts have refractory lithophile element ratios akin to terrestrial basalts but are invariably depleted in volatile and moderately volatile elements by a factor of $\sim 5$  (for alkali metals) to $\sim 50$ (highly volatile elements like Zn, Ag and Cd) relative to their terrestrial counterparts \citep{Wolf_Anders_1980, Ringwood_1987, ONeil_1991, Lodders_Fegley_2011, Day_Moynier_2014}. This depletion occurs in concert with enrichment in the heavy isotopes of moderately volatile elements including Zn \citep{Paniello_2012, Kato_et_al_2015}, K \citep{Wang_Jacobsen_2016, Tian_et_al_2020}, Cl \citep{Sharp_2007, Boyce_2015, Gargano_et_al_2020}, Ga \citep{Kato_Moynier_2017},  and Rb \citep{Pringle_Moynier_2017}, indicative of an evaporative origin for volatile element loss from the Moon. Conversely, the enrichment of the lighter isotopes of Cr in lunar magmatic rocks compared to Earth's mantle suggests devolatilisation occurred near equilibrium and under low temperatures ($\le$ 1800 K) \citep{Sossi_Moynier_VanZuilen_2018_}, far below those expected during a giant impact event. \\

Even under the extreme temperatures reached as a result of a giant impact (from $\sim 2000$ to $\sim 10000$ K \cite{Canup_2004, Cuk_Stewart_2012, Nakajima_Stevenson_2014}, the removal of volatile elements from the Earth-Moon system by thermal escape is inhibited because of (i) the Earth's substantial gravitational field and  (ii) the high molecular masses of the moderately volatile elements ($\ge 20$ g/mol).  Both cases are aggravated by the low abundance of hydrogen in the proto-Earth, preventing loss through aerodynamic drag \citep{Nakajima_Stevenson_2018}. As such, recent models attempt to devolatilize Moon material prior to, or during, its assembly close to the Roche Limit within the post-impact debris disk (the "protolunar disk", a partially- to fully- vaporized disk). In the protolunar disk, volatile depletion by incomplete condensation and liquid-vapour partitioning may result from viscous separation of vapour from liquid \citep{Charnoz_Michaut_2015}, or gravitational torque from the proto-Moon exerted on the vapour \citep{Canup_2015} or separation of growing melt droplets from the vapour due to gas-drag \citep{Lock_2018}. However, these scenarios are critically dependent on the structure of the protolunar disk, which remains poorly understood, mostly due to its multi-phase nature and the unknown mixing states between gas and condensed phases \citep{Thompson_Stevenson_1988, Machida_Abe_2004, Ward_2012, Gammie_2016, Charnoz_Michaut_2015, Salmon_Canup_2012, Lock_2018}. In some of these models, the protolunar disk is contained within the Earth's Roche Limit, whereas numerical simulations of giant impacts show that most material ($> 70 \%$ by mass) is ejected beyond the Roche Limit \citep{Nakajima_Stevenson_2014} permitting volatile depletion beyond this point. The Roche Limit is the distance within which orbiting material cannot accrete into a single body due to disruption by the planets' tides and is located at 3 Earth radii for silicate material orbiting the Earth. \\

Here, we present an alternative scenario for atmospheric loss from the proto-Moon assembling at Earth's Roche limit. It is underpinned by the result that the influence of Earth's tides lower the kinetic energy required for a gas to thermally escape from the proto-Moon (see next section), relative to that for the Moon considered in isolation in space. As a conceptual investigation of this process, model approximations were made in order to illustrate the first-order influence of tidally-assisted atmospheric escape on the chemical composition of the Moon. For all realistic model parameters, intense hydrodynamic escape is found to be inevitable from the proto-Moon, or its building blocks, provided it accreted at a distance of a few Earth's radii such that Earth's tides were sufficiently strong to lower the Moon's escape velocity. We describe tidally-assisted thermal escape and show that, if the gas at the surface of the proto-Moon is hot enough to accelerate upwards to its Lagrange points, then this gas would be lost from the Moon, either by ejection into space, or by injection into orbit around the Earth. We explore two end-member scenarios, a 'wet' and a 'dry' model of the proto-lunar atmosphere, and show that, in both cases, the gas accelerates sufficiently through tidally-assisted thermal expansion to be able to escape. The gas velocity slows as the Moon's orbital expansion causes the escaping flux to diminish. Over time, atmospheric loss via hydrodynamic escape produces a net depletion in the volatile element abundances in the Moon, consistent with measurements of lunar mare basalts.


The paper is organized as follows. In section \ref{section_considerations} we examine, using simple arguments, under what conditions material may escape from the proto-Moon surface under the effect of tides. In section \ref{section_physics_hydro_escape} we investigate two models of hydrodynamic escape assisted by tides and show that under all reasonable assumptions, escape is inevitable. In section \ref{section_orbital_expansion} we discuss the orbital evolution of the proto-Moon, which sets an upper time limit for hydrodynamic escape to occur. In section \ref{section_core_text_evaporation_section} we compute the composition of the exsolved gas and the resulting melt composition. We show that it is possible to explain Na and K abundances in lunar records. We discuss our results and conclude in section \ref{section_discussion_conclusion}.

\section{A conceptual framework for atmospheric loss from the proto-Moon at the Roche Limit}
\label{section_considerations}

 Models of lunar assembly suggest that the Moon's building blocks coalesce just beyond the Roche Limit \citep{Canup_2015,Takeda_2001}. Traditionally the Roche Limit for a fluid body $R_{\rm R}$, is  \citep{Roche_1873, Canup_Esposito_1995} :

\begin{equation}
    R_{\rm R}=2.456 \times R_{\rm p} \left( \frac{\rho_p}{\rho_{\rm m}} \right)^{1/3}
\end{equation}

   where $R_{\rm p}$ is the planet radius, $\rho_p$ is the planet density, and $\rho_{\rm m}$ is the orbiting material density. For the case of the Earth, $\rho_p=5500~{\rm kg/m}^3$, and assuming the orbiting material has the same average density as the present-day Moon, we choose $\rho_{\rm m}=3300~{\rm kg/m}^3$. These numbers give $R_{\rm R} \simeq 3 R_\oplus$. Aggregates can begin to form at the edge of the proto-lunar disk around $3 R_\oplus$ but the fully-grown Moon might appear further away, around $5R_{\oplus}$ \citep{Canup_2015,Salmon_Canup_2012}. The Moon's gravitational attraction sphere is called the Hill Sphere and its radius is  $R_{\rm h}=a(M_{\rm m}/3M_\oplus)^{1/3}$ ($a$ is the Earth-Moon distance, $M_{\rm m}$ and $M_\oplus$ are the masses of the Moon and Earth, respectively). We assume that the proto-Moon is hot and is surrounded by gas evaporated from its surface. If the proto-Moon's atmosphere is contained within its Hill sphere, then it may be hydrostatically stable, like it is for the present-day atmospheres of the terrestrial planets. However, when the proto-Moon (or its building blocks) is located close to the Roche Limit, its Hill sphere has a size comparable to its physical radius ($R_{\rm h} \simeq 2500$ km for a Moon located at $3 R_{\oplus}$), meaning the atmosphere is hydrostatically unstable, and thereby liable to escape. Specifically, the Earth's tidal pull lowers the minimum energy required for a particle to escape the proto-Moon's surface (relative to the case for the Moon considered in isolation). The computation of the potential energy at the Moon's surface, under Earth's tidal field is detailed in Appendix \ref{SOM_potential_energy}. As the L1 and L2 Lagrange points are the points on the Hill's sphere closest to the Moon's surface (Appendix \ref{SOM_potential_energy}), escape of the gas is most readily achieved via passage through L1 and L2, because it requires the least energy \ref{Figure_energy}). The kinetic energy required can be converted into a gas temperature  \ref{Figure_T_escape}. For this fiducial case, we assume a molar mass equal to $20$ g/mol, as a proxy for an atmosphere consisting of sodium \cite{Visscher_Fegley_2013}) . A well-mixed, static and transparent atmosphere in thermal equilibrium with the Moon's hot surface at temperature $T_{s}$ (which is treated as a free parameter) is assumed, for simplicity, to produce a temperature profile that scales with $T_{s}(r/R_{m})^{-1/2}$, where $r$ is the distance to the Moon's center and $R_{\rm m}$ is the radius of the Moon (red lines in Figure \ref{Figure_T_escape}). This scaling is simply obtained by equalizing the cooling power of a gas parcel, $P_{cool}=\sigma T^4$ to the radiative power of a hot sphere at at distance $r$ : $P_{rad}=L/(4 \pi r^2)$ where L is the luminosity of the hot sphere (this is, for example, a common scaling in passive protoplanetary disk studies).

   Under these simplistic assumptions, we find that, if $T_s > 1600K$ the atmosphere is hot enough to escape beyond the Hill sphere when the proto-Moon is located at distance $\lesssim 5 R_\oplus$. Moreover a smaller proto-Moon (0.5 lunar mass), or its constituents, are more liable to atmospheric loss under the same conditions at same surface temperature (Figure\ref{Figure_T_escape}. b).

 Whether the escaping atmosphere is permanently lost hinges upon the dynamics of the material after leaving the Moon's Hill sphere. Classical celestial mechanics \citep{Murray_Dermott_SS_Dynamics} constrain outgoing particles to pass through either the L1 or L2 Lagrange points before being ejected into orbits around the Earth. The fate of this escaping material depends on the nature of the physical processes acting upon it. If the escaping material is gaseous, it does not re-accrete onto the Moon. That is, any dissipative process like gas viscosity or radiative cooling prevents the gas from re-entering the Hill Sphere because of a loss of mechanical energy. This is illustrated in Figure \ref{fig_orbits} (left) and Figure \ref{figure_potential_gas_drag} where trajectories of particles leaving the Moon in a fictive gaseous disk were tracked and computed. It is observed that they do not fall-back onto the Moon, but rather migrate inward or outward, leaving the Moon permanently.

 Alternatively, if the Earth's surface is sufficiently hot during assembly of the proto-Moon ($> 3000$ K) \citep{Canup_2004, Cuk_Stewart_2012, Nakajima_Stevenson_2014} with a photosphere around 2000 K, then black-body emission may induce radiation pressure on micrometer-sized particles (in addition to heating the near-side of the proto-Moon). This leads to their ejection from the Earth-Moon  system \citep{Hyodo_2018}  (Figure \ref{fig_orbits}, right and Figure \ref{figure_potential_rad_press}). This scenario of ejection strongly depends on the physical size of dusty-grains. Here it is noted that escape is efficient for particles in the micrometer range from a Moon surface heated to $>$ 1500K following a giant impact (Appendix \ref{SOM_RAD_PRESSURE}).

We conclude from the first-order considerations detailed above that it is reasonable to expect that tidal effects would (1) facilitate the escape of material from the Moon's surface and (2) prevent its return to the lunar surface due to 3 body-effects. Although the above-mentioned scenarios (dissipative gas disk, radiation pressure) may prevent the return of escaping material onto the Moon's surface, the "bottleneck" is to understand how material can be transported from the proto-Moon's surface (i.e., the locus of its evaporation) up to the L1/L2 Lagrange points at which this material can escape.

\section{Tidally-assisted hydrodynamic escape}
\label{section_physics_hydro_escape}
Mechanisms capable of carrying gas from the molten proto-lunar surface to the escape points are a product of the thermochemical structure of its atmosphere. We investigate the mode of atmospheric escape occurring under the influence of the tidal pull of the Earth, and derive expressions that permit calculation of the escaping flux. To do so, we formulate two end-member scenarios using a 1D approach and take into account (1) evaporation from a magma ocean at surface temperature $T_{\rm s}$ , and (2) the mutual gravitational attraction between the Earth and the Moon. Contrary to the abundant literature on planetary atmospheric escape, the atmosphere we consider here is not heated from above by an incoming XUV flux (see SI \ref{XUVearth}), but rather from its base by a hot magma ocean, considered to be at constant temperature $T_{\rm s}$. In the first "dry" model we assume a purely adiabatic gas evaporated from the magma ocean and neglect condensation during its adiabatic expansion (Section \ref{SOM_hydro_escape_adiabatic}). The second "wet" model assumes that individual species comprising the atmosphere may condense as liquid droplets depending on its pressure-temperature structure, and can be subsequently dragged along with the prevailing outflow (Section \ref{hydro_escape_wet_model}). Each of these scenarios involves simplifying assumptions to render calculations tractable, and the reality
of atmospheric escape probably lies in between these two extremes. However, we use these two end-member scenarios as a proof-of-concept that tidally-driven thermal degassing is unavoidable under all reasonable assumptions, provided the proto-Moon assembles close to the Roche Limit.

In the following, it is assumed that the proto-Moon (or its building blocks) is surrounded by an atmosphere. The surface of the proto-Moon is assumed to be always liquid, and, in contact with the gas. We examine under which conditions the atmosphere can escape from its Hill sphere by solving the hydrodynamic equations, and considering the Earth's tidal force.

\subsection{Dry expanding atmosphere}
\label{SOM_hydro_escape_adiabatic}
Due to the intrinsic complexity of the physical process we aim to investigate, it is not feasible to treat the full problem self-consistently. We are mindful that, due to the temperature diminution with altitude, some fraction of the gas may recondense, and thus may not behave adiabatically. This case is treated in Section \ref{hydro_escape_wet_model}. However, it is of primary importance to first understand the physics of hydrodynamic escape above the lunar magma ocean by solving the fully adiabatic approximation, as it is the original (and natural) framework of the theory of hydrodynamic escape \citep{Parker_1963, Parker_1965}, and hence the crux of the present paper. Since the atmosphere is thin (at most 100 Pa) and composed largely of monatomic gases, it is reasonable to start with the adiabatic approximation, since acceleration is strong at the base of the atmosphere. However, for the sake of completeness, isothermal solutions are also presented in Appendix \ref{isothermal_appendix}.

\subsubsection{Mathematical treatment}
The derivation of the mass flux in the dry model is solved within the adiabatic approximation, thus we ignore here any condensation process during the escape of the gas from the proto-Moon's surface, as well as heat transfer with the surroundings. We solve the Euler equations (conservation of mass and momentum + adiabatic equation of state) and calculate the 1D flow along the line joining the lunar surface to the Lagrange L1 or L2 points (located at the Hill radius). The gas follows the adiabatic equation of state, so that
\begin{equation}
P=\alpha \rho^{\gamma}
\label{eq_adiab_P}
\end{equation}
where P is the total gas pressure, $\alpha$ is a positive constant, $\gamma$ is the adiabatic index ( $\simeq 1.5$, corresponding to a gas with a mixture of monatomic -$\gamma$ = 5/3- and diatomic species $\gamma$ = 7/5), where P is given by the sum of partial pressures of a gas produced in equilibrium with a magma of Bulk Silicate Earth (BSE) - like composition \citep{Palme_Oneil_2003}.  These properties are related to the temperature via the adiabatic relation:
\begin{equation}
    T=\frac{\mu \alpha \rho^{\gamma-1}}{k}
    \label{Eq_temp_adiabatic}
\end{equation}
where $\mu$ and k are the mean molecular weight and Boltzmann's constant, respectively. The molar mass of the gas is assumed to be $\simeq 35g/mol$ (so $\mu \sim 5.81\times 10^{-26} kg)$. This corresponds roughly to a mixture of Na, O$_2$ and K that are the dominant species for $T<2000K$. At steady state, the conservation of mass reads :

\begin{equation}
\frac{\partial \rho v r^2 }{\partial r}=0
\label{eq_adiab_ass}
\end{equation}

Conservation of momentum for a steady-state flow reads:
\begin{equation}
\rho v \frac{\partial v }{\partial r}=-\frac{\partial P}{\partial r}+\rho A(r)
\label{eq_adiab_momentum}
\end{equation}

where A(r) is the local acceleration in the non-inertial frame rotating with the Moon (Equation \ref{equation_accel} in Appendix \ref{SOM_potential_energy}).
Equations \ref{eq_adiab_P} to \ref{eq_adiab_momentum} form a closed set of ODEs. The famous Bernoulli equation for an adiabatic gas is recovered by integrating equation \ref{eq_adiab_momentum} from the surface (subscript S) to the Hill sphere (subscript H) to obtain:

\begin{equation}
    \frac{V_S^2}{2}+\frac{\alpha\gamma}{\gamma-1}\rho_S^{\gamma-1}+Ep_S=\frac{V_H^2}{2}+\frac{\alpha\gamma}{\gamma-1}\rho_H^{\gamma-1}+Ep_H
\end{equation}

where $Ep$ is the potential energy in the frame rotating with the Moon. In the above Bernoulli equation, V and $\rho$ are linked through the mass conservation equation so that $V_S r_S^2\rho_S=V_H r_H^2\rho_H$. Replacing $\rho_H$ in the Bernoulli equation one finally obtains:

\begin{equation}
    \frac{V_S^2}{2}+K\rho_S^{\gamma-1}+Ep_S=\frac{V_H^2}{2}+V_H^{1-\gamma}K(V_S\rho_Sr_S^2/r_H^2)^{\gamma-1}+Ep_H
    \label{eq_adiab_vs_vh}
\end{equation}

where $K=\alpha\gamma/(\gamma-1)$. Note that Equation \ref{eq_adiab_vs_vh} can also be written as follows :
\begin{equation}
    \frac{V_S^2}{2}+\frac{\gamma}{\gamma-1} \frac{kT_S}{\mu}+Ep_S=\frac{V_H^2}{2}+\frac{\gamma}{\gamma-1} \frac{kT_H}{\mu}+Ep_H
    \label{eq_adiab_vs_vh_T}
\end{equation}
with $T_S$ and $T_H$ standing for the gas temperature at the surface and at the Hill sphere, by inserting Equation \ref{Eq_temp_adiabatic} into Equation \ref{eq_adiab_vs_vh}. It shows that the velocity field is \textit{independent} of the surface pressure and is controlled only by the surface temperature and the gravity field. So for a given surface velocity $V_S$ and temperature $T_S$, the velocity at the crossing of the Hill sphere $V_H$ can be determined by numerically solving equation \ref{eq_adiab_vs_vh_T}. This is a well known problem for the case of a point-source gravity field and has been extensively treated by E.N. Parker in a series of papers on solar wind outflows (see e.g. Parker 1963).

The Bernoulli equation gives rise to several families of solutions extensively discussed in \cite{Parker_1963} and \cite{Parker_1965}. The only physical solutions are those which start from the surface with flow velocities smaller than the sound speed. Then the velocity increases with height, passing through the \textit{transonic point} beyond which the flow velocity becomes supersonic (but without a shock). All other solutions correspond either to inflow solutions (subsonic at large distance, supersonic on the ground) or do not connect the surface with infinity, or have finite pressure at infinity (see Parker 1965 for a detailed discussion).

The location of the transonic point is found as follows: inserting equation \ref{eq_adiab_ass} into equation \ref{eq_adiab_momentum} and replacing $\partial P/\partial r$ by $\partial P/\partial\rho \times \partial\rho/\partial r$, and using Equation \ref{Eq_temp_adiabatic} one obtains:

\begin{equation}
\frac{\partial v }{\partial r}=\frac{2 \gamma T k/(\mu r) +A(r)}{v(1- \gamma T k/(\mu v^2))}
\label{eq_adiab_dvdr_sonic}
\end{equation}

For v(r) to be physical, the denominator and numerator must cancel simultaneously (this is the transonic point). Finding the zero of the denominator gives v at the sonic point ($v_{transonic}$), and finding the zero of the numerator, and noting that $v^2/2+\gamma/(\gamma-1) kT/\mu+Ep(r)=cst$ along a current line, gives the location ($r_{transonic}$) and temperature of the sonic point. Once $r_{transonic}$ and $v_{transonic}$ are determined, the velocity at the Moon's surface is determined by solving equation \ref{eq_adiab_vs_vh} between the sonic point and the surface, and the velocity at the Hill sphere is obtained by solving the Bernoulli equation between the sonic point and the Hill sphere. One unknown still needs to be determined: the gas density at the sonic point. In order to do so, we prescribe the surface temperature. Since the velocity field is independent of density (Equation \ref{eq_adiab_vs_vh_T}) V can be computed everywhere starting from the sonic point. Once the velocity field is found, the density and pressure at the surface of the magma ocean must be determined (see next section), in order to obtain the net escaping flux. 
\\
\subsubsection{Determining the gas density at the evaporating surface}
\label{vs_vt}

We now focus on the determination of the surface flux, which depends on the evaporative flux at the ocean magma surface (+) and on the return flux back to the magma ocean (-), with the difference being the net flux away from the surface. The evaporative flux can be determined by the Hertz-Knudsen-Langmuir (HKL) equation (e.g. Knudsen, 1909) integrated over the surface area of the Moon, and is proportional to the equilibrium partial pressure of the gas ($P_{sat}$) that can be calculated thermodynamically. The subsequent transport of this gas away from the surface occurs by molecular motion through two distinct mechanisms; advection and diffusion, or a combination thereof. If this transport rate is slow relative to the evaporation rate, there is a 'return flux' to the surface, which, in turn, results in a build up of the partial pressure (or density) at this surface ($P_{s}$) owing to mass conservation (\cite{Richter_et_al_2002, Young_2019, Tang_Young_2020}). This is a classical feature of the HKL equation (the $P_{sat} - P_{s}$ term), though it does not explicitly prescribe how $P_{s}$ is calculated.
An expression for calculating the return flux (and thus the $P_{s}$ term) is proposed in \cite{Young_2019, Tang_Young_2020}, following the \cite{Richter_et_al_2002} approach. Because \cite{Young_2019, Tang_Young_2020} assume there is a surface boundary layer directly above the magma ocean that is well-mixed and in hydrostatic equilibrium, diffusion is the only possible mode of mass transport within this layer.  In this approach, diffusive transport is derived from Fick's Law, in which the diffusion coefficient for a binary gas is given by the Chapman-Enskog equation. This model always results in $P_{s}/P_{sat}$ values that are very close to unity (i.e., equilibrium between the magma and the gas) because diffusion is exceedingly slow when integrated over the Moon's surface area.  While self-consistent, such an approach is not applicable for cases in which advection occurs at the surface, as described here. A suitable expression for an advected gas is derived in Appendix \ref{Appendix_Pressure_surface}, where we compute the return flux, as well as pressure and density of a gas at the surface of an evaporating surface, taking into account advection. We use a kinetic approach very similar to the original Hertz-Knudsen theory. We show that surface pressure and density depend only on the ratio of the gas velocity at the surface $V_S$ to the local thermal velocity $V_t$. It is found that the gas density above the fluid surface ($\rho_s$) is (Equation \ref{equation_Ps_Psat} in appendix \ref{Appendix_Pressure_surface}) :

\begin{equation}
\rho_s/\rho_{sat} = \left(  \frac{V_S}{V_t} \sqrt \pi(2- erfc(\frac{V_S}{V_t})) + e^{\frac{-V_S^2}{V_t^2}}
\right)^{-1}
\label{equation_rhos_rhosat}
\end{equation}

where $\rho_{sat}$ is the gas density at saturation and at temperature $P_{sat}$ is the saturating vapour pressure of a magma of BSE composition at temperature $T_S$ given in \cite{Visscher_Fegley_2013}. We start by computing the velocity profile using the Bernoulli equation. This permits the surface velocity to be calculated, and the density and pressure at the surface are then computed using Equation \ref{equation_rhos_rhosat}. The escaping flux is given as $F=\rho_sV_sR_{\rm m}^2$. Here we assume for simplicity that the solid angle of gas emission is 1 steradian (compared to $4\pi$ for a full sphere). Finally, one obtains a velocity profile with a mass flux that is perfectly conserved, allowing a precise estimate for the return flux to the surface to be made.

\subsubsection{Results}
Figure \ref{Figure_Vsurface_adiab_1Mm} shows the ratio of surface velocity to thermal velocity, for different surface temperatures, as a function of the Earth-Moon distance. We see that $V_s/V_t<0.4$ for all cases considered and thus the surface pressure is always $> 0.6  P_{sat}$ (see appendix \ref{Appendix_Pressure_surface}), but for $T<1800K$ (our preferred case), $V_s/V_t$ never exceeds $0.2$, meaning the surface pressure is always $> 0.8 P_{sat}$. Therefore, the actual surface pressure is found to lie close to gas-liquid equilibrium.  In our case, and contrary to comets, the atmosphere expands at velocities much lower than the thermal velocity. Values reported in Figure \ref{Figure_Vsurface_adiab_1Mm} also imply that the net escaping gas flux from the proto-Moon is small compared to the free evaporative flux from the magma ocean (the ratio of the two being displayed in Figure \ref{figure_flux_ratio}), apart from cases in which $T>2000K$ and $a<3.5 R_{\oplus}$ where the escaping flux becomes about half of the free evaporative flux.

The resulting escape fluxes are displayed in \ref{Figure_escape_adiab_1Mm}. The flux of escaping gas drops with increasing Earth-Moon distance up to the point where escape becomes negligible because of the diminishing gravitational influence of the Earth that otherwise promotes escape. For a lunar surface temperature of 2200 K, the atmosphere of the proto-Moon is efficiently lost provided the proto-Moon is located  inward of 5$R_{\oplus}$, with a flux varying between $8\times10^{-5}$ Moon masses/year at to the Roche limit, down to $10^{-8}$ Moon mass/year at 5.5$R_{\oplus}$. For $T=1600$ K the flux is much lower, between $3 \times 10^{-7}$ and $6 \times 10^{-9}$ Moon mass/year at 3.7 $R_{\oplus}$ only, due to the lower internal energy of the escaping gas. Note that the fluxes reported above may be underestimated, as we assume (for simplicity) the gas emitting surface is only 1 steradian, one order of magnitude smaller than a fully degassing spherical surface ($4\pi$ steradians). This conservative lower bound is justified due to the uncertainty of the areal extent of a lunar magma ocean, together with the fact that the mathematical treatment outlined above strictly applies only along the Earth-Moon line (L1 and L2 Lagrange points).

If we consider potential lunar buildings blocks with half a lunar mass, the flux is more intense by about a factor 2 to 5 (Figure \ref{Figure_escape_adiab_05Mm}), and, interestingly escape is arrested, on average, at a distance about twice as far from the Earth compared to a fiducial lunar mass. Thus, hydrodynamic escape from a 0.5 lunar mass proto-Moon at 1600 K occurs up to about 6.4 $R_{\oplus}$ (compared to 3.7 $R_{\oplus}$ for a 1 lunar mass object), and beyond 10  $R_{\oplus}$ for a surface temperature of 2000K.

\subsection{Wet model with condensation}
\label{hydro_escape_wet_model}

The dry model presented above neglects any condensation or drag of silicate and oxide droplets. Condensation may affect the escape flux due to internal energy transfer within the atmosphere associated with the latent heat of condensation. To account for this, we have developed a \textit{wet} model that includes gas condensation.  We calculate the composition of the vapour phase presuming it is in equilibrium with a liquid of Bulk Silicate Earth (BSE) composition whose renewal timescale at the surface of the magma ocean is very short ( $< 50$ days, see Appendix \ref{magma_renewing}) relative to the cooling timescale. 

\subsubsection{Mathematical treatment}
We introduce a method to compute the velocity field and composition of the escaping atmosphere. The treatment is not fully self-consistent as we treat each species as pure (i.e. they have the same speciation in both the condensed and the gas phase), however this approach still represents a significant advance compared to the state-of-the-art, as we consider a multicomponent atmosphere in dynamical flow following a moist-adiabat. Whereas the computation of a steady-state atmospheric structure taking into account a moist adiabat is common, the present work is, to our knowledge the first work that explicitly computes the composition of the atmosphere during its escaping trajectory and for a mixture of species.

\subsubsection{Calculation of the (P,T) structure}
During gas expansion leading to escape, we assume that the mixture (comprised of gas and condensates) outgassed from the Moon's surface follows a \textit{moist adiabat}, that is, the mixture cools adiabatically with condensates remaining bound to the gas (as justified in Appendix \ref{SOM_drag_droplets}), leading to a specific (P,T) or ($\rho$,T) relation that depends only on the initial entropy and composition of the mixture at the Moon's surface. Here we calculate the (P,T) and ($\rho$,T) relations for a given mixture composition and temperature at the Moon's surface. We consider a mixture including N species that can be either in gaseous or condensed form. We consider two stages during hydrodynamic escape of the gas. First, the temperature drops isentropically from that of the magma ocean, $T_S$, to some temperature, $T_0$ , with no heat exchange with the surroundings. When the gas has cooled sufficiently to reach a lower threshold temperature $T_0$, we assume that the mixture of gas and dust is heated by the Sun in order to maintain a constant temperature. $T_0$ is therefore simply assumed to be the equilibrium temperature at the location of the Earth-Moon system in the absence of greenhouse heating: 250 K. This is an approximation; in order to compute the true temperature self-consistently, we would need to couple our hydrodynamic/chemical system to a radiative transfer model which is beyond the scope of this work.
This 250 K can be considered as a lower bound for the wet model, because it neglects heating from the Moon's hot surface and from radiative heating by the Earth. The "upper bound" is given by the purely isothermal case in which $T_0$ = $T_S$. This case is described in Appendix \ref{isothermal_appendix}. During the isentropic stage (from $T_S$ down to $T_0$), the decrease of temperature leads to a variation in the partial pressures of every species, as well as phase-changes. For each species, $i$, the number of moles ($n_i$) present in one volume unit of the system at any location is $n_i=(\rho_{i,g}+\rho_{i,c})/\mu_i$ where  $\rho_{i,g}$ and $\rho_{i,c}$ are the densities of species $i$ in gaseous and condensed form. $\mu_i$ is the molar mass.

Let consider a transformation of this mixture. The total entropy variation of one volume unit is  $dS=\Sigma(dS_i)$ where $dS_i$ is the entropy variation of species $i$. Each $dS_i$ is given as \citep{Ingersoll_1969, Atmos_Thermo_1998, Bechtold_2009}:

\begin{equation}
    \text{if species $i$ is not saturated\\}
dS_i=Cp_{i,g}\rho_{i,g}\frac{dT}{T}-\rho_{i,g}\frac{R}{P_i}dP_i
    \label{Eq_ds_not-saturared}
\end{equation}

\begin{equation}
\text{if species $i$ is  saturated\\}
dS_i=Cp_{i,g}\rho_{i,g}\frac{dT}{T}-\frac{ \rho_{i,g} L_i}{T^2} dT-\frac{L_i}{T}d\rho_{i,c}+\rho_{i,c}Cp_{i,c}\frac{dT}{T}
\label{Eq_ds_saturared}
\end{equation}

where $Cp_{i,g}$  and $Cp_{i,c}$ are the heat capacity of species $i$ at constant pressure for its gaseous and condensate form respectively, and $L_i$ is the latent heat. Note that these three quantities depend on the local temperature. A species is considered as saturated if $P_i > P_{sat,i}$ where $P_{sat,i}$ is the vapour pressure of species $i$ at the local temperature T. When this condition is met, we set $P_i= P_{sat,i}$ and the density of formed condensates $d\rho_{i,c}$, is found by solving $dS=\Sigma(dS_i)=0$. Since we assume (for simplicity) that each species has the same speciation in both the condensate and the gas (neglecting solid solutions or complex solid phases) $P_{sat,i}$ only depends on the local temperature. This is a necessary simplification for the moment to make the calculation tractable when coupled to the computation of the gas-velocity field (next section).

For a given temperature change, $dT$, the new partial pressure $P_i$ and condensate density $\rho_{i,c}$ of each species $i$ is found by solving $dS=\Sigma(dS_i)=0$. To close the system of equations we assume that the mole fraction of every species $i$ is constant with T and height above the surface. We include O$_2$, Zn, Fe, Cu, SiO, Mg and Na. All thermodynamic values are taken from the python library \textit{Thermo}  (Chemical properties component of Chemical Engineering Design Library (ChEDL)
https://github.com/CalebBell/thermo). The calculation is performed as follows. A surface temperature $T_S$ and a gas surface composition are calculated in section \ref{section_core_text_evaporation_section}, in which all species are present in gaseous form (by definition) with partial pressures $P_i$ at the surface. Then temperature is incremented by a small quantity dT ($dT << 1K$ in general) and the new partial pressures and densities are found by solving Equations \ref{Eq_ds_not-saturared}, \ref{Eq_ds_saturared} simultaneously (using a Newton-Raphson method) and constraining the molar fraction of every species to be constant. With this procedure isentropic (P,T) and ($\rho$, T) curves are tabulated. A final check of entropy variations shows that the total entropy of the system changes by less than $10^{-3}$ relatively (at worst) over the course of the calculation.

In the isothermal region of the atmosphere ( when T crosses $T_0$) the gas is assumed to depressurise isothermally (due to geometrical expansion of the gas), and the new composition is simply computed by assuming that the total mole fraction of each species (including gas and condensate phases) is constant.

An example of gas transformation following the above thermodynamic path is provided in Figure \ref{Figure_PT_adiababatic}, for a mixture with the elemental composition of a vapour derived from the BSE at 1600 K (see section \ref{section_core_text_evaporation_section}). One observes that, during the isentropic cooling process from 1600 K down to 250 K, Cu and Fe are the first to condense. At $\sim 575K$ Zn condenses, then Na at about $\sim 490K$. K condenses in the isothermal regime for T = 250 K and $P<10^{-3}$ Pa. 

\subsubsection{Calculation of the velocity field}

The velocity field of the escaping atmosphere is computed as follows: the atmospheric gas (composed of different species, either in gaseous or in solid form) has a total local density $\rho(r)$, a total pressure P(r) and a local temperature T(r). Starting from an initial composition at the lunar surface at temperature $T_S$ ( Appendix \ref{section_core_text_evaporation_section}) we assume that the gas follows a moist adiabat along its escaping trajectory such that the mixture follows the P(T) relation computed in the previous section. In spherical geometry the steady-state conservation of mass reads :
\begin{equation}
    \frac{1}{r^2}\frac{\partial}{\partial r}\left( \rho v r^2\right)=0.
    \label{Eq_flux}
\end{equation}

The conservation of linear momentum gives:
\begin{equation}
    \frac{1}{r^2}\frac{\partial \rho v vr^2}{\partial r} = \rho v \frac{\partial v}{\partial r}=-\frac{\partial P}{\partial r}+\rho A(r)
    \label{Eq_momentum}
\end{equation}
The gas velocity field depends on its surface velocity $V_S$. The two above equations are solved simultaneously along the line joining the Earth-Moon's centers, while considering the $P(\rho)$ function to be isentropic. It is solved using a simple finite difference method. To find the transonic solution (the one that has zero pressure at infinity) we adopt a shooting method: we perform an iterative search on $V_S$. Starting from very low $V_S$ we progressively increase its value and determine when the the solution to Equations \ref{eq_flux} and \ref{Eq_momentum} switches abruptly from the "breeze" solution (low $v$ and finite pressure at infinity) to the transonic solution (supersonic $v$ and zero pressure at infinity). This procedures allow to find $V_S$ within $10\%$ error quite rapidly ($<$ 100 iterations). However the computation of the full velocity, pressure and temperature profiles is much more demanding. Indeed the transonic solution is the threshold solution between the four different families of solutions of the Parker Wind. To obtain a reasonable velocity profile after integration of the partial differential equations, $V_S$ must be determined to very high precision, which requires more than $10^3$ and sometime more than $10^4$ iteration steps on $V_S$.

Note that for the wet model the decrease in surface pressure relative to that at equilibrium caused by the hydrodynamic wind (see Appendix \ref{Appendix_Pressure_surface}) is not taken into account. To consider this effect properly, the iterative search on the surface velocity (to find for the transonic solution) must be also coupled to a search on the surface gas density (to take into account the density correction due to flux of material returning to the surface). This would necessitate a 2D iterative search rather than a 1D search. As the determination of surface velocity is computationally intensive (see above), this 2D search could not be conducted. This results in an overestimation of the net escaping flux of the order of the surface velocity divided by the thermal velocity (Appendix \ref{Appendix_Pressure_surface}) for the wet model. Some posterior checking reveals that for $T_s> 2000K$ and for A=Earth-Moon distance $< 3.5 R_{\oplus}$ our net flux is overestimated by less that $30\%$, and falls to less than $10\%$ for  $A > 5 R_{\oplus}$. For $T_s$ in the range of 1600-1800K (our preferred range) the flux is overestimated by less than $15\%$ and falls to less than $2\%$ for  $A > 5 R_{\oplus}$. This error is small, and compensated for, by other approximations in the development of our model. In particular, we consider that the evaporating surface of the Moon is only 1 steradian in solid angle (rather than $4\pi$ at maximum), because of lack of knowledge on the actual state of the Moon's surface.

\subsection{Results for the wet model}

The resulting atmospheric structure for a 1 lunar mass proto-Moon located at $4 R_{\oplus}$ is displayed in Figure \ref{Figure_atmo_struc} for $T_S=1600K$ and  $T_S=2000K$. While the $T_S=2000K$ velocity profile is typical of a transonic adiabatic flow with a linear increase of velocity with altitude, the $T_S=1600K$ shows more peculiar features. A strong velocity bump is visible at 1.8 $R_m$, corresponding to $T \sim 575K$. Figure \ref{Figure_PT_adiababatic} reveals that this is the effect of Zn condensation, that leads to a step dP/dT, and thus strong dP/dr (because dP/dr=(dP/dT) $\times$ (dT/dr) ), leading to strong gas acceleration. A similar, but more pronounced effect is also visible at R=2.1 $R_m$ and $T\sim 490 K$ corresponding to Na condensation, due to its being the most abundant gas species. Condensation causes a steep pressure drop that, in turn, induces an acceleration of the gas and leads to a higher flux. This process was already identified in atmospheric dynamics, where \cite{Makarieva_et_al_2013} suggest that gas condensation acts to release of some internal potential energy that then becomes available to accelerate the gas.

The gas composition as a function of altitude is displayed in Figure \ref{gas_compo_1600K_wet_model_escape}, for $T_S=1600K$ and for a proto-Moon located at 4 Earth Radii. In our model atmosphere, most of the gas species present at the surface re-condense at higher levels in the atmosphere prior to reaching the Hill Sphere (2.45 Moon radii). Only potassium and oxygen remain in gaseous form when crossing the Hill Sphere. Although condensation does occur along the \textit{moist adiabat}, they remain in the atmosphere and are nevertheless dragged outward with the gas-flow provided the grains or droplets into which they condense remain small. This is justified because, when the condensates form by momentum conservation, they do so together with the velocity vector of the surrounding gas. By inertia they continue their travel upward and do not fall immediately to the ground. Later, as they are less subject to buoyancy forces than the gas itself, they decelerate more rapidly than the surrounding gas and gas-drag may occur. Our analysis in Appendix \ref{SOM_drag_droplets} shows that the droplets stay well coupled to the gas and are efficiently dragged outward during hydrodynamic escape, provided their size is $< 1$ mm  and that the condensate mass fraction relative  $<0.5$ at the surface of the magma ocean. This is because the Moon's gravitational field drops with altitude, enabling more efficient coupling of particles with the gas at higher altitudes, despite the gas' decreasing density.

The escaping fluxes for different surface temperatures and various Earth-Moon distances are displayed in Figure \ref{fig_flux_vs_Moon_distance_wet_model}, a key result of our study. It illustrates that escaping fluxes drop with increasing Earth-Moon distance, and increase with surface temperature. Comparing this behaviour with the dry adiabatic case (Figure \ref{Figure_escape_adiab_1Mm}) shows that the fluxes for the wet model are, on average, 10- to 50 times higher. This difference is due to the release of latent heat during condensation, which provides additional energy to drive the escape. Neglecting the return flux to the magma-ocean may amount to only about a $20\%$ error at most in the calculated fluxes for a Moon < $4R_{\oplus}$. For a proto-Moon at 3.5 $R_{\oplus}$ and $T_S=1600K$ the flux is about $ 8 \simeq 10^{-5}$ Moon mass/year but drops almost to 0 beyond $6R_{\oplus}$, while for  $T_S=2000K$ it is still about 
from $ 10^{-4}$ Moon mass/year at $6R_{\oplus}$.
These fluxes may be under under-estimated considering (1) that the solid angle of the escaping flux might increase up to $4\pi$ and (2) that a partially-formed Moon, or moons building blocks with individual masses smaller than the fully grown Moon would experience more efficient atmospheric loss because of lower mass. 

Several simplifications were made in reaching this result (Appendix \ref{SOM_approximations}). Most importantly, neither radiative heating from the lunar surface nor radiative cooling were considered, as this would require the development of a full radiative transfer model with a chemical kinetics code. Note that calculation of the atmospheric optical depth (Appendix\ref{SOM_optical_depth}) shows that the atmosphere may become transparent for  $T_{\rm s} < 2000$ K. This would facilitate heating (thereby increasing the escaping flux) of the Moon's surface on its near-side by thermal emission from the Earth as well as heating and cooling of the expanding atmosphere. Figure \ref{Figure_T_escape} shows that, even for exact radiative balance of droplets in a transparent atmosphere, the minimum escaping temperature is always reached. So either the escape is fully dry (our dry model) or implies recondensation (our wet model), we find in these two scenarii that intense hydrodynamic escape of the atmosphere above a lunar magma ocean occurs. Future studies should incorporate a self-consistent Gibbs Free Energy minimisation code with a more realistic condensate mineralogy. This would notably include feldspars and olivine that are shown to be stable phases for a cooling nebular gas. That these are not considered in the present study means that the condensation behaviour of the elements in the wet model should only be taken as instructive rather than definitive.
Because of Earth-Moon tides, the proto Moon will inevitably migrate outwards, lowering the escaping flux over time, a phenomenon that is discussed in the next section.

\section{Effect of the Moon's orbital expansion}
\label{section_orbital_expansion}
The tidal evolution of the early Moon is a matter of debate. Here we consider a simplified approach, where the Moon's semi-major axis, $a$, depends only on the Earth's  $k_{2e}/Q_e$, where $k_{2e}$ is the Earth's Love Number, and $Q_e$ is the Earth tidal dissipation quality factor. Both numbers are unknown for a molten Earth after the lunar giant impact. For simplicity, we neglect the Moon's eccentricity, so that its orbital evolution is driven solely by the Earth's dissipation. We solve the time evolution of the early Moon following (see e.g. \cite{Goldreich_Soter_1966}):


\begin{equation}
  \frac{da}{dt}=\frac{3 k_{2e} M_{\rm m} R_e^5 \sqrt{G}}{Q_e \sqrt{M_\oplus a^{11}}}
   \label{Equ_tidal_evol}
\end{equation}

The resulting semi-major axis evolution of the Moon is reported in figure Fig \ref{Figure_tidal_evol}. We assume that the Moon starts at the Earth's Roche limit. Several values of $Q_e$ are explored from 10 to 1000, as in \cite{Cuk_Stewart_2012, Chen_Nimmo_2016}, noting that the present-day value is $Q_e \simeq 10 $ (\citep{Goldreich_Soter_1966}). The $Q$ of a molten Earth is not known, but is expected to have been larger than today's $Q$, because of lower internal dissipation.

As the Moon's orbit expands, so does its Hill sphere, making hydrodynamic escape increasingly difficult and thereby inducing a decline of the escaping mass flux. Atmospheric escape shuts down between 3.5 to 5 $R_\oplus$ for the dry adiabatic regime for surface temperatures between 1400 K and 2200 K (Figure \ref{Figure_escape_adiab_1Mm} and Figure \ref{Figure_escape_adiab_05Mm}). For the wet model, the escaping flux is still about $10^{-7}$ Moon masses/year, for surface temperatures between $\sim 1600$ K and $\sim 1800$ K in the same distance range. For conservative estimates of an entirely liquid Earth, its tidal dissipation factor is large ($Q>100$) and the Moon reaches $7.5 R_\oplus$ in $10^{4-5}$ years. Moon-disk interaction shortens this timescale by only 1000 years \citep{Salmon_Canup_2012}. This timescale is sufficient to allow atmospheric mass loss on the order of a percent of the Moon's mass. We note also that, as the Moon's orbit expands (or, alternatively, as the lunar surface cools down), degassing could switch to a Jeans escape rather than a hydrodynamic escape regime. Indeed the Knudsen number, $Kn$  (the molecular mean-free path divided by the atmosphere scale height; Jeans escape occurs for $Kn>1$), is found to increase as the Moon's orbit expands and temperature drops (Figure \ref{Figure_atmo_struc}, left column, third line). Therefore, while hydrodynamic escape would cease at larger distances, some residual atmospheric loss via Jeans escape, particularly for lighter molecules, may still occur. 
The escape process is, however, more efficient for an isothermal atmosphere. In the end-member case in which the atmospheric temperature is kept constant at the Moon's surface temperature is displayed in Appendix \ref{isothermal_appendix}. Here the net escape flux varies smoothly with distance and is comparable to fluxes in the adiabatic case for an Earth-Moon distance about 3$R_{\oplus}$, that is, a very high value. The absence of a radiative transfer code to treat opacity increases due to droplet formation, prevents an accurate treatment of the thermal profile at this stage. In reality, a scenario falling in between the two end-member models adiabatic (lowest flux) and isothermal (highest flux) models is likely to have occurred.

\section{Behaviour of moderately volatile elements during evaporation of the Moon }
\label{section_core_text_evaporation_section}

\subsection{Vapour pressures of metal-bearing gases above the silicate Moon}

The range of fluxes and tidal expansion timescales computed above permits evaluation of how the residual composition of the Moon evolves as a result of atmospheric escape. Given that the initial bulk composition of the Moon is likely similar to that of the bulk silicate Earth (BSE) \citep{Ringwood_1987}, and presuming that the composition of the escaping vapour is that of the vapour at the liquid-gas interface (neglecting molecular diffusion in the outgoing flux, that is negligible in a hydrodynamic wind, see \cite{Tang_Young_2020}), the effect of atmospheric loss on the composition of the Moon may be estimated.

The composition of the gas at the surface of the magma is computed using a thermodynamic approach (Appendix \ref{SOM_THERMO}) taking advantage of recent laboratory measurements of chemical activities of melt oxide species (\cite{Sossi_Fegley_2018, sossi2019}). The equilibrium partial pressure may be calculated for any gas species containing a metal, \(M\), according to the generalised congruent vaporisation reaction:

\begin{equation}
\left( M^{x+n}  O_ \frac{x+n}{2} \right) ({l})=\left( M^{x} O_\frac{x}{2} \right) ({g})+\frac{n}{4}O_2 ({g})
\label{Eq_reaction_MO_vap}
\end{equation}

where $l$ denotes the liquid phase, $g$ the gas phase, $x$ the oxidation state of the metal in the gas phase and $n$ the number of electrons exchanged (an integer value). At equilibrium, the partial vapour pressure of species $M^x O_{x/2}$ (where the melt oxide species is designated as species $i$ for short) is obtained:

\begin{equation}
p \left( M^x O_{\frac{x}{2}} \right)=\frac{K X(i) \gamma(i)}{f(O_2)^{n/4}}
\label{Eq_partial_pressure_equil}
\end{equation}
where $K$ is the equilibrium constant of reaction, $X(i)$ is the mole fraction of species $i$ (see Table \ref{table_fO2}), $\gamma(i)$ its activity coefficient and $f(O_2)$ the oxygen fugacity. The $f(O_2)$ is calculated according to the model of \cite{Visscher_Fegley_2013} for their BSE composition and lies close to the fayalite-magnetite-quartz (FMQ) buffer, as indicated by thermodynamic and experimental studies for the evaporation of silicates \citep{Costa_et_al_2017, Sossi_Fegley_2018}. Activity coefficients are calculated according to enthalpies of solution determined in experimental studies (\cite{sossi2019, charles1967}). It is emphasised that the activity coefficients of trace elements in silicate melts are accurate only to within a factor of ~2 to 3 for basaltic melts (\cite{sossi2019}). Moreover, because these compositions are similar, but not identical to those expected for a lunar magma ocean, this introduces an additional uncertainty in their application to the evaporation of lunar compositions. Therefore the associated partial pressures of these elements may be conservatively estimated to vary by up to an order of magnitude. This translates into an equivalent temperature uncertainty of about $\pm 100$ K. However, the relative partial pressures of these elements are more robust as systematic errors in the measurement of activity coefficients cancel. Therefore, the total mass loss from the Moon required to account for elemental depletion is less well constrained than are the conditions leading to the \textit{relative} chemical fractionation between them. Thus, element ratios of Na, K and Zn are more exacting than are their abundances for constraining the conditions of volatile loss. Keeping these caveats in mind, the partial pressures of the gas species are computed for all elements for which laboratory measurements in lunar rocks are available. The resulting partial pressures of the major gas species are plotted in Figure \ref{Figure_partial_pressure}, and show that Na dominates the composition of the vapour phase with Zn, Cu and K as minor metal-bearing gas species. The major components, Mg, Si and Fe, despite their high mole fractions in the BSE, constitute a relatively smaller fraction of the gas phase owing to their low volatilities over the temperature range 1500 to 2000 K  (\cite{Sossi_Fegley_2018}). The total pressure obtained (solid black line) is very close to the one provided by \cite{Visscher_Fegley_2013} (dashed line) using their MAGMA code.

 \subsection{Influence of evaporation on the Na, K and Zn abundances in the Moon}
 \label{SOM_Evaporation_K_Zn}
Highly- and moderately volatile elements are particularly exacting records of lunar volatile loss, because they still remain in abundances that are measurable in lunar rocks \citep{Wolf_Anders_1980}. Such volatile metals in the Moon define two distinct plateaus relative to the BSE (Figure \ref{Fig_Earth_Moon_abundances}). The moderately volatile elements (including the alkalis and Ga) show only modest depletion (a factor of 5) whereas highly volatile elements in the Moon (e.g. Zn, Cd, In, Bi) are depleted by a factor of ~50, and, importantly, define constant chondrite-normalised abundances, leading some authors \citep{ONeil_1991,Taylor_et_al_2006,Wolf_Anders_1980} to suggest that they were accreted later than the alkali metals, by addition of chondritic material. However, the heavy stable isotopic composition of Zn and K attest instead to their evaporative loss at an early stage during lunar formation \citep{Wang_Jacobsen_2016, Paniello_2012}. The alkali metals and Zn, in particular, are unequivocal probes of vapour loss because they do not appreciably partition into the metallic phase during lunar core formation. Indeed, the chemical similarities among alkali metals have been used to constrain likely conditions of volatile loss from the Moon \citep{Kreutzberger_et_al_1986}. \\

Although the temperature of evaporation is often treated as a free parameter in lunar formation models \citep{Canup_2015, Lock_2018}, the similarity of the abundance and isotopic composition of lithium in lunar mare basalts relative to terrestrial basalts, in contrast to the near-quantitative loss of Na in lunar rocks, sets limits on the likely  temperatures under which volatile depletion occurred \citep{ONeil_1991, Taylor_et_al_2006, Magna_et_al_2006}.
 In order to vaporise 80 percent of the Na whilst preserving 90 percent of the Moon's Li budget requires temperatures below 1800 K, considering estimates for the activity coefficients of these two elements in silicate melts \citep{Sossi_Fegley_2018, ONeil_1991}. These temperatures are in good agreement with those deduced from Cr isotope systematics between lunar and terrestrial basalts. The light composition of the former require temperatures $>1600$ K (to vaporise sufficient Cr) but $<1800$ K to produce sufficient isotope fractionation \citep{Sossi_Moynier_VanZuilen_2018_} where near-equilibrium conditions had to have prevailed during vaporisation. In addition, the Li/Na and Cr constraints yield similar temperatures to those obtained by thermal models of the near-surface melt  (Tang and Young 2020).

Here we exploit calculations of equilibrium partial pressures to determine whether the observed relative depletion factors between Na, K and Zn are consistent with a hydrodynamic escape scenario over a range of temperatures between 1200 and 2200 K. Considering a conservative possible range of lunar abundances for Na, K, Zn relative to the BSE from Figure \ref{fig_visscher}, reported in Table \ref{Table_minmax_K_NA_ZN}, we investigate the temperature at which the modelled abundances are consistent with observations.

\begin{table}[h]
\begin{tabular} {|l|l|l|l|}
  \hline
  \textbf{Moon/BSE abundance}  & \textbf{Na} & \textbf{K} & \textbf{Zn}  \\
  \hline
  \textbf{Maximum} & 0.63   &  0.8 & 0.093  \\
    \hline
  \textbf{Minimum} & 0.11   & 0.11  &  0.02 \\
  \hline

\end{tabular} \\
\caption{Minimum and maximum values for the mass fraction of each element measured in the silicated Moon divided by that in the silicate Earth ($f$ in the main text). Values extracted from Figure \ref{fig_visscher} \citep{Visscher_Fegley_2013}. In Figures \ref{Figure_vaporize_to_Na}, \ref{Figure_vaporize_to_K} and \ref{Figure_vaporize_to_Zn} we show that we can match simultaneously the range of measurements for Na and K, but never for Zn}
\label{Table_minmax_K_NA_ZN}
\end{table}

Given the mixing ratios of these three elements in the atmosphere, it is possible to uniquely determine the mass of vapour that must have escaped from the Moon, $M_v^L$, at a temperature T, in order to match the  abundance of a given element in lunar mare basalts (and by extension the bulk Moon).  Due to the simplicity of the atmospheric structure assumed here, these calculations are designed to provide first-order estimates. Consider a element, whose Lunar abundance / BSE abundance is denoted, $f$, then, the mass that must be evaporated and subsequently lost from the Moon is computed as follows (see derivation in Appendix \ref{SOM_escaped_mass}):

  \begin{equation}
      M_v^L=\frac{(f-1)M_M}{1-\mu_i^v/\mu_i^E}
     \label{eq_mass_removed}
  \end{equation}
 where $M_M$ is the mass of the Moon, $\mu_i^v$ and $\mu_i^E$ are the mass fractions of element $i$ in the vapour phase and in the BSE respectively.
 We assume that the escaping gas has the composition of the vapour in equilibrium with the magma at the lunar surface (Figure \ref{Figure_partial_pressure}) at temperature T. Assuming that the proto-Moon starts with the composition of the BSE we consider two modes of evaporation:

 \begin{itemize}
     \item \textit{Equilibrium evaporation}, the magma vapour composition does not change as material is removed, so $\mu_i^v$ is constant with time in Equation. \ref{eq_mass_removed}.
     \item \textit{Fractional evaporation}: the vapour composition changes progressively as vapour is removed from the system at steps of $10^{-6}$ lunar masses per increment. After each step of vapour removal, the compositions of the magma and the vapour are re-calculated.
 \end{itemize}

Applying the two procedures described above, the Na (Figure \ref{Figure_vaporize_to_Na}), K (Figure \ref{Figure_vaporize_to_K}), and Zn abundances (Figure \ref{Figure_vaporize_to_Zn}) are treated successively as target quantities, with the other two elements calculated at the temperature listed in the figure caption.

The total mass of material that must escape from the Moon in order to match the mean abundance of K measured in lunar mare basalts is in listed Table \ref{table_evap_K} for different surface temperatures. The same calculation is presented but for Na in table \ref{table_evap_Na}. For a temperature around 2000 K, both abundances could be matched almost simultaneously. However, there is a range of measured values for K and Na that allows a larger range of possible temperatures matching both abundances, reported as blue boxes in  \ref{Figure_vaporize_to_Na} and \ref{Figure_vaporize_to_K}. In these figures we see that it is possible to match both Na and K composition (within measured ranges) via either equilibrium or fractional evaporation between 1600 K and 2000 K. Inspection of Tables \ref{table_evap_K} and \ref{table_evap_Na} shows that it requires from $0.1\%$ to $0.6\%$ of a lunar mass to evaporate to reach the observed depletion. For a small escaping surface flux $\sim 10^{-6}$ lunar mass/year (Figure \ref{fig_flux_vs_Moon_distance_wet_model}), this requires between $\sim 1000$ and $\sim 6000 $ years. For an  escaping surface flux $\sim 10^{-4}$ lunar mass/year, these timescales are reduced by a factor of 100.

However all three of Na, K and Zn are never adequately fit simultaneously because Zn is quantitatively vaporised for even minor evaporation of Na and K, in disagreement with their abundances in lunar basalts. Possible solutions to this conundrum may be deviations in activity coefficients of these elements relative to those measured in experimental studies. Alternatively, evaporation occurred at higher oxygen fugacities than predicted according to \cite{Visscher_Fegley_2013}.  Higher \(fO_2\) promotes alkali vaporisation relative to Zn because Zn vaporises according to an $n$ = 2 stoichiometry compared to $n$ = 1 for the alkali metals (Equation \ref{Eq_reaction_MO_vap}). A similar exercise performed with a greater suite of elements, combined with a more detailed atmospheric model would further aid in constraining the conditions of vapour loss from the Moon, but is beyond the scope of this work.

\begin{table*}
\begin{tabular} {|l|l|l|l|l|}
  \hline
  \textbf{Temperature (K)}  &  \textbf{1600} & \textbf{1800} & \textbf{2000}  & \textbf{2200} \\
  \hline
  \textbf{Lunar mass fraction evaporated}  &  0.088 \% & 0.197 \% & 0.33\% & 0.55 \%    \\
  \textbf{for equilibrium evaporation}      & & &  &  \\
    \hline
  \textbf{Lunar mass fraction evaporated} &  0.083 \% & 0.115 \% & 0.154\% & 0.223 \%    \\
  \textbf{for fractionated evaporation/}      & & &  &  \\
  \hline

\end{tabular} \\
\caption{Mass that must be evaporated to reproduce the mean lunar K abundance, for different temperatures of evaporation.}
\label{table_evap_K}
\end{table*}

\begin{table*}
\begin{tabular} {|l|l|l|l|l|}
  \hline
  \textbf{Temperature (K)}  &  \textbf{1600} & \textbf{1800} & \textbf{2000}  & \textbf{2200} \\
  \hline
  \textbf{Lunar mass fraction evaporated}  &  0.33 \% & 0.48 \% & 0.58\% & 0.73 \%    \\
  \textbf{for equilibrium evaporation}      & & &  &  \\
    \hline
  \textbf{Lunar mass fraction evaporated} &  0.24 \% & 0.23 \% & 0.24\% & 0.27 \%    \\
  \textbf{for fractionated evaporation/}      & & &  &  \\
  \hline

\end{tabular} \\
\caption{Mass that must be evaporated to reproduce the mean lunar Na abundance, for different temperatures of evaporation.}
\label{table_evap_Na}
\end{table*}

 \subsection{Isotopic fractionation during hydrodynamic escape}
 
 The observation that the stable isotope compositions of moderately volatile elements, such as K, Zn and Cr, are distinct in lunar mare basalts compared to their terrestrial equivalents \cite{Wang_Jacobsen_2016,Paniello_2012,Sossi_Moynier_VanZuilen_2018_} may be used to place constraints on the potential conditions under which volatile loss occurred from the Moon. While K and Zn isotopes are heavier in the Moon than in the Earth, Cr isotopes are lighter, implying that, should this variation be due to volatile loss, the vapour reached equilibrium with the lunar surface. Moreover, there could not have been any process in the atmosphere capable of separating isotopes from one another prior to escape. For atmospheric loss driven by a hydrodynamic wind, as detailed here, there is a bulk outflow of mass in which no mass separation can occur in the atmosphere. Therefore, no isotopic fractionation results during this process.\\
 
 The other possible locus at which isotope fractionation could develop is at the magma ocean-atmosphere interface. In section \ref{vs_vt}, the effect of a non-stationary gas on the pressure at the surface of the magma ocean relative to the equilibrium partial pressure ($P_s/P_{sat}$) was quantified, and shown to be \textless 1. This quantity is important in determining the degree of isotope fractionation that could result during the evaporation process, through the relation (\cite{Richter_et_al_2002}):
 
\begin{equation}
^{i/j}M\alpha_{gas-liq, net} = ^{i/j}M\alpha_{gas-liq, eq}+ ((M_j/M_i)^{0.5}-1)(1-P_s/P_{sat}))
\label{Eq_isotopes}
\end{equation}

Where $i$ and $j$ are the two isotopes of element with molar masses $M$, and for the Moon, liq = lunar magma and gas = lunar atmosphere, the subscripts 'eq' and 'net' denote the equilibrium isotope fractionation and net isotopic fractionation, respectively. As the former is proportional to 1/T$^2$ and temperatures are relatively high ($ > 1000 K$), $\alpha_{gas-liq, eq}$ is expected to be of the order of ~ 0.9999 for vaporisation reactions between minerals and monatomic gases at 1500 K (see \citep{Sossi_et_al_2020} and references therein). However, precise values for the relevant melt-oxide species of K and Zn and gas remain uncalibrated, precluding an accurate estimate of $\alpha_{gas-liq, eq}$ at present. This expression for calculating the fractionation factor neglects the effect of diffusion, because it is not the predominant mode of mass transport during hydrodynamic escape of the protolunar atmosphere, rather, it is advection. That is, the Peclet number of the flow is \textgreater \textgreater 1 and diffusion is unimportant. As such, the lower the value of $P_s/P_{sat}$, the larger the fractionation factor between vapour and liquid. The apparent fractionation factors between vapour and liquid deduced for Zn and K in lunar basalts with respect to the Earth's mantle can be used to place limits on $P_s/P_{sat}$. These values are obtained by assuming a Rayleigh fractionation process during evaporation and using the measured \textit{f} of Zn and K in the bulk Moon together with their isotopic compositions (see \citep{Sossi_et_al_2020}, their Table 4), assuming initial values reflected those of the present-day BSE. Substituting in the values obtained by this calculation, $(^{66/64}Zn)\alpha_{gas-liq}$ = 0.9996 $\pm{0.0001}$ and $(^{41/39}K)\alpha_{gas-liq}$ = 0.9998 $\pm{0.0001}$ for evaporation from the Moon into \ref{Eq_isotopes} yields values of $P_s/P_{sat}$ of 0.98$\pm0.01$ for Zn and 0.99$\pm0.01$ for K, both illustrating the necessity of near-equilibrium conditions to have prevailed between the liquid and the atmosphere. Inspection of Fig.\ref{Figure_Vsurface_adiab_1Mm} shows that this precludes scenarios in which $V_s/V_t$ $>$ 0.03 (see also \ref{equation_rhos_rhosat}) that are favoured during high temperatures and/or for low Earth-Moon distances. Thus, isotopic constraints require relatively low temperatures at moderate Earth-Moon distances such that the advective velocity does not exceed 1 \% of the gas thermal velocity.  This configuration is indeed achieved for T in the range 1600-1700 K and for Earth-Moon distances beyond $\sim 3.7 R_{\oplus}$ (Figure \ref{Figure_Vsurface_adiab_1Mm}) for the wet model. This combination of temperatures, $V_s/V_t$ values and Earth-Moon distances constrain maximum loss rates to be $\leq 10^{-5}$ (wet) or $\leq 10^{-7}$ (dry) lunar masses/yr, generally exceeding those calculated by \citep{Tang_Young_2020}, $7\times 10^{-8}$ lunar masses/yr.

\section{Implications for the early evolution of the Moon}
\label{section_discussion_conclusion}
Combining the low temperature range implied by isotopic data with best fits to the observed depletion factors of Na, K and Zn at $T = 1600 - 1800 K$, we find that between 0.1\% and 0.6\% of a lunar mass must have been evaporated and lost to explain its present composition.  Whereas higher surface temperatures promote higher escaping fluxes, Moon orbital expansion leads to a  decrease of flux distance. The rate of lunar orbital expansion increases with decreasing tidal $Q$ parameter, which is unknown for a molten Moon. So to explain the observed K and Na abundances in the Moon, both surface temperatures and Q must be sufficiently high to allow escape for an extended period of time. Solidification of the magma's ocean surface also acts to dampen escape, as its solidification may occur in about $\sim 1 Kyr$. To consider each possibility simultaneously, we have performed coupled tidal-evolution (with different Q) and evaporation simulations (using the wet model with constant surface temperature). In a first set of runs we assume that the magma ocean never solidifies, and that only the effect of the increasing Earth-Moon distance with time arrests hydrodynamic escape. The simulation ran for $10^5$ years, so that the tidal evolution of the Moon would always result in a distance sufficiently far from the Earth to shut down hydrodynamic escape. The final abundance of Na and K resulting from equilibrium or fractional evaporation in the Moon at the end of the run are then compared to observations.
We find that the only combinations of (T,Q) that can simultaneously match lunar K and Na range from (T=1580K, Q=215) to (T=1690K, Q=46) for fractional vaporisation, whereas equilibrium vaporisation never leads to a satisfactory match within the allowed ranges. Within these ranges of T and Q, the resulting lunar abundances range from $\sim$ 250  to $400 \mu g/g$ for Na and from  $\sim$ 30  to $50 \mu g/g$ for K. All other runs results either in too much or too little mass loss of Na, or K or both.  

Finally, we consider an alternative case where the formation of a stagnant lid at the surface of the magma ocean abruptly shuts down escape. Indeed, while it may take about 200 Myrs  for the complete solidification of the lunar magma ocean \citep{Meyer_et_al_2010, Elkins-Tanton_ET_AL_2011}, crystallisation of a plagioclase lid may occur over much shorter timescales, around 1000 years \citep{Elkins-Tanton_ET_AL_2011}.
Using our simulation with constant surface temperature, we compute the time at which the stagnant lid must form for resulting in Na and K abundances compatible with lunar measurement (Figure \ref{fig_stagnant_lid_timescale}). We find a wider range of solution in terms of Q, but still a narrow temperature range. Keeping only solutions for which the stagnant lid forms \textgreater 500 years and \textless 3000 years \citep{Meyer_et_al_2010, Elkins-Tanton_ET_AL_2011, Tang_Young_2020} we find $100<Q<1000$ and $1600<T_{surf}<1700$.  The remainder of mantle cooling under a conductive anorthositic lid may take $> 100$ Myrs because of Earth-Moon tidal heating \citep{Meyer_et_al_2010, Elkins-Tanton_ET_AL_2011} by which time the Moon would have migrated beyond any proximal $R_\oplus$ values, rendering loss by hydrodynamic escape ineffective.

In summary, independent of whether the Moon's expansion or the formation of stagnant lid induce a shuts down in hydrodynamic escape,  we find that the surface temperature must be close to $1600-1700 K$ to explain the Na and K abundances, with $40 < Q < 300$. Interestingly, this range of temperatures is also compatible with the chromium isotopic fractionation observed in lunar basalts, if such an isotopic signature were due to the loss of $CrO_{2}$ vapour in equilibrium with the melt \cite{Sossi_Moynier_VanZuilen_2018_}. It is therefore remarkable that a variety of independent observations and constraints (isotopic fractionation, amount of mass evaporated to reproduce K and Na, formation timescales of the stagnant lid and the tidal Q factor) all converge upon the conclusion that the surface temperature of during evaporative loss was in the range 1600-1700K. The ostensible coherence of these results may indicate that tidally-driven hydrodynamic escape is a key process governing the loss of volatile elements from the Moon.

\section{Differences with previous work and conclusions}

Our calculated lunar loss rates, between $10^{-5}$ to $10^{-7}$ lunar masses/yr, are on the upper end- or higher than those presented in \cite{Tang_Young_2020} (TY20 hereafter), $7\times 10^{-8}$ lunar masses/year. On this basis, TY20 claim that hydrodynamic escape from the early Moon is insufficient to account for neither the chemical depletion of moderately volatile elements, nor their isotopic fractionation in the Moon. The discrepancy in loss rates arises due to the very different physical descriptions of the atmospheric structure imposed. In the present work, the escaping atmosphere is described as a single layer, globally accelerated upward. In detail, the gas begins to accelerate away from just above the liquid surface (z = 0), with an upward-directed velocity at z = 0. The atmospheric structure is then calculated self-consistently using Euler's hydrodynamic equation, but neglecting radiative transfer to make the computation tractable. \\
In contrast, in TY20, the atmospheric structure is assumed to be stratified just above the surface, inspired by results on hydrostatic atmospheres (\cite{Lebrun_et_al_2013}). In their case the atmospheric structure is not resolved as a single entity, rather, escaping fluxes are computed for each layer individually and balanced to determine a steady-state. Just above the magma ocean there is a hydrostatic layer, separated from an isothermal, hydrodynamically-escaping (i.e., non-hydrostatic) layer above it by the homopause. In TY20 hydrodynamic escape takes places \textit{above} the hydrostatic layer, which means that the advective flow is zero at the surface of the magma ocean. 
TY20 state (correctly) that the net escaping flux at the transonic point ($F_E$)  must be equal to the free evaporative flux at the liquid's surface ($F_F$) minus the return flux to the liquid ($F_R$) (see our Appendix \ref{Appendix_Pressure_surface} for more details). To compute the return flux down to magma ocean they use the formalism of R02 (their Equation 6 precisely). It describes the transport of the evaporating species by molecular diffusion within a globally static surrounding gas (the R02 equation is simply Fick's law in spherical coordinates). In the R02 approach the net evaporative flux is inversely proportional to the gas pressure, meaning transport is more efficient when the pressure is low. When the diffusive transport flux from the hydrostatic layer is fast enough (proportional to 1/P) to balance the escape flux by hydrodynamic escape at the transonic point (proportional to P) the 'cross-over pressure' at the homopause (which is at a height above the melt surface given by that at which the adiabatic temperature in the diffusive layer declines to the skin temperature calculated by $2^{-Cp/(4R)}$) is obtained, which is equal to $10^{-8}$ bar in TY20. This value is far lower than that calculated for a BSE-derived gas at 2000 K ($10^{-3}$ bar) and hence results in a significantly lower escaping flux. \\
Our hypothesis differs in that we assume that the atmosphere starts to accelerate at the magma ocean surface (there is no hydrostatic layer), so that, in our study, the vapour is advected rather than diffused in the thin layer just above the magma ocean.  Advective transport allows the density of the evaporating species to remain high because there is no need for a concentration gradient to sustain the flux - the flux is simply sustained by inertia. Our return flux, which is computed while taking advection into account, is detailed in Appendix \ref{Appendix_Pressure_surface} by using an approach very similar to the original Hertz-Knudsen calculation. The effective pressure at the surface is given by Equation \ref{equation_Ps_Psat}, and only depends on the ratio of the velocity at the surface to the thermal velocity ($V_s/V_t$). We show that when the surface velocity is small compared to the thermal velocity, the pressure above the surface remains very close to $P_{sat}$ (about $10^{-3}$ bar, depending on temperature, see \ref{Figure_partial_pressure}) while \textit{allowing a substantial net escaping flux}. Thus, the surface pressure is coupled to the velocity at the surface through Equation \ref{equation_Ps_Psat}.
The pressure at the transonic point is then obtained by solving Euler's hydrodynamic equations (as we do in both the 'wet' and 'dry' models) that describe advective transport. In our dry adiabatic model, these two equations are fully coupled, and the flux is perfectly conserved (including the return flux to the magma ocean). In our wet model this coupling could not be done because of computer limitations, but we show that the error on the flux is small ($20\%$ in the worst case, and $<1\%$ in general) because of small surface velocities compared to thermal velocities. At the transonic point, we calculate pressures between $10^{-2} $ and $10^{2}$ Pa ($10^{-7} $ and $10^{-3}$ bar), that are much larger than $10^{-7}$ Pa ($10^{-12}$ bar) obtained in TY20, which results in a much higher escaping flux. Nevertheless, solutions that are consistent with the lack of isotopic fractionation of moderately volatile elements in the Moon require near-equilibrium values of $P_{s}$ at the surface, and relatively low temperatures, such that total loss rates were likely no higher than $10^{-7}$ lunar masses/yr for the 'dry' model and $10^{-5}$ lunar masses/yr for the 'wet' model. Such loss rates, combined with the total lunar mass loss required to explain the Na and K depletion (0.1 to 0.6 $\%$) suggest that hydrodynamic escape in the tidal field of the Earth \textit{is} a plausible mechanism for atmospheric loss from the early Moon over $<1$ kyr timescales. \\

The main uncertainty in the present work is the temperature profile of the escaping gas. In order to answer this question, the next step would be to couple a model for radiative transfer and a physical model for droplet growth to the existing chemical and hydrodynamic model. In the current work, we have considered two end-member temperature profiles: adiabatic and isothermal, with or without condensation (wet vs. dry). We have shown that in all cases considered, even for the "coolest" model (adiabatic expansion for T \textgreater 250 K and isothermal for T \textless 250 K) hydrodynamic escape always occurs (it is stronger in the isothermal case than in the adiabatic case).

Assuming that the acceleration of the atmosphere starts at the surface of the magma ocean, we conclude that tidally-assisted hydrodynamic escape from an evaporating proto-Moon is an unavoidable outcome of lunar assembly and can explain the measured lunar abundances of K and Na (at least). However, under the oxygen fugacities presumed here, for any reasonable losses of Na and K, Zn is entirely evaporated, indicating either different accretion histories for the moderately- and highly-volatile elements, or different thermodynamic conditions than those modelled here. Although several simplifications were made in order to understand the magnitude and mechanisms of hydrodynamic escape from the Moon, that tides lower the energy required for thermal escape according to the proximity of the Moon from the Earth's Roche Limit is founded on robust physical principles. The present paper is a first attempt to quantify this process. It is an independent and complementary process to devolatilisation through incomplete condensation occurring within a protolunar disk (developed in recent lunar accretion models \citep{Thompson_Stevenson_1988, Canup_2015, Charnoz_Michaut_2015, Lock_2018}). Our study should encourage the development of more refined models in the future, including radiative transfer with full coupling with chemistry and magma ocean model, to precisely quantify tidally-assisted Moon devolatilisation and to better asses the thermal structure of the escaping atmosphere.




\section{Acknowledgements}
SC acknowledges the support from IPGP and by the UnivEarthS Labex program of Sorbonne Paris Cit\'e, project ExoAtmos (ANR-10-LABX-0023, ANR-11-IDEX-0005- 02) and by IdEx Université de Paris ANR-18-IDEX-0001. PAS was supported by the European Research Council under the H2020 framework program/ERC grant agreement 637503 (PRISTINE) at IPGP and by an SNF Ambizione Fellowship (180025) at ETH Zürich. YNL acknowledges funding from the Einstein young scholar fellowship program (MOST 108-2636-M-003-001) and the MOE Yushan young scholar program. R.H. acknowledges the financial support of JSPS Grants-in-Aid (JP17J01269, 18K13600). We also would like to thank E. Marcq for useful discussions about modelisation of multi-species adiabatic atmosphere, P. Tremblin for discussions on radiative transfer, as well as our reviewers (B. Fegley and E. Young) for their thoughtful comments that improved the quality of this paper.

\bibliography{bibliography}
\bibliographystyle{plainnat}


 \newpage
 \section{Figures}
\begin{figure}[h]
\begin{center}
   \includegraphics[width=.9\linewidth]{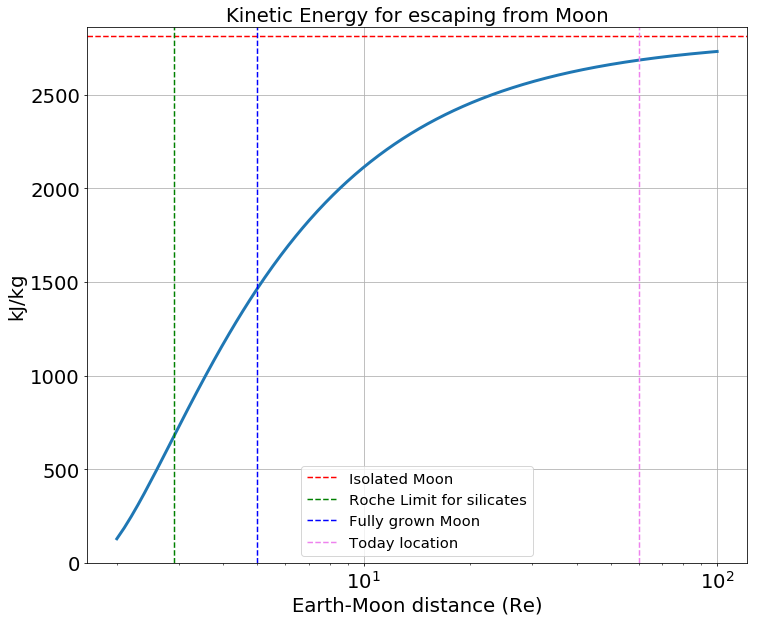}
   \caption{Blue solid line : Kinetic energy per mass unit necessary for escaping Moon's surface, as a function of the Earth-Moon distance in units of the Earth's radius (Appendix \ref{SOM_potential_energy}). Red dashed : Escape kinetic energy when the Moon is isolated. Green, blue and violet dashed: locations at 3,5, and 60 $R_\oplus$ respectively}
    \label{Figure_energy}
\end{center}
\end{figure}

\newpage

\begin{center}
   \begin{figure}[h]
   \includegraphics[width=.5\linewidth]{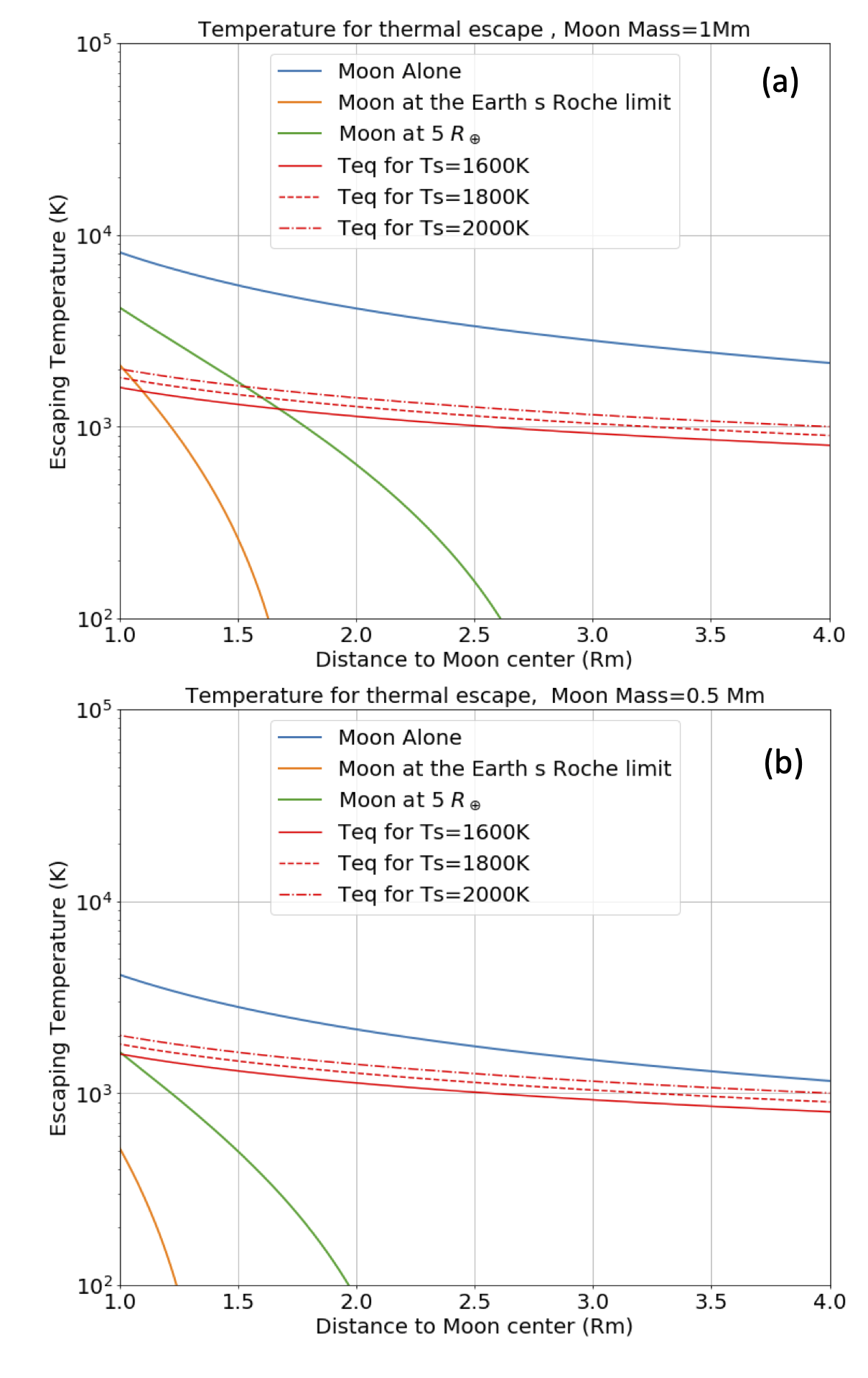}
   \caption{Gas temperature needed for escaping the Moon's gravitational field, as a function of the distance to the Moon's center and for different Earth-Moon distances. Here a simple model of a static atmosphere is used, comparing the gas thermal velocity to the escape velocity. Blue: Moon at infinity from the Earth; Orange: Moon orbiting the Earth at the Roche Limit. a) the proto-Moon with 1 lunar mass; b) the proto-Moon with 0.5 lunar masses. Red lines display the equilibrium temperature of a static atmospheric gas, assuming that it is heated by the Moon and cooled by black-body radiation for different lunar surface temperatures (dot-dashed: 2000 K, dashed : 1800 K, solid: 1600 K ). When the proto-Moon is sufficiently close to the Earth ($\lesssim 5 R_\oplus$), the gas temperature (red) exceed the escape temperature.
   }
   \label{Figure_T_escape}
   \end{figure}
\end{center}

\newpage

\begin{center}
\begin{figure}[h]
   \includegraphics[width=1.\linewidth]{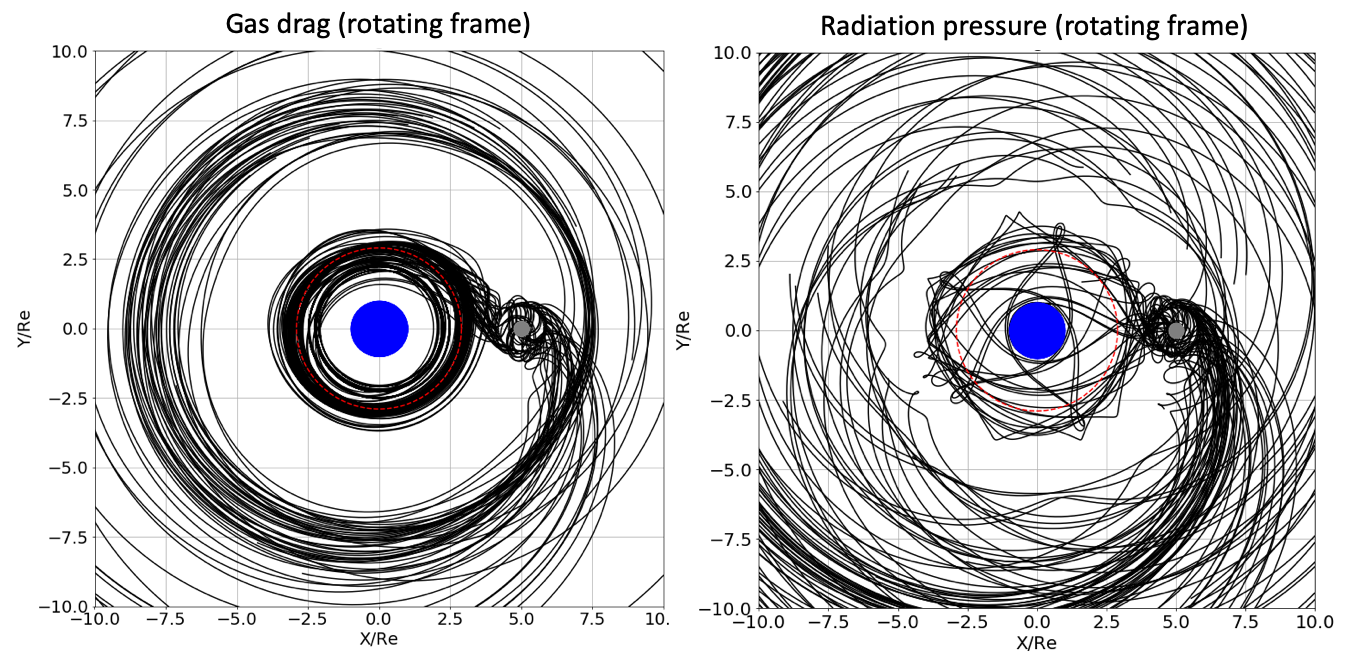}
   \caption{Examples of orbital paths of particles launched from the Moon's Hill sphere (black) displayed in the frame rotating with the Moon (grey disk) and centered on the Earth (blue disk).  The X and Y axes are in units of the Earth's radius. The Moon is located at $5 R_\oplus$, red-dashed circle shows the Earth Roche Limit. (Left) Particles evolve under the Earth and Moon's gravity and experience a drag force with a fictitious gaseous disk around Earth. The gas-drag coupling-time is set to 5 orbits (as an example). Notice that particles do not return to the Moon after escaping and instead remain in orbit around the Earth. (right) Orbits of particles experiencing the Earth and Moon gravity- and radiation pressure forces (with $\beta_{R}=0.2$ (assuming an Earth's surface at 2300 K and a dust particle density around $3~ {\rm g/cm}^3$), as an example, see Appendix  \ref{SOM_RAD_PRESSURE} . $\beta_{R}$   is the ratio of the radiation force to the Earth's acceleration). In the gas drag case, orbits are rapidly circularized (left), whereas they are not in the radiation pressure-only case (right). Nevertheless, in both cases, once a particle has left the Moon there is no active mechanism for its return. Cylindrical frame representations of these trajectories, superimposed on a potential energy map are displayed in Figures \ref{figure_potential_gas_drag} and  \ref{figure_potential_rad_press}.}
    \label{fig_orbits}
\end{figure}
\end{center}

\newpage

 \begin{figure} [h]
 \begin{center}
   \includegraphics[width=.9\linewidth]{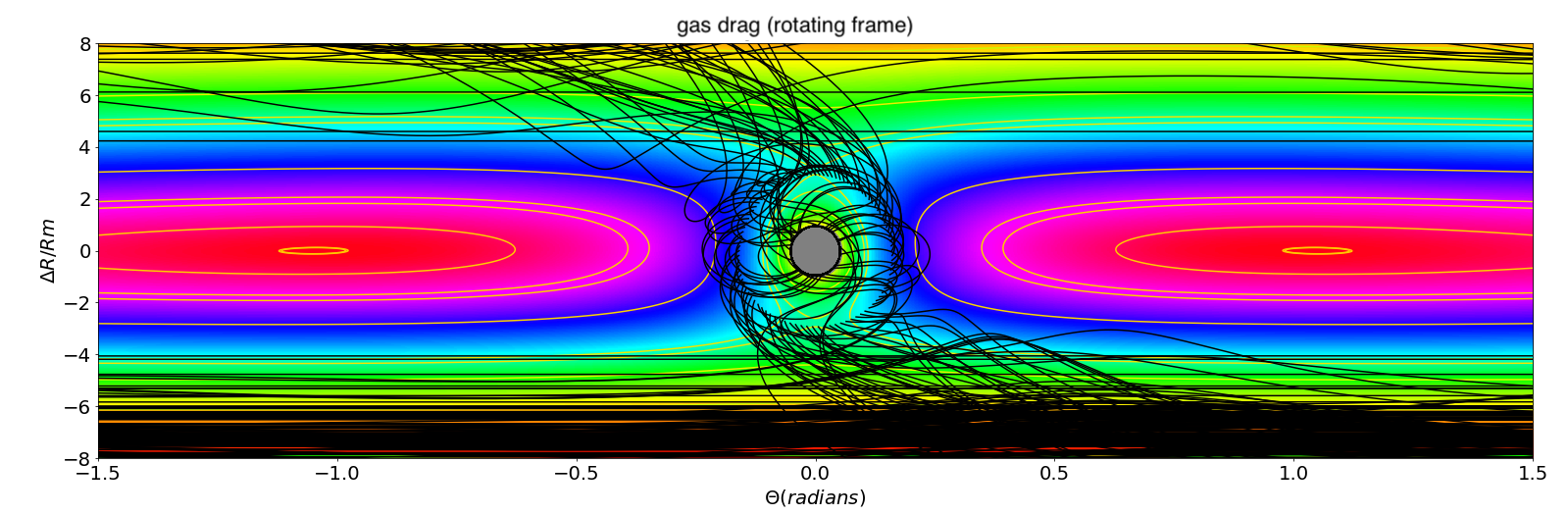}
   \caption{Potential map (colors) with over-imposed examples particles trajectories (black solid lines). The Moon is the grey disk and is located at $5 R_\oplus$. Here the particles suffer gas-drag, see legend of Figure \ref{fig_orbits}.  }
   \label{figure_potential_gas_drag}
\end{center}
\end{figure}

\newpage

\begin{figure} [h]
\begin{center}
  \includegraphics[width=.9\linewidth]{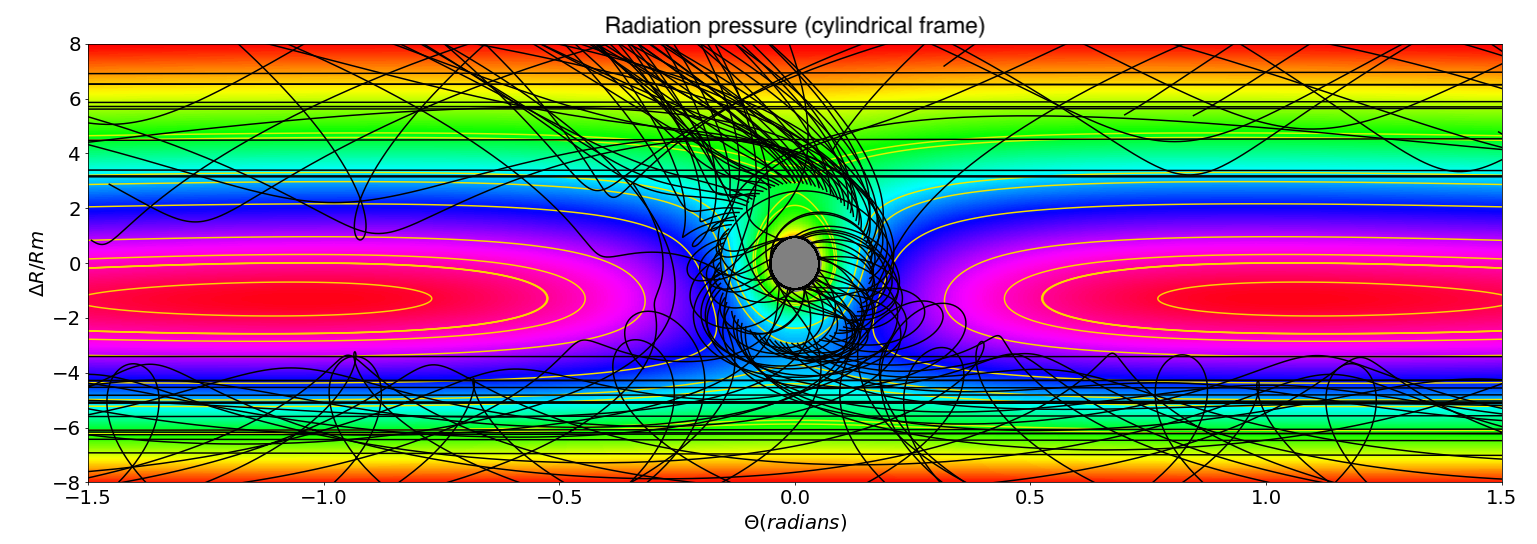}
  \caption{Potential map (colors) with overimposed examples particles trajectories (black solid lines). The Moon is the grey disk and is located at $5 R_\oplus$. Here the particles suffer radiation pressure with $\beta=0.2$, see legend of Figure \ref{fig_orbits}. Here the potential map was modified to take into account the effect of radiation pressure by multiplying the Earth's gravitational potential by $(1-\beta)$.  }
  \label{figure_potential_rad_press}
\end{center}
\end{figure}

\newpage

\begin{figure}[h]
\begin{center}
   \includegraphics[width=.9\linewidth]{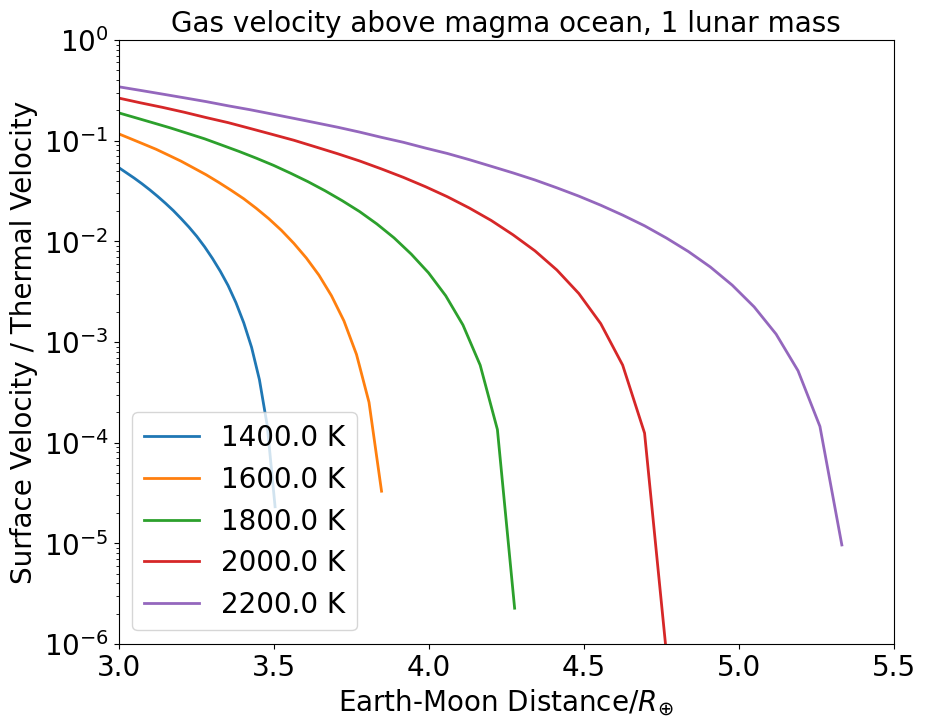}
   \caption{Gas velocity above the magma ocean (in units of thermal velocity), as a function of the Earth-Moon distance (in units of Earth radii) and for different surface temperatures (coloured curves). Here the proto-Moon mass is set to 1 lunar mass.}
    \label{Figure_Vsurface_adiab_1Mm}
\end{center}
\end{figure}

\newpage

\begin{figure}[h]
\begin{center}
   \includegraphics[width=.9\linewidth]{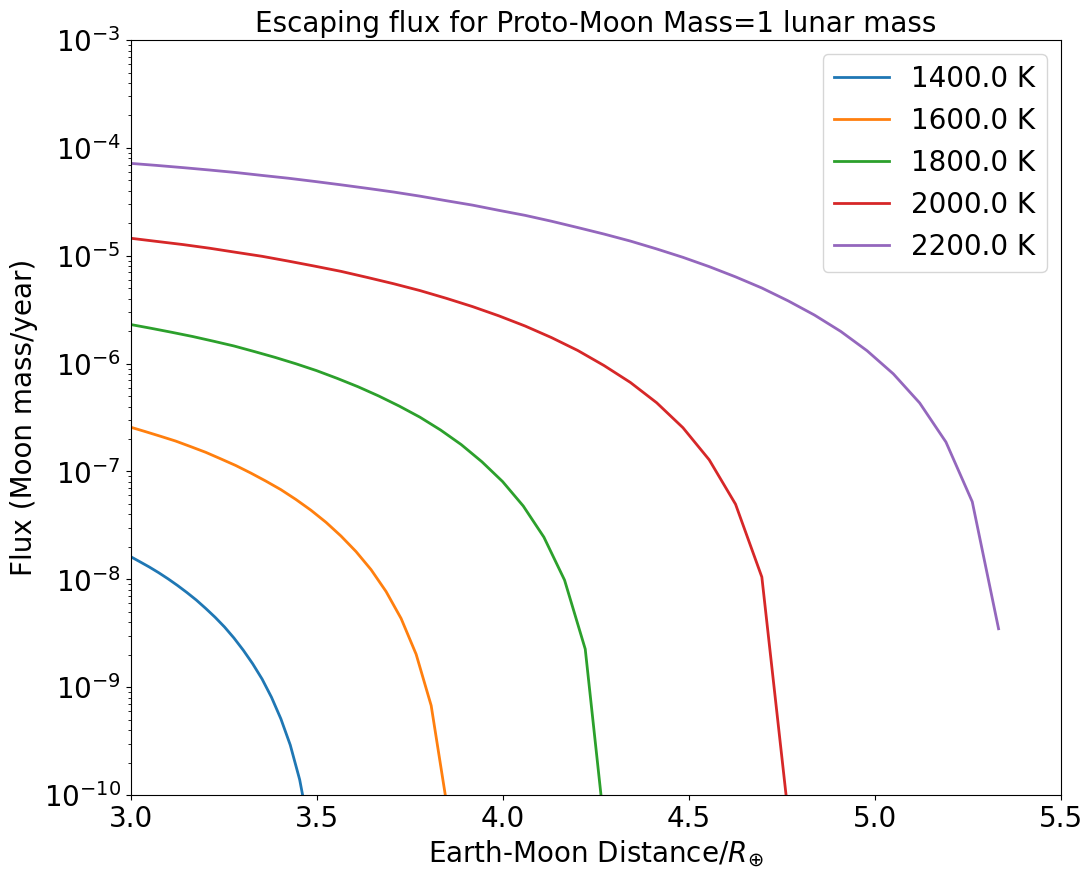}
   \caption{Surface flux ($F=\rho_sV_sR_{\rm m}^2$) as a function of the Earth-Moon distance (in units of Earth radii) and for different surface temperatures (coloured curves). Here the proto-Moon mass is set to 1 lunar mass.}
    \label{Figure_escape_adiab_1Mm}
\end{center}
\end{figure}

\newpage

\begin{figure}[h]
\begin{center}
   \includegraphics[width=.9\linewidth]{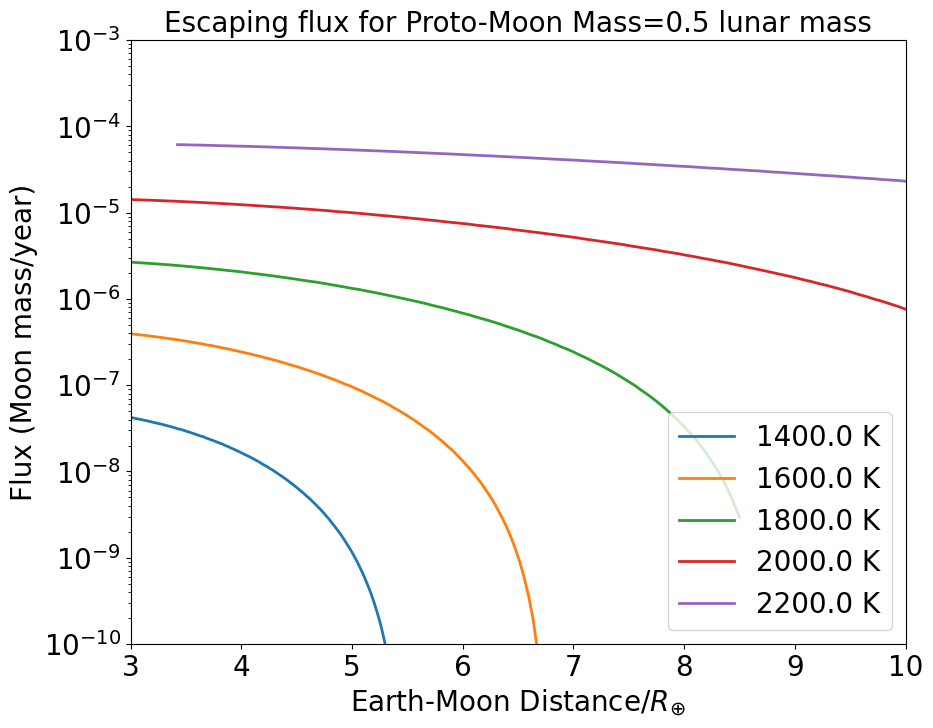}
   \caption{Surface flux ($F=\rho_sV_sR_{\rm m}^2$) as a function of the Earth-Moon distance (in units of Earth radii) and for different surface temperatures (coloured curves). Here the proto-Moon mass is set to 0.5 lunar mass.}
    \label{Figure_escape_adiab_05Mm}
\end{center}
\end{figure}

\newpage

\begin{figure}[h]
\begin{center}
   \includegraphics[width=.9\linewidth]{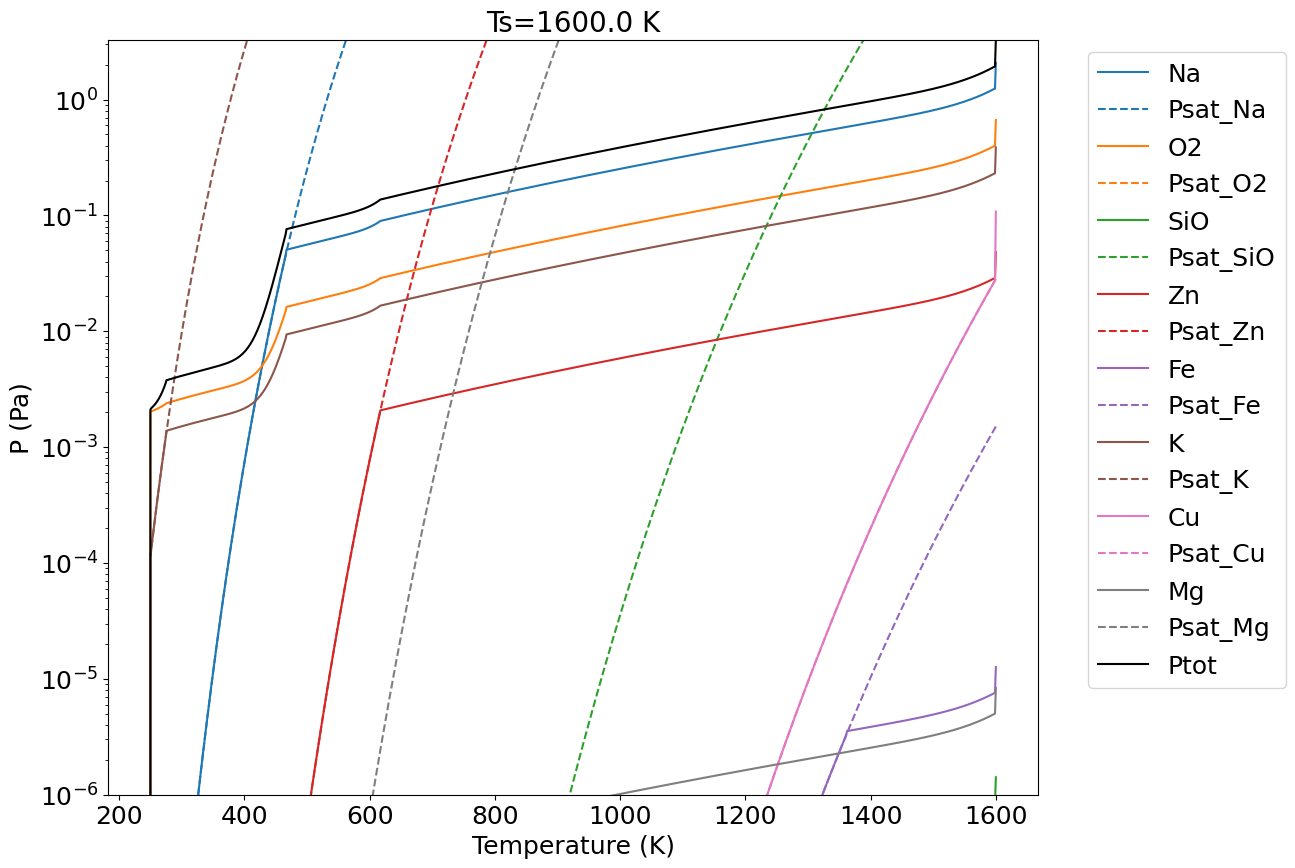}
   \caption{Partial pressures in an isentropic gas (from 1600K down to 250K), then isothermal gas (at T=250K) with the composition of a degazed gas at the Moon surface at 1600K. Note that at 250K the gas goes-on to depressurize isothermally, this is why the pressure drops to 0. The solid lines correspond to the partial pressure of every species, and dashed-lines display the saturating vapor pressure of every pure species. The relative entropy variation from 1600K down to 250K is $ \sim 3.2 \times 10^{-4}$.
   }
    \label{Figure_PT_adiababatic}
\end{center}
\end{figure}

\newpage

\begin{figure}[h]
\begin{center}
   \includegraphics[width=.7\linewidth]{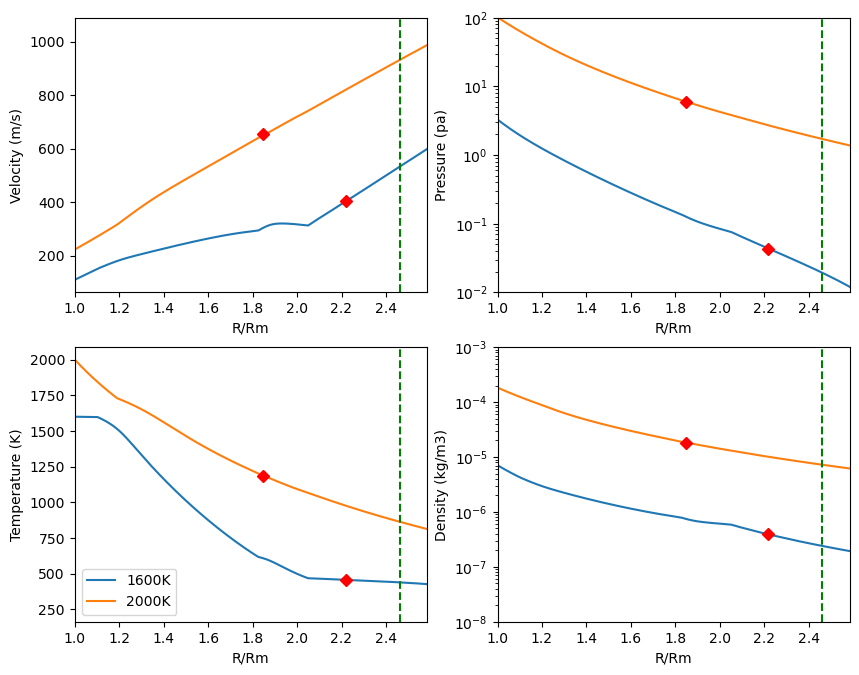}
   \caption{Atmospheric structure when the Moon is located at $4 R_\oplus$ for the wet model using moist adiabat. The color stands for the surface temperature $T_{s}$ (blue: 1600K, orange: 2000K). The red diamond shows the location of the transonic point, and the dashed-green line shows the location of the Hill sphere.Upper-Left: gas radial velocity, upper-right: local gas pressure, lower-left: local temperature, lower-right: local total density (gas+condensates). Rm is the Moon's radius.}
    \label{Figure_atmo_struc}
\end{center}
\end{figure}

\newpage

\begin{figure}[h]
\begin{center}
   \includegraphics[width=.8\linewidth]{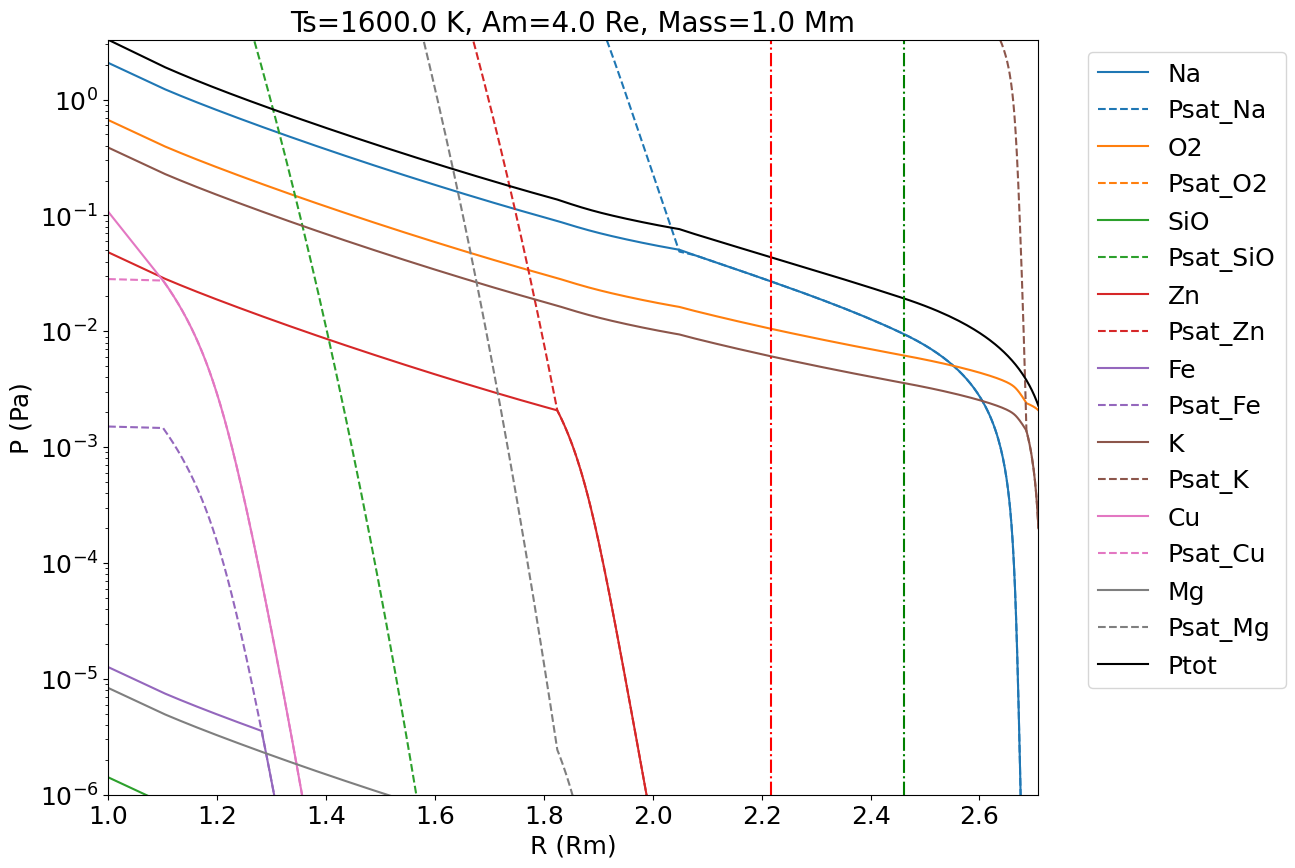}
   \caption{Partial pressures of the components of the escaping mixture, versus distance to the Moon's center. The proto-Moon has a surface temperature = 1600K, and is located at 4 Earth-Radii.The doted-dashed-green vertical line shows the location of the Hill's sphere, and the doted-dashed-green vertical line shows the location of the transonic point.}
   \label{gas_compo_1600K_wet_model_escape}
\end{center}
\end{figure}

\newpage

\begin{figure}[h]
\begin{center}
   \includegraphics[width=.8\linewidth]{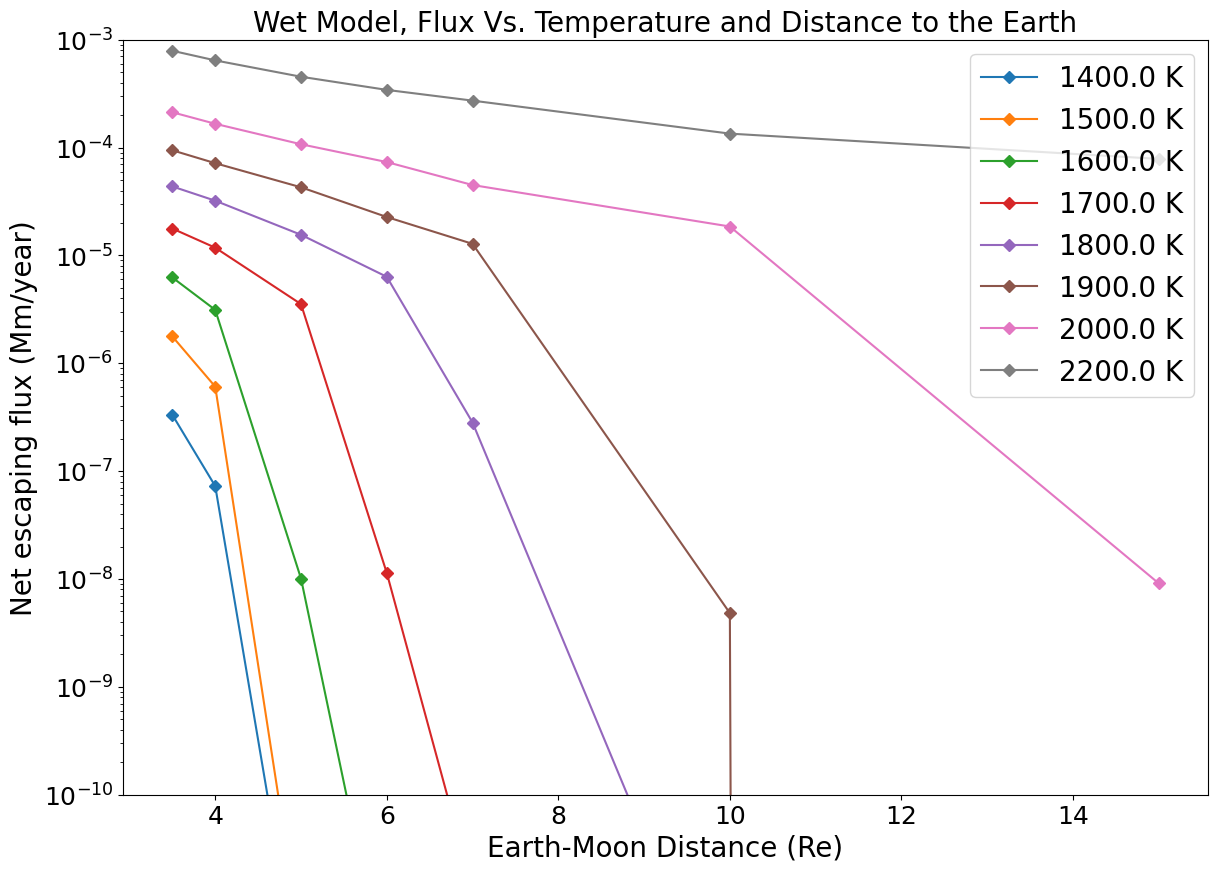}
   \caption{Surface gas flux (Y axis) as a function of surface distance (X axis) and for different Moon surface temperatures (colors) and for a proto-Moon with 1 lunar mass. Fluxes are in Moon mass/year. Earth-moon distances are given in units of the Earth radius ($R_e$).}
    \label{fig_flux_vs_Moon_distance_wet_model}
\end{center}
\end{figure}

\newpage

\begin{figure}[h]
\begin{center}
   \includegraphics[width=.6\linewidth]{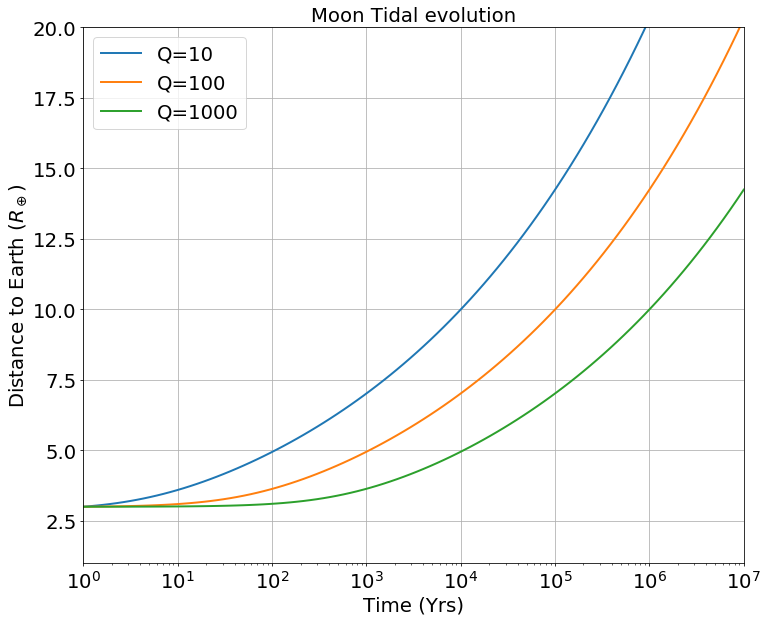}
   \caption{Moon semi-major axis (in units of the Earth Radius) plotted as function of time for different values of the Earth's dissipation factor $Q$.}
    \label{Figure_tidal_evol}
\end{center}
\end{figure}

\newpage

\begin{figure}[h]
\begin{center}
   \includegraphics[width=.8\linewidth]{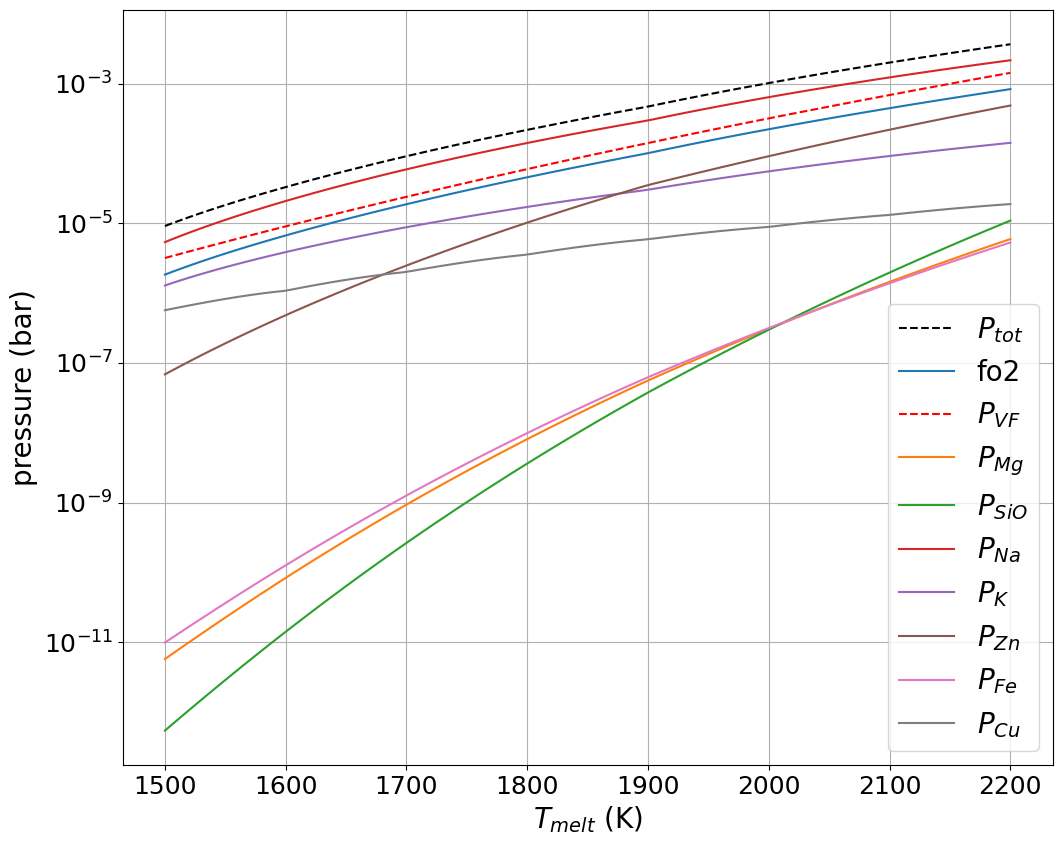}
   \caption{Partial pressures obtained with Equation \ref{Eq_reaction_MO_vap} and the thermodynamic data listed in \ref{SOM_THERMO}. The total saturating vapor pressure is in dashed-black line, and the BSE saturating vapor pressure by Visscher \& Fegley (2013) is plotted in dashed red for reference.}
    \label{Figure_partial_pressure}
\end{center}
\end{figure}

\newpage

\begin{figure}[h]
\begin{center}
  \includegraphics[width=.7\linewidth]{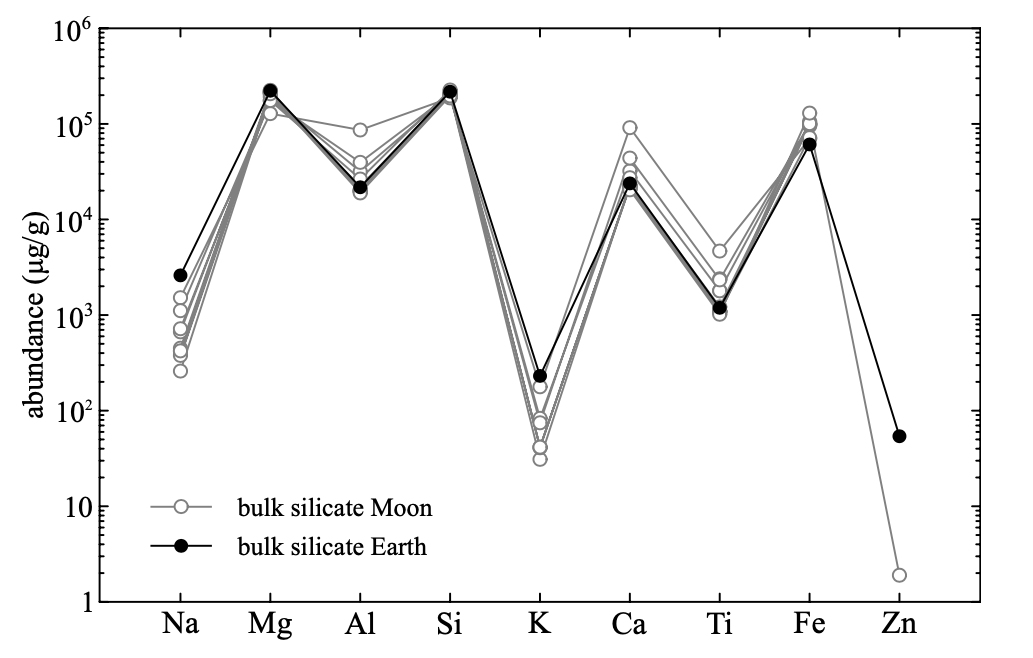}
  \caption{Figure adapted from Visscher \& Fegley (2013). Elemental abundances in the bulk silicate Moon (gray lines with open symbols) and bulk silicate Earth (BSE; black lines with filled symbols). Lunar abundances are taken from multiple literature sources (e.g., Morgan et al. 1978; Ringwood 1979; Taylor 1982; Wänke \& Dreibus 1982; Ringwood et al. 1987; Buck \& Toksöz 1980; Delano 1985; Jones \& Delano 1989; O’Neill 1991; Warren 2005; Lodders \& Fegley 2011); BSE abundances are taken from Palme \& O’Neill (2003).}
   \label{Fig_Earth_Moon_abundances}
   \label{fig_visscher}
\end{center}
\end{figure}

\newpage

\begin{figure}[h]
\begin{center}
   \includegraphics[width=.6\linewidth]{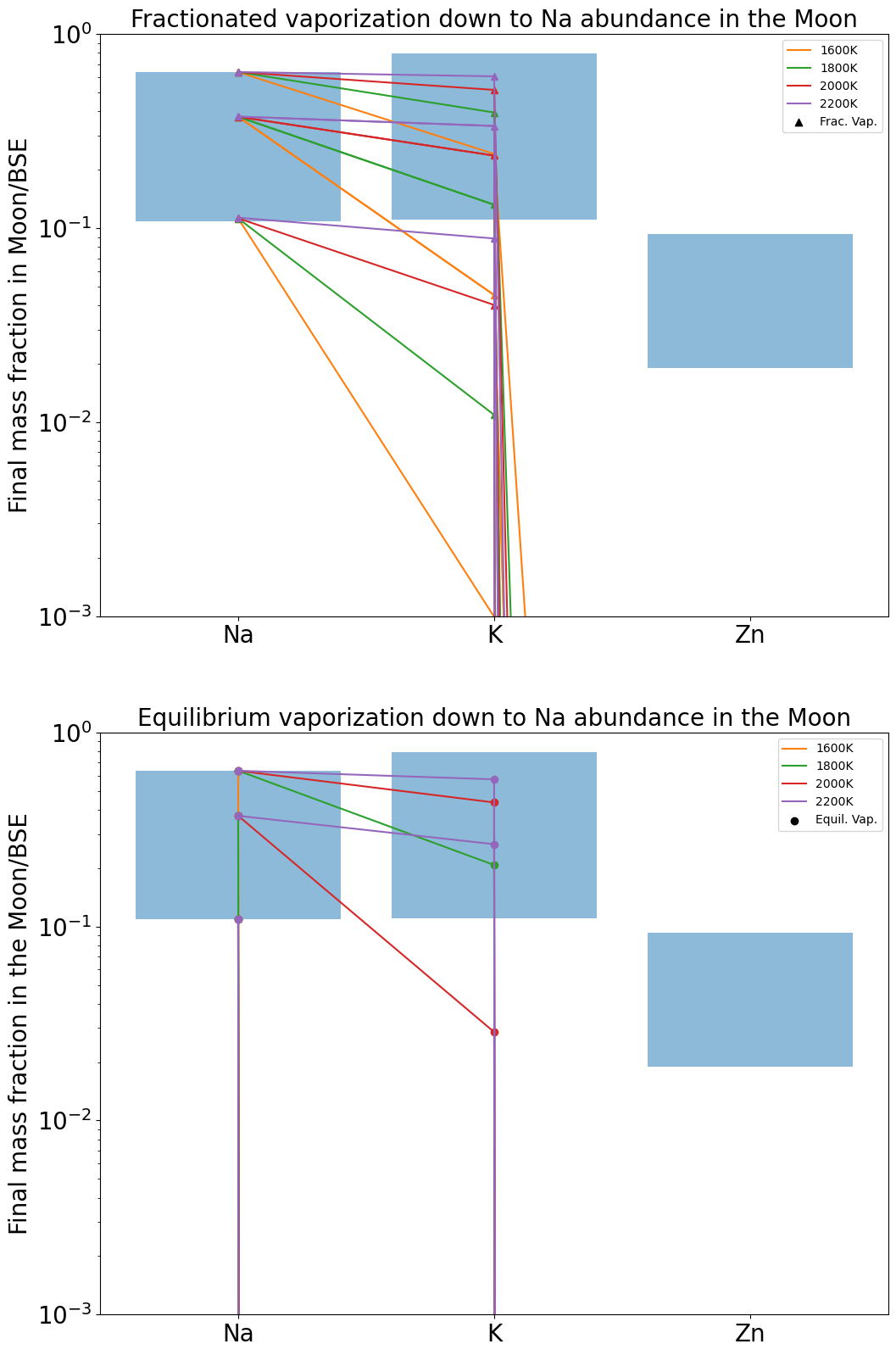}
   \caption{Final composition of the melt for different temperatures (colored lines) of the magma ocean in the case of equilibrium evaporation (top) and fractionated evaporation (bottom). Here the ammount of material removed is tuned to match either the maximum, the average or the miminimum abundance of Na in lunar samples. Blue boxes displays the acceptable ranges for Na,K and Zn.}
    \label{Figure_vaporize_to_Na}
\end{center}
\end{figure}

\newpage

\begin{figure}[h]
\begin{center}
   \includegraphics[width=.6\linewidth]{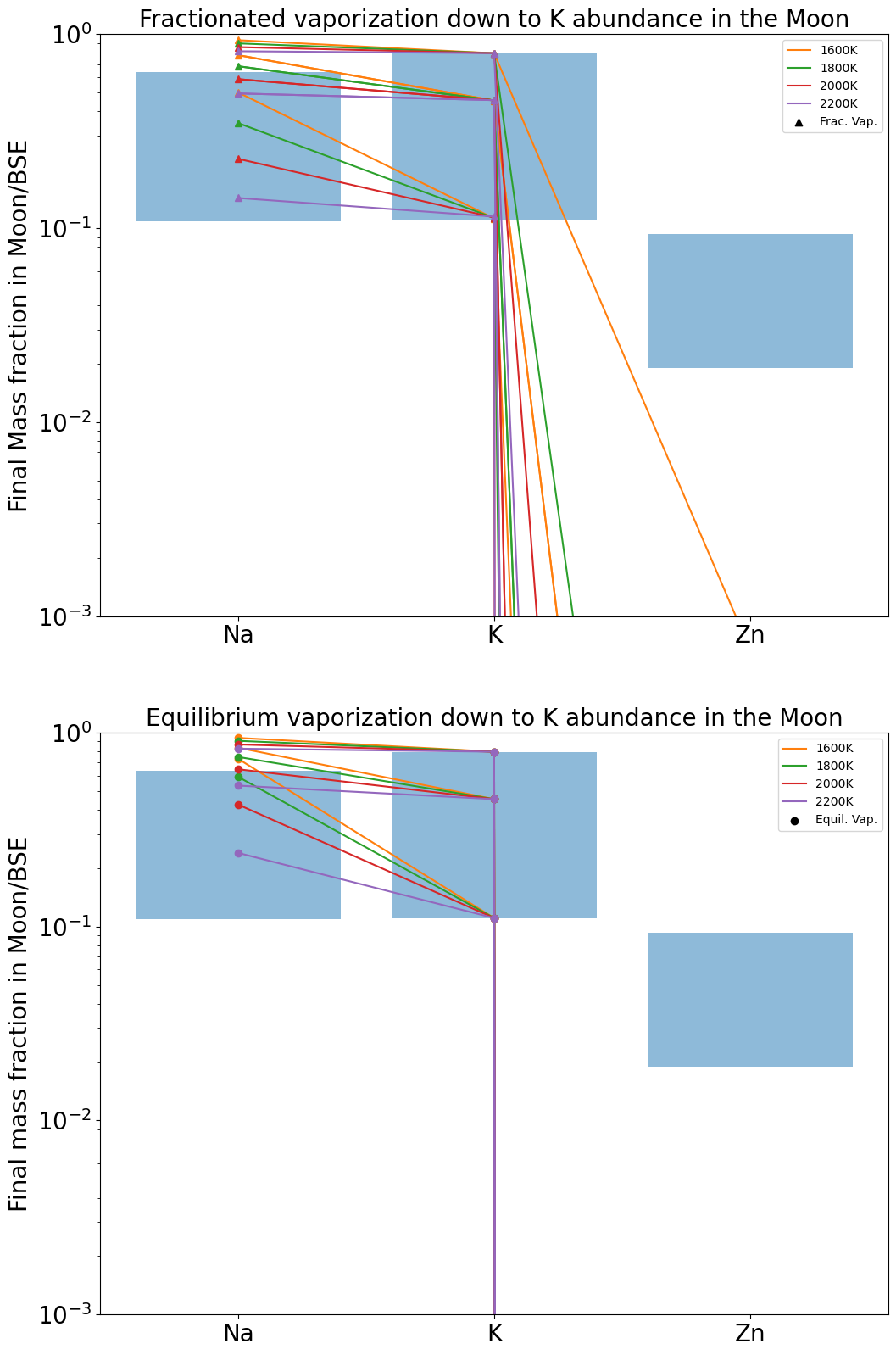}
   \caption{Final composition of the melt for different temperatures (colored lines) of the magma ocean in the case of equilibrium evaporation (top) and fractionated evaporation (bottom). Here the ammount of material removed is tuned to match either the maximum, the average or the miminimum abundance of K in lunar samples. Blue boxes displays the acceptable ranges for Na,K and Zn.}
    \label{Figure_vaporize_to_K}
\end{center}
\end{figure}

\newpage

\begin{figure}[h]
\begin{center}
   \includegraphics[width=.6\linewidth]{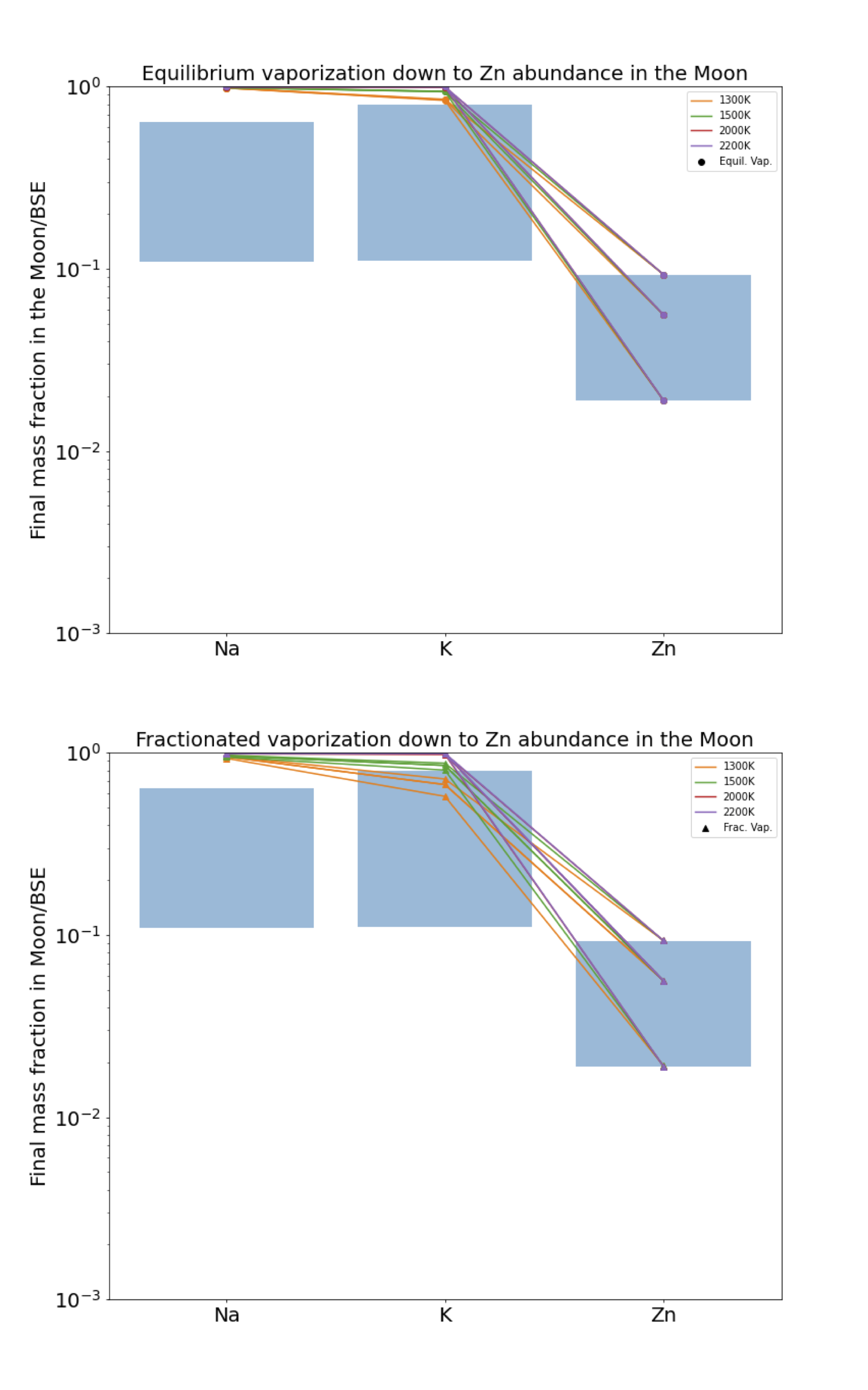}
   \caption{Final composition of the melt for different temperatures (colored lines) of the magma ocean in the case of equilibrium evaporation (top) and fractionated evaporation (bottom). Here the ammount of material removed is tuned to match either the maximum, the average or the miminimum abundance of Zn in lunar samples. Blue boxes displays the acceptable ranges for Na,K and Zn.}
    \label{Figure_vaporize_to_Zn}
\end{center}
\end{figure}



\newpage

\begin{figure}[h]
\begin{center}
   \includegraphics[width=.6\linewidth]{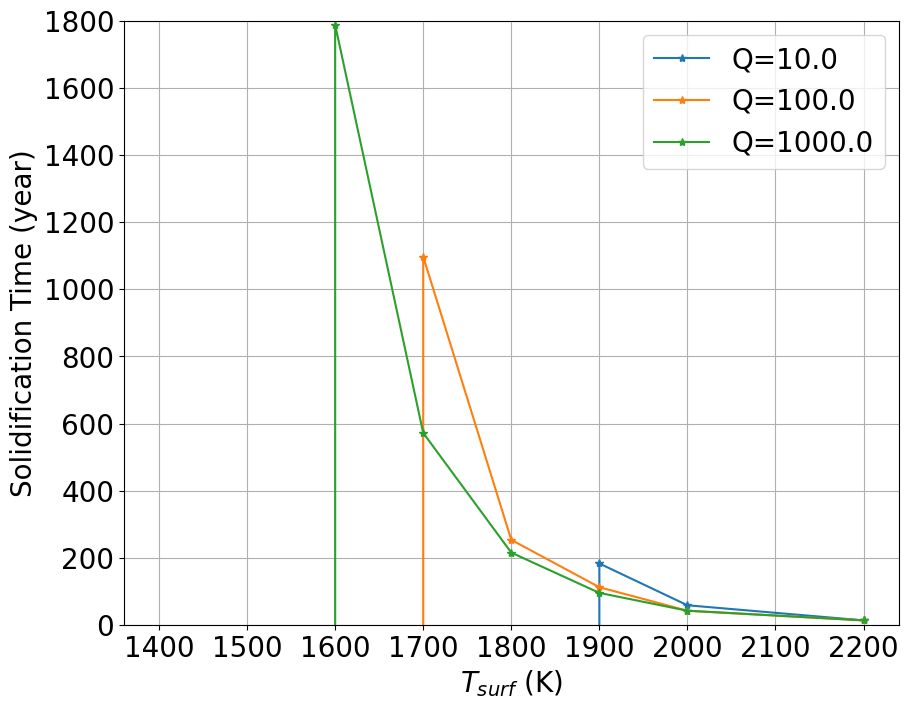}
   \caption{Formation timescale of the stagnant lid (as a function of Q and surface temperature) for matching the Na an K abundance in lunar moon samples. When the stagnant lid forms we assumes that the outgoing flux drops to 0. Previous studies suggest that the lid may form in $\sim 1000$ years only \citep{Elkins-Tanton_ET_AL_2011, Tang_Young_2020} }
    \label{fig_stagnant_lid_timescale}
\end{center}
\end{figure}

\newpage

\begin{appendix}

\section{Moon gravitational potential in the Earth's tidal field}
\label{SOM_potential_energy}

We consider a molten Moon located at distance $a$ from the Earth, and orbiting on a circular orbit at keplerian frequency :
\begin{equation}
    \Omega_k=\sqrt{\frac{G(M_\oplus+M_{\rm m})}{a^3}}
\end{equation}
With $G$, $M_\oplus$, and $M_{\rm m}$ standing for the universal gravitational constant, the Earth's mass and the Moon's mass respectively. We work in a rotating frame with $X-$axis pointing from the Earth to the Moon, and with origin  at the Earth's center. We compute the acceleration on a line joining the Earth and the Moon. The acceleration felt by a particle is then :

\begin{equation}
    \label{equation_accel}
    A(r)=\frac{-GM_{\rm m}}{r^2}-\frac{GM_\oplus}{(a+r)^2}+(a+r)\Omega_k^2-\frac{GM_{\rm m}}{a^2}
    -2\overrightarrow{\Omega}\times \overrightarrow{V}
\end{equation}

with $r$ is the abscissa of the particle measured from the Moon's center, $\overrightarrow{V} $ is the velocity in the rotating frame and $\overrightarrow{\Omega}$ is the rotation vector of the  frame. The first two terms are the gravitational accelerations, while the last three terms are the inertial accelerations. The first term corresponds to the Moon's gravitational attraction, the second is the Earth's gravitational attraction. The third term is the centrifugal force, the fourth is the opposite of the acceleration on the the center of the system (the Earth) because we work in an accelerated frame. The last term is the Coriolis acceleration, but which has a null contribution onto the axis joining the Earth-Moon's centers. The corresponding potential energy is :
\begin{equation}
    E_p(r)=\frac{-GM_{\rm m}}{|r|}-\frac{GM_\oplus}{|a+r|}-\frac{(a+r)^2\Omega_k^2}{2}-\frac{G M_{\rm m} x}{a^2}
    \label{Eq_Ep}
\end{equation}
where $x$ is the projection of the Earth-centered position vector onto the line joining the Earth to the Moon's center.

Under the Earth tidal field, two potential maxima appear on both sides of the Moon: the L1 and L2 Lagrange points. Their absolute distance to the Moon's center is given, approximately, by the Hill Radius $R_{\rm h}$:

\begin{equation}
    R_{\rm h} \simeq a \left( \frac{M_{\rm m}}{3M_\oplus} \right)^{1/3}
    \label{Eq_Rh}
\end{equation}

Note that for $|r| < R_{\rm h}$ the acceleration is directed toward the Moon and for $|r| > R_{\rm h}$ the acceleration is directed opposite to the Moon's center. So any material bounded to the Moon's surface must be closer than $R_{\rm h}$ to the Moon's center.

%


\section{Radiation pressure around the Earth}
\label{SOM_RAD_PRESSURE}
In this subsection, we discuss the ratio ($\beta$) of radiation pressure force to gravitational force on a dust grain. The effect of radiation pressure is discussed in a context of Martian moons-forming gas/dust disk \citep{Hyodo_2018} produced by a giant impact on Mars \citep{Hyodo_2017}. Here, following their simple approach, we estimate $\beta$ values of condensed small particles as a function of their sizes and temperature of the Earth. $\beta$ values is given by
\begin{equation}
    \beta = F_{\rm RP}/F_{\rm grav}
    \label{eq_beta}
\end{equation}
where $F_{\rm RP}$ and $F_{\rm grav}$ are the radiation pressure force and gravitational force respectively. $F_{\rm PR}$ is given by \citep{Burns_1979}
\begin{equation}
    F_{\rm RP} = \bar{Q}_{\rm RP} \frac{S}{c} \times \sigma_{\rm col}
    \label{eq_Frp}
\end{equation}
where $\bar{Q}_{\rm RP}$ is the Plank mean of the radiation pressure efficiency averaged over the spectrum. $c$ is the speed of light. $\sigma_{\rm col}=\pi d^2$ is the cross section of a particle whose radius is $d$. $S$ is the radiation flux density at distance $r$ to the Earth and given by
\begin{equation}
   S = \frac{\sigma_{\rm SB}T_\oplus^4 \times 4\pi R_\oplus^2}{4\pi r^2}
    \label{eq_flux}
\end{equation}
where $\sigma_{\rm SB}$ is the Stefan–Boltzmann constant. $T_\oplus$ is black-body temperature of the Earth. Since $F_{\rm grav}=GM_\oplus/r^2$, the $\beta$ is rewritten as \citep{Hyodo_2018}
\begin{equation}
    \beta = F_{\rm RP}/F_{\rm grav} = \frac{9}{16\pi} \bar{Q}_{\rm RP} \times \frac{\sigma_{\rm SB}T_\oplus^4}{cGR_\oplus d \rho_\oplus\rho_{\rm d}}
    \label{eq_beta_final}
\end{equation}
where $\rho_\oplus$ and $\rho_{\rm d}$ are the densities of the Earth and that of a condensed dust, respectively. In order to calculate $\bar{Q}_{\rm RP}$, we follow a simple approach discussed in the previous work \citep{Hyodo_2018}. $\bar{Q}_{\rm RP}$ is given by
\begin{equation}
    \bar{Q}_{\rm RP} (\lambda_{\rm cri},T_\oplus) = \int_0^{\lambda_{\rm cri}} B(\lambda,T_\oplus) Q_{\rm RP}(\lambda) {\rm{d}}\lambda
    \label{Qrp}
\end{equation}
where $B(\lambda,T_\oplus)$ is the normalized Planck function at a wavelength of $\lambda$ and $Q_{\rm RP}(\lambda)$ is the radiation pressure efficiency as a function of $\lambda$ where we assume that $Q_{\rm RP}(\lambda)$ is unity for particles whose characteristic length $2\pi d$ is larger than wavelength and zero vice-versa (thus $\lambda_{\rm cri}=2\pi d$). Note that, the above equation indicates that the value of $\beta$ does not depend on the radial distance to the central planet.

\begin{figure}[!t]
\begin{center}
   \includegraphics[width=.8\linewidth]{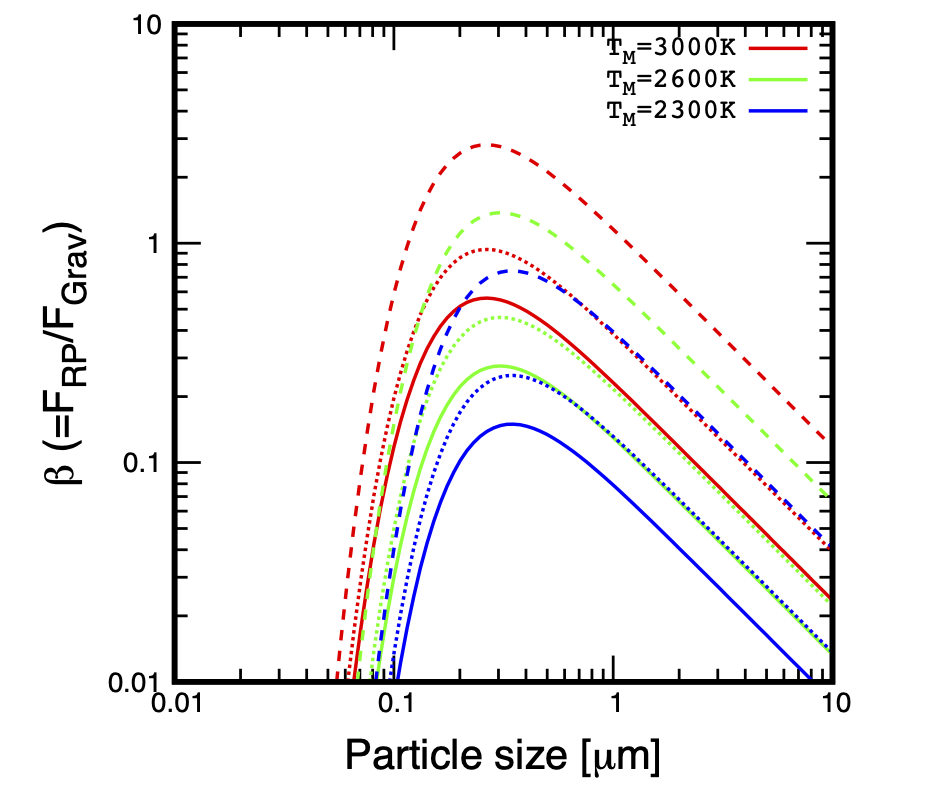}
   \caption{$\beta$ values around the Earth as a function of particle size ($d$) for different temperatures of the Earth ($T_\oplus$) and particle densities ($\rho_{\rm d}$). Red, green, and blue lines represent cases where $T_\oplus=3000, 2600, 2300$ K, respectively. The solid, dotted, and dashed lines represent cases where $\rho_{\rm d} = 5.0$, $3.0$, and $1.0$ $g/cm^{3}$, respectively.}
    \label{Figure_beta}
\end{center}
\end{figure}

Figure \ref{Figure_beta} shows the $\beta$ values around the Earth as a function of particle size. As the temperature of the Earth increases or as the density of particles becomes smaller, the $\beta$ values become larger. The critical value of $\beta$ for a particle whose orbit is circular to be blown-off by radiation pressure is $\beta_{\rm cri} > 0.5$. If a particle orbit is an eccentric orbit, the critical value of $\beta$ depends on radial distance to the central planet and $\beta_{\rm cri}$ can be larger/smaller than 0.5 \citep{Hyodo_2018}. Figure \ref{Figure_beta} indicates that if the black-body temperature of the Earth is $\sim 2300$ K and the density of particle is $3.0$ g cm$^{-3}$, then $\beta > 0.1$ , which is enough for particles not coming back to the Moon (see Figure \ref{fig_orbits}).

\section{Drag of liquid droplets }
\label{SOM_drag_droplets}
We have assumed that the droplets stay tightly coupled to the gas. Of course there should be a maximum size above which this assumption must be invalid. The coupling time of a dust-grains with the surrounding gas, $\tau_c$ is in the Stokes regime\citep{Birnstiel_2011}:

\begin{equation}
    \tau_{\rm c}=\frac{4 \rho_{\rm s} a^2 \sigma}{9 \mu Cs}
    \label{Eq_Taus}
\end{equation}

where $\rho_s$ is the condensate density (3000 kg/m$^3$ here), $a$ is the condensates radius, $\sigma$ is the molecular cross section ($=\pi r_m^2$ with $r_m$ standing for the molecular radius, here $r_m=180\times10^{-12}$ m, valid for Na atoms), and $C_s$ is the pure gas sound speed.

\begin{figure}[!t]
\begin{center}
   \includegraphics[width=.9\linewidth]{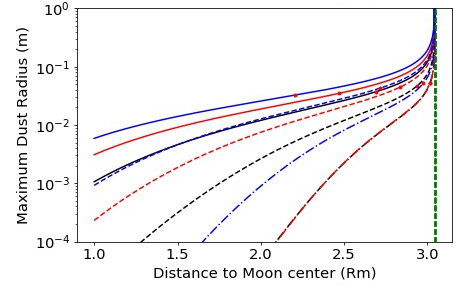}
   \caption{Maximum size of droplets  for tight gas-coupling ($a_{\rm max})$. Here the Moon is located at 5$R_{\oplus}$. The color stands for the surface temperature $T_{\rm s}$ (black:1500 K, red:2000 K, blue:2500 K), and the line-style stands for the liquid fraction at the surface $X_{\rm s}$ (solid:$X_{\rm s}=0.2$, dashed:$X_{\rm s}=0.5$, dotted-dashed:$X_{\rm s}=0.9$). The red dot shows the location of the sonic point. The green-dashed line is the location of the Hill sphere.}
    \label{Figure_amax}
\end{center}
\end{figure}

As the droplets are formed in the gas traveling upward, they form keeping an upward directed velocity. During their ascension, they are slowed-down by the Moon gravity field, while being accelerated upward by the friction with the gas. Thus, we compare the  slowing-down time due to the gravity field ($\tau_s=v(r)/|A(r)|$) to the gas coupling time. We define the coupling parameter as follows : $\kappa=\tau_c / \tau_s$. If $\kappa < 1$ then condensates stay tightly coupled to the gas, while if $\kappa > 1$ the condensates start to fall before they couple to the gas. From condition $\kappa < 1$ we can derive  a maximum droplet radius for coupling $a_{\rm max}$ can be derived for every r:

\begin{equation}
    a_{max}=\sqrt{\frac{9 \mu Cs v(r) }{4 \rho_{\rm s} |A(r)| \sigma}}
\end{equation}

Finally the Epstein gas-drag  regime does not play a role here. As shown on Figure Appendix\ref{Figure_amax} particles are mostly decoupled from the gas at the base of the atmosphere, due to the strong gravitational pull from the Moon and the resulting short acceleration timescale. In this region the gas is dense, so that the Epstein regime does not apply. At the top of the atmosphere, when the gas density drops severely, particles simply follow ballistic trajectories: they are not efficiently decelerated by the Moon's pull.

Concerning the growth of droplets during their journey upward, it depends on the turbulence state of the flow which is unknown for the moment.

\medskip

\section{What is the net evaporative flux, and
pressure,  above the magma ocean ?}
\label{Appendix_Pressure_surface}

In \cite{Tang_Young_2020} (TY20 hereafter) it is argued that the pressure of the evaporating material above the lunar magma ocean is about $10^{-7}$ bar, far below its saturated vapour pressure ($\sim$ $10^{-3}$ bar at $\sim  2000 K$). 
This conclusion is reached by finding the pressure at which the escaping flux at the top of the atmosphere (namely, the point of escape; the Bondi radius) yields an escape rate that can be sustained by diffusive transport of the evaporated material through the atmosphere. These two processes have opposing dependencies on total pressure, in which escape rate increases linearly with pressure, whereas mass transport rates (by diffusion) decrease linearly with pressure. In Figure 6 of TY20 it should be noted that the x-axis is incorrectly designated as the "surface pressure",  but it actually corresponds to the "far-field" pressure as introduced in \cite{Richter_et_al_2002} and re-used in \cite{Young_2019}. In TY20 the "far-field pressure" is also designated as the "ambient pressure" in their section 3. The concept of far-field pressure was used in \cite{Richter_et_al_2002} to designate the pressure far (or at infinity) from the surface of an evaporating bead of CAI-like composition, immersed in a surrounding static gas ($H_{2}$ in Richter et al., 2002). Inside this static gas, the only mode of mass transfer is diffusion, which proceeds through binary collisions between the evaporating species and the surrounding gas. Indeed Equation 6 of \cite{Richter_et_al_2002} is simply Fick's law of mass diffusion, applied to a dilute species (the evaporating gas) diffusing through an ambient medium (a gas, in this case) with a fixed diffusion coefficient that remains constant with distance from the surface. The analytical solution to this equation is reported in Equation 6 of \cite{Young_2019} and Equation 4 of \cite{Tang_Young_2020}.
\cite{Richter_et_al_2002} shows that at steady-state ($t \rightarrow + \infty$), the far-field pressure (as defined above) is lower than the vapour pressure ($P_{sat}(T)$) of the evaporating species at the evaporative surface. As a consequence, the net evaporative flux is lower than the free evaporative flux. This leads \cite{Tang_Young_2020} to the conclusion that the net escaping flux is very low. The far-field pressure is found to be about $10^{-6} P_{sat}$, a pressure at which hydrodynamic escape rates and evaporation rates are found to be equal.
However, the use of the Richter et al. (2002)'s approach, is motived by TY20's assumptions that the escaping flow does not start at the surface of the magma ocean, but starts above a convectived layer, just above the magma ocean. So there is no net advection at the surface of the magma ocean (by hypothesis).
In our case the upward advection of the atmosphere starts at the bottom of the atmosphere. So the R02 appraoch  is not suited for in our case  for the following reasons :
\begin{itemize}
    \item The evaporating species are not transported in the vertical direction through diffusion alone, but through advection, with a non-zero net upward velocity. The acceleration is provided essentially by the strong pressure gradient between the surface and infinity. In this context, the classical Euler hydrodynamic equations (conservation of mass, momentum and energy as used here) must be considered to describe properly the atmospheric structure, rather than Fick's law only.
    \item The evaporating species is not a minor constituent as assumed in Richter et al. (2002), and, again its transport cannot be simply described by diffusive transport: its presence affects the local flow velocity and pressure field, and thereby requires an equation of state (the ideal gas law) to yield a self-consistent solution.

\end{itemize}
So our description of the connection of the magma ocean in our model and in TY20 model differ largely. For the above reasons, we do not follow the approach of \cite{Tang_Young_2020}. In addition, the concept of far-field pressure, whereas properly defined in the context of Richter et al.(2002), becomes ill- defined in our case, because we do not know "where" to take it. If it is the pressure at infinity (as defined in Richter et al. 2002) , then it should be zero for a Parker Wind, like considered in \cite{Tang_Young_2020}. But the Parker Wind do solve for the conservation of mass, momentum and energy fluxes, that is inconsistent with the Richter et al. (2002)'s approach.

However, a point still holds and must be discussed here: \cite{Richter_et_al_2002}, \cite{Young_2019} and \cite{Tang_Young_2020} argue correctly that there should be a "return flux" ($F_R$) of evaporating material back to the magma ocean. This return flux is opposed to the free evaporative flux ($F_F$), and thus the net escaping flux ($F_E$) is the difference between the two. The net evaporative mass flux in a static gas ($F_{E, static} $)is given by the Hertz-Knudsen equation:
\begin{equation}
    F_{E, static}=F_F-F_{R,static}
\end{equation}

\begin{equation}
F_{E, static}= \frac{m_i P_{sat}(T)}{(2\pi m_i R T)^{1/2}}-\frac{m_i P_i}{(2\pi m_i R T)^{1/2}}
\end{equation}
with $P_i$ and $m_i$ standing for the actual pressure and molar mass of evaporating specie $i$.
At thermodynamic equilibrium
$F_F=F_{R,static}$ and so the net escaping flux is zero. However the above equations are not valid for a non-static gas, i.e a gas that is globally advected away from the evaporating surface, like an expanding atmosphere above a magma ocean. Whereas this does not affect the free evaporative flux, the return flux must be reconsidered.


We present here an approach very similar to the original Hertz-Knudsen theory, but generalized to a moving gas to compute the return flux. We show that in a system in which the net advective gas velocity is non-zero, the surface pressure remains comparable to $P_{sat}(T)$ as long as the advective velocity is small compared to the thermal velocity. We proceed as follows: consider a liquid surface located at Z=0 and temperature T. We assume that the gas has a non-zero advective velocity above the surface called $V_a$ and directed upward. This non-zero velocity is due to the atmospheric structure that gives rise to a vertical acceleration due to the pressure gradient, allowing it to escape from the system. The free evaporative mass flux per surface unit may be simply written (replacing pressure by density, using ideal gas equation, in the fluxes equations above):

\begin{equation}
    F_F=\frac{\rho_{sat} V_t}{2 \sqrt \pi}
\end{equation}
where $\rho_{sat}$ is the gas density at saturation at temperature T ($=P_{sat}(T)
\rho_{sat}kT/m_i$ ) where 
. $V_t$ is the "thermal velocity" meaning precisely the most problable velocity in the Maxwell Boltzman velocity distribution $V_t=(2R T/m)^{1/2}$. Let $F_R$ the return flux to the melt. We assume also the gas density at the surface, $\rho_s$, is unknown. We can write, in a general way that:

\begin{equation}
    F_R=\rho_s g_s(T,V_a)
\end{equation}
where $g_s(T,V_a)$ has the dimension of a velocity, and can be seen also as the flux of particles returning back to the liquid, in gas unit volume above the surface. For a static gas, for example in a closed system $g_s(T,V_a=0)=V_t/(2 \sqrt \pi)$ and equation 3 of TY20 is recovered. However, for an atmosphere in motion, we require the general expression of $g_s(T,V_a)$. Consider a parcel of gas above the liquid, at a few particles mean-free path above the surface. There, the velocity distribution of particles follows a Maxwell-Boltzmann distribution at temperature T but with a non-zero net-vertical velocity with $V_z=V_a$. So in the reference frame of the liquid's surface the 3D velocity distribution of gas particles is not centered on the velocity vector $(V_x=0, V_y=0,V_z=0)$ (as in a static gas), but rather centered on $(V_x=0, V_y=0, V_z=+V_a)$, because the evaporating gas is globally accelerated upward in this open system. The particle 3D velocity distribution $P(V_x,V_y,V_z)$ is then  :
\begin{equation}
    P(V_x,V_y,V_z)=f_{MB}(V_x,0) \times f_{MB}(V_y,0) \times f_{MB}(V_z,+V_a)
\end{equation}

where $f_{MB}(v,V_a)$ is the decentered Maxwell Boltzmann velocity distribution :

\begin{equation}
    f_{MB}(v,V_a)=\frac{1}{V_t\sqrt{\pi}} e^{(-\frac{(v-V_a)^2}{V_t^2})}
\end{equation}

\begin{figure}[!t]
\begin{center}
  \includegraphics[width=.9\linewidth]{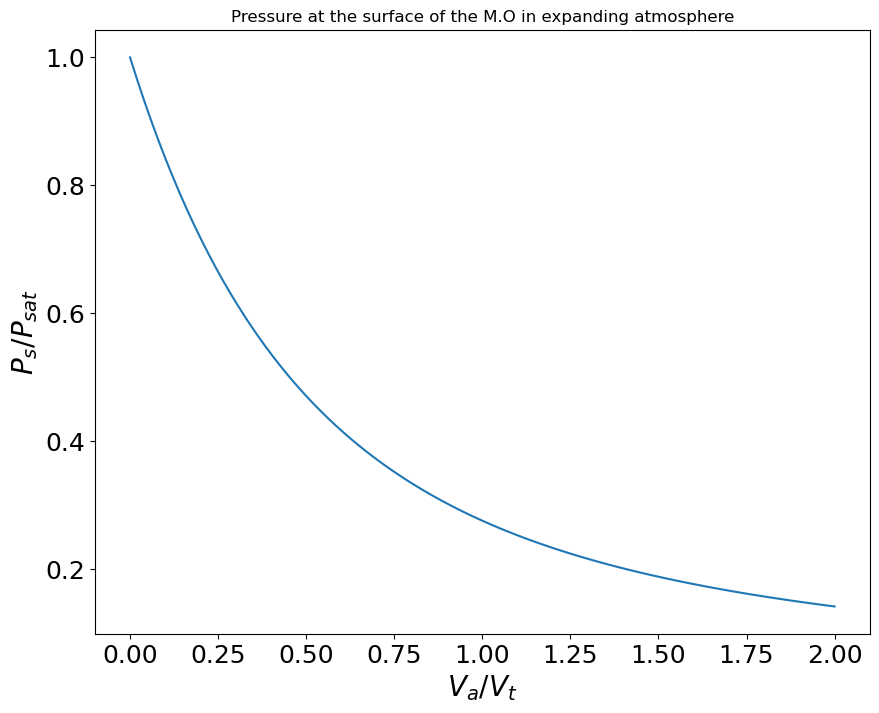}
  \caption{Ratio of surface pressure ($P_s$) to saturating pressure ($P_{sat}$) at the surface of the magma ocean as a function of the gas vertical velocity at the surface ($V_a$) and thermal velocity ($V_t$). Note that in the case $V_a=0$ we recover the result of \cite{Richter_et_al_2002}: at equilibrium the surface pressure equals the saturating vapour pressure and the the net outflux ($=\rho_s V_a$) is zero. }
  \label{figure_PsPsat}
\end{center}
\end{figure}

\begin{figure}[!t]
\begin{center}
  \includegraphics[width=.9\linewidth]{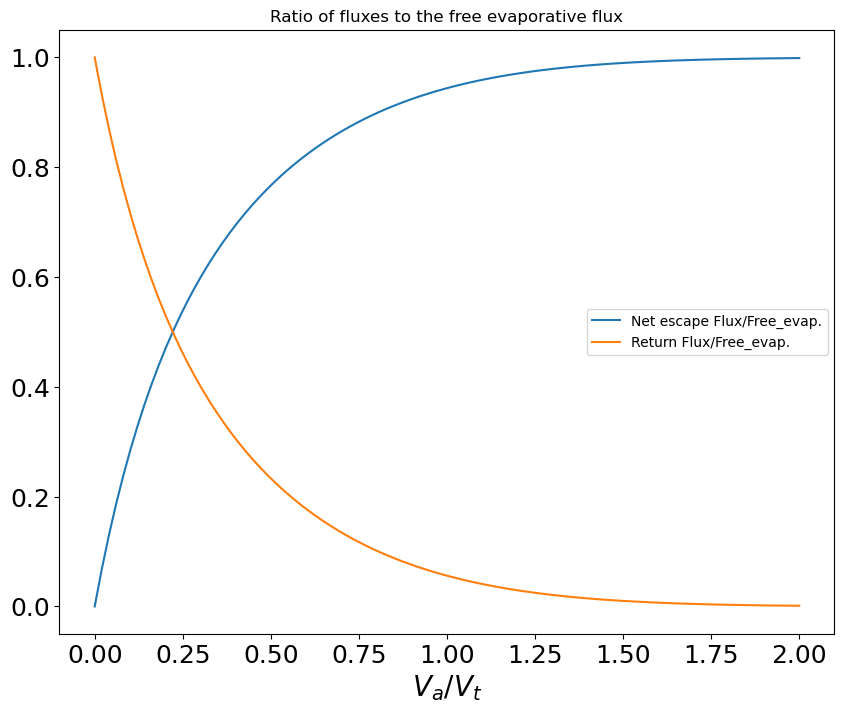}
  \caption{Ratio of fluxes at the surface as a function of $V_a/V_t$: Orange : return flux to the surface divided by the free evaporative flux ($F_R/F_F$), blue : Net total escaping flux  divided by the free evaporative flux ($F_E/F_F$). }
  \label{figure_flux_ratio}
\end{center}
\end{figure}

\medskip

\medskip

Then $g_s(T,V_a)$ can be computed : it is the flux of particles crossing the surface, that is the flux of particles with $V_z < 0$ (in the reference frame of the liquid  surface). Counting the flux positively for particles with $V_z < 0$, so $dg_s=-V_z P(V_x,V_y,V_z) dV_x dV_y dV_z$ which gives :

\begin{equation}
    g_s(T,V_a)=-\int_{-\infty}^{+\infty} f_{MB}(V_x,0)  \mathrm{d}V_x \times
    \int_{-\infty}^{+\infty} f_{MB}(V_y,0)  \mathrm{d}V_y  \times
    \int_{-\infty}^{0} V_z f_{MB}(V_z,+V_a) \mathrm{d}V_z
\end{equation}

And we get, after integration :

\begin{equation}
    g_s(T,V_a)=\frac{-V_a erfc(\frac{V_a}{V_t})+V_t e^{\frac{-V_a^2}{V_t^2}}}{2 \sqrt \pi}
\end{equation}

where $erfc(x)$ is the complementary error-function ($=1-erf(x)$). So the net return flux is :
\begin{equation}
    F_r=\rho_s \frac{-V_a erfc(\frac{V_a}{V_t})+V_t e^{\frac{-V_a^2}{V_t^2}}}{2 \sqrt \pi}
 \end{equation}

Which we rewrite for easy manipulation

 \begin{equation}
    F_r=\rho_s V_t C(V_t/V_a)
\end{equation}
With C(x) :

\begin{equation}
    C(x)=\frac{-x erfc(x)+e^{-x^2}}{2 \sqrt \pi}
 \end{equation}

Note that in the limit case of a static gas (like in a closed chamber) $V_a=0$ and we recover  the return flux as reported in \cite{Richter_et_al_2002, Young_2019, Tang_Young_2020} valid for a static-vapour.
The gas density (and pressure) above the surface is obtained by considering mass conservation.
The relation between the escape-flux ($F_E$), the free evaporative and return flux is : $F_E=F_F-F_R$ that reads :

\begin{equation}
\rho_s V_a= \frac{\rho_{sat} V_t}{2 \sqrt \pi}-\rho_s V_t C(V_t/V_a)
\end{equation}
Thus
\begin{equation}
\rho_s/\rho_{sat} = P_s/P_{sat} = \frac{1}{2 \sqrt \pi} \frac{1}{V_a/V_t+C(V_a/V_t)}
\end{equation}
 or
\begin{equation}
\rho_s/\rho_{sat} = P_s/P_{sat} = \left(  \frac{V_a}{V_t} \sqrt \pi(2- erfc(\frac{V_a}{V_t})) + e^{\frac{-V_a^2}{V_t^2}}
\right)^{-1}
\label{equation_Ps_Psat}
\end{equation}

Since we assume that the liquid and the gas are at the same temperature  at the surface $\rho_s/\rho_{sat}=P_s/P_{sat}$. $P_s/P_{sat}$ is plotted in Figure \ref{figure_PsPsat}  as function of $V_a/V_t$. It shows that for $V_a=0$ the surface pressure is equal to the saturating vapor pressure. For  $V_a>0$ the surface pressure diminishes slowly, which allows a net and non-zero escaping flux. When the advective velocity (and the net escaping flux) increases, the surface pressure may drop significantly. For our present paper we find surface velocities that are in the range $0<V_a/V_t< 0.2$ (for magma ocean at 2200 K and for a Moon orbiting at 3.5 Earth radii only) so $0.7 <P_s/P_{sat} < 1$ that is comparable, but still lower, than the saturating vapor pressure. So liquid-vapor equilibrium is almost reached at the surface of the magma ocean and in no-way is the surface pressure as low as $10^{-5} P_{sat}$, nor is it this low at the escaping Bondi radius. The origin of this discrepancy is due to the reliance on diffusion in TY20 as the dominant mode of mass transport. 

For the sake of completeness we provide below the useful quantities, $F_E/F_F$ and $F_R/F_F$

\begin{equation}
    \frac{F_E}{F_F}=\frac{(V_a/V_t)}{(V_a/V_t)+C(V_a/V_t)}
\end{equation}

\begin{equation}
    \frac{F_R}{F_F}=\frac{C(V_a/V_t)}{(V_a/V_t)+C(V_a/V_t)}
\end{equation}

They are plotted in Figure \ref{figure_flux_ratio} .

\section{Isothermal models}
\label{isothermal_appendix}
In the core section of the text we have assumed that escape was always adiabatic. As the gas pressure is quite low (< 100 Pa at the base of the atmosphere), a thin and transparent atmosphere may indeed behave adiabatically.  However, as gas condense into droplets, the optical depth may increase, leading to stronger absorption. So it is also interesting to present isothermal solutions for the hydrodynamical escape.

\begin{figure}[!t]
\begin{center}
  \includegraphics[width=.7\linewidth]{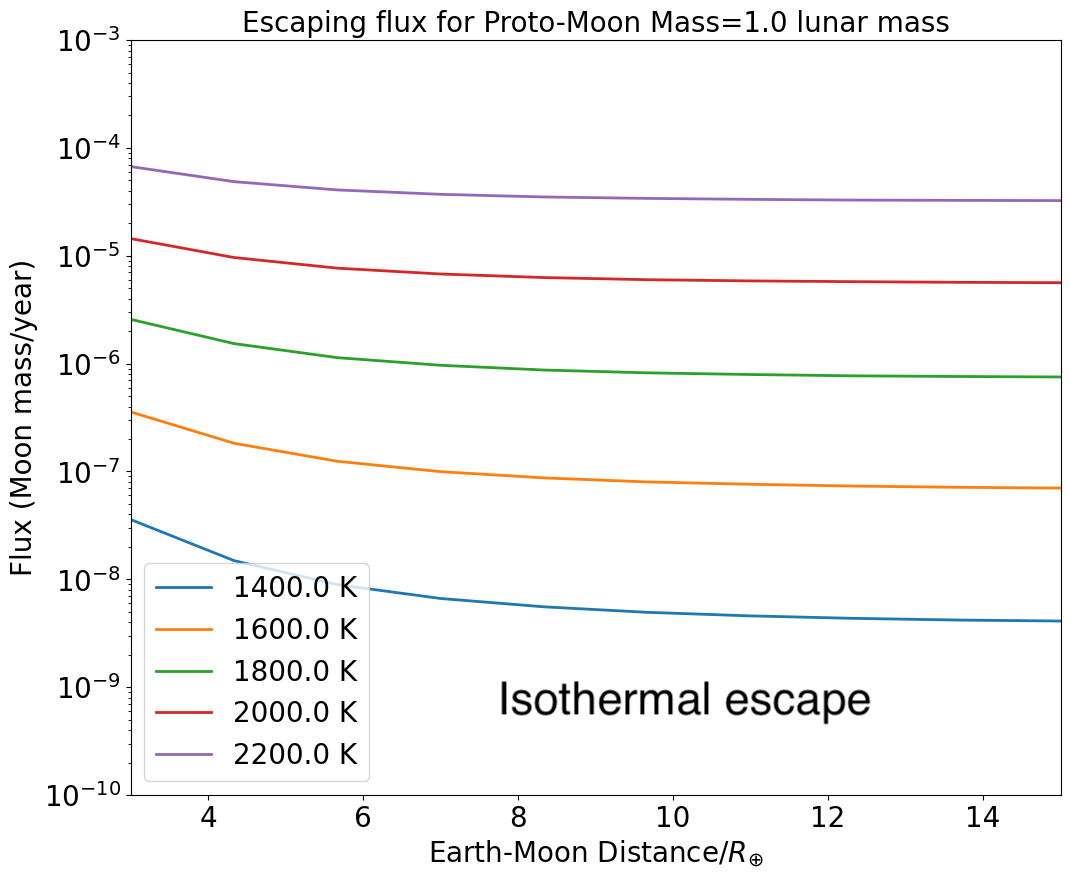}
  \caption{Flux versus Earth-Moon distance and vs Moon's surface temperature for the dry model and isothermal case (the escaping atmosphere temperature is kept at the surface temperature of the Moon.) }
  \label{figure_flux_dry_isothermal}
\end{center}
\end{figure}

For the  dry model (without recondensation), we now have $P=\alpha \rho$ and the isothermal Bernouillie relation is rewritten :
\begin{equation}
    \frac{V_S^2}{2}+\alpha Ln(\rho_S)+Ep_S=\frac{V_H^2}{2}+\alpha Ln(\rho_H)+Ep_H
    \label{eq_dry_bernouuillie_isothermal}
\end{equation}
The resulting flux is displayed in Figure \ref{figure_flux_dry_isothermal}.

\begin{figure}[!t]
\begin{center}
  \includegraphics[width=.7\linewidth]{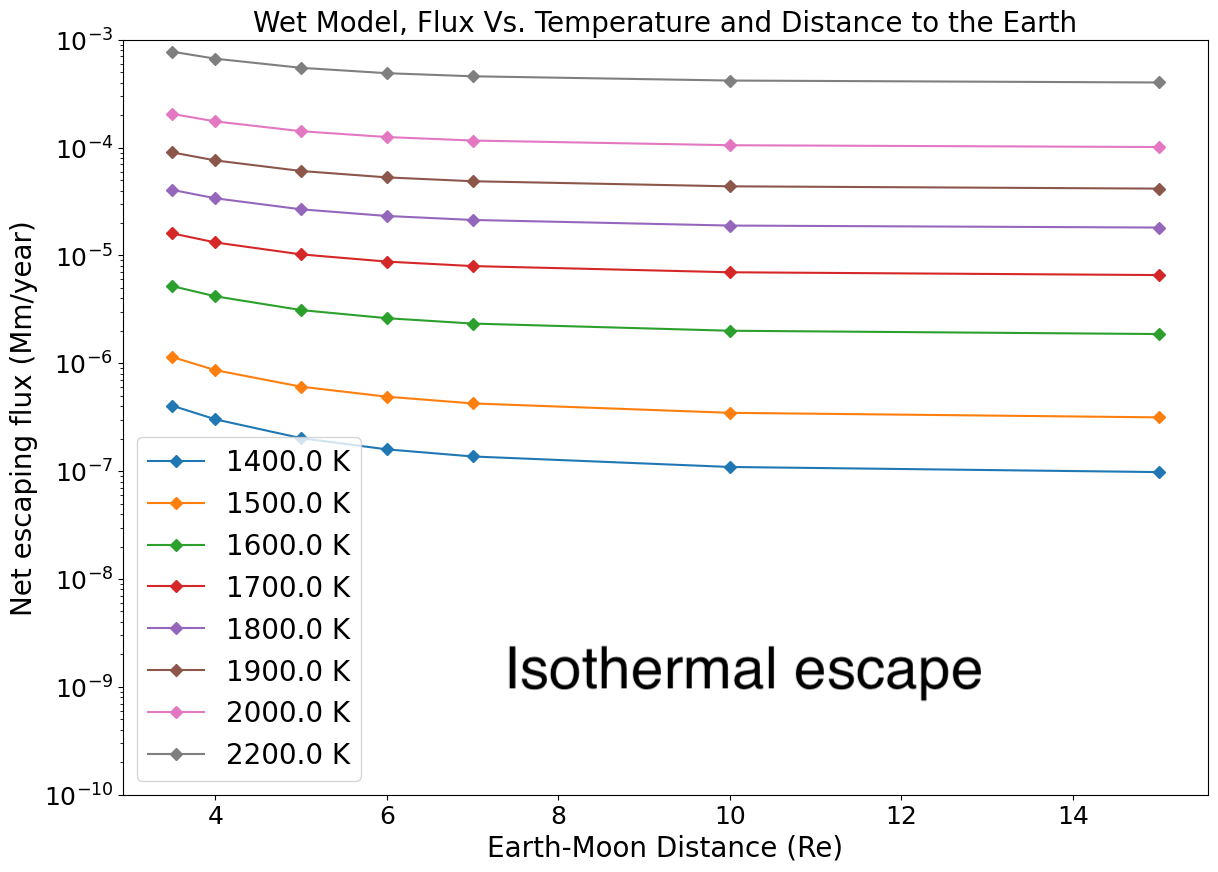}
  \caption{Flux versus Earth-Moon distance and vs Moon's surface temperature for the wet model and isothermal case (the fluid temperature is always at the surface temperature of the Moon) }
  \label{figure_flux_wet_isothermal}
\end{center}
\end{figure}

For the wet model, following the procedure described in the core text (see section \ref{hydro_escape_wet_model}), we have recomputed the fluxes, but only keeping the isothermal part, keeping $T=T_S$. The isothermal wet fluxes are displayed in Figure \ref{figure_flux_wet_isothermal}.
In bothe cases we see that the isothermal fluxes drops slowly with distance. This can be understood by examining Equation \ref{eq_dry_bernouuillie_isothermal} where the dependency on $\rho$ is logarithmic in the isothermal case. So $Ln(\rho)$ almost do not play a role during the escape. As a results only $V^2$  balances the variation of potential energy , which varies only by a factor of 3 between 3.5 and 10 Earth radii (see Figure \ref{Figure_energy}). In consequence V, as well as the flux, varies only by a factor $\sim \sqrt(3)=1.7$ between 3.5 and 10 Earth radii. This corresponds approximately to what is observed in Figures \ref{figure_flux_dry_isothermal} and  \ref{figure_flux_wet_isothermal}.

\medskip

\medskip

\medskip

\section{Renewing of the magma ocean material through convection}
\label{magma_renewing}

In the model described above the magma ocean is simply considered as a "source" function, that provides gas to the atmosphere at that maintains the base of the escaping atmosphere at constant temperature, $T_{\rm s}$, and ensuring thermodynamical equilibrium of the gas with the liquid. It was also assumed that the magma droplets dragged by the gas have a BSE composition. This implies that during the degassing, there is a constant renewing of the magma surface material, rapid enough to maintain its surface with BSE composition, despite of the enrichment in refractory material due to degassing. We now estimate the typical timescale of such a renewing.

When the magma ocean is fully-liquid, it convects vigorously, in a turbulent regime. The convective heat flux that escapes from this turbulent magma ocean scales as \citep{Kraichnan1962,Siggia1994}

\begin{equation}
F \simeq 0.089 \frac{k \Delta T}{H} Ra^{1/3}
\end{equation}
where
\begin{equation}
Ra = \frac{\alpha g \Delta T H^3}{\kappa \nu}
\end{equation}
is the Rayleigh number, $k$ is the thermal conductivity, $\Delta T$ the temperature difference across the magma ocean, $H$ the magma ocean depth, $\alpha$ is the thermal expansion coefficient, $g$ the gravity at the Moon surface, $\kappa$ the thermal diffusivity, $\nu$ the kinematic viscosity. Using $\alpha\simeq 5\times 10^{-5}$ K$^{-1}$, $\kappa \simeq 8\times 10^{-7}$ m$^{2}~$s$^{-1}$ and $\nu \simeq 5\times 10^{-6}$ m$^{2}~$s$^{-1}$ for fully-liquid silicates \citep{Bottinga1983,Karki2010,Eriksson2003}, and $g\simeq 1.62$ m$^2$s$^{-1}$ at the Moon surface, $H\simeq1700$ km close to the Moon radius, and $\Delta T$ in the range $500-1000$ K \citep{Solomatov2000}, we obtain $Ra\simeq 10^{28} - 10^{29}$ and $F$ in the range $2\times 10^5-5\times 10^5$ J$~$m$^{-2}~$s$^{-1}$.

The velocity for turbulent convection scales as \citep{Kraichnan1962}
\begin{equation}
U \simeq 0.6 \left( \frac{\alpha g F H}{\rho C_p}\right)^{1/3},
\end{equation}
where $F$ is the heat flux from the ocean to the atmosphere, $C_p$ the thermal capacity of liquid silicates, $\rho$ the magma ocean density. Using the above estimate for $F$, and assuming $\rho \simeq 3340$ kg$~$m$^{-3}$ for the Moon and $C_p\simeq10^3$ J$~$kg$^{-1}~$K$^{-1}$ \citep{Solomatov2000}, we find convective velocities $U\simeq 1 - 1.5 $ m$~$s$^{-1}$. The associated turnover time scale in the magma ocean is $t_{turn} = H/U\simeq 14-20$ days. This implies that the silicates below the Moon surface are renewed in about $14-20$ days.

\section{Silicate volume brought to the surface}
Assuming volatiles exsolve from the magma ocean as gas bubbles \citep{Hamano2013,Salvador2017}, the volume of liquid silicates that can interact with the atmosphere, before the magma ocean solidifies at time $t_{sol}$, is
\begin{equation}
V_{\rm sil} \simeq 4\pi R_{\rm m}^2 U t_{\rm sol}\simeq4\pi R_{\rm m}^3  \frac{t_{\rm sol}}{t_{\rm turn}}
\end{equation}
where $R_{\rm m}$ is the Moon radius. Assuming a solidification time ranges from $200$ years\citep{Tang_Young_2020} to $1000$ yrs\citep{Solomatov2000}, we obtain
\begin{equation}
4 \times 10^{3}  V_{\rm moon} <V_{\rm sil} < 8 \times 10^{4}  V_{\rm moon},
\end{equation}
where $V_{\rm moon}$ is the Moon volume. In this case, $V_{\rm sil}$ is orders of magnitude larger than the volume of the Moon. Hence, the whole magma ocean has time to interact with the atmosphere. We therefore expect that including the dynamics of the magma ocean will not affect the leading-order conclusions of this paper.

\section{Location of the photosphere}
\label{SOM_optical_depth}
\begin{figure}[!t]
\begin{center}
   \includegraphics[width=.8\linewidth]{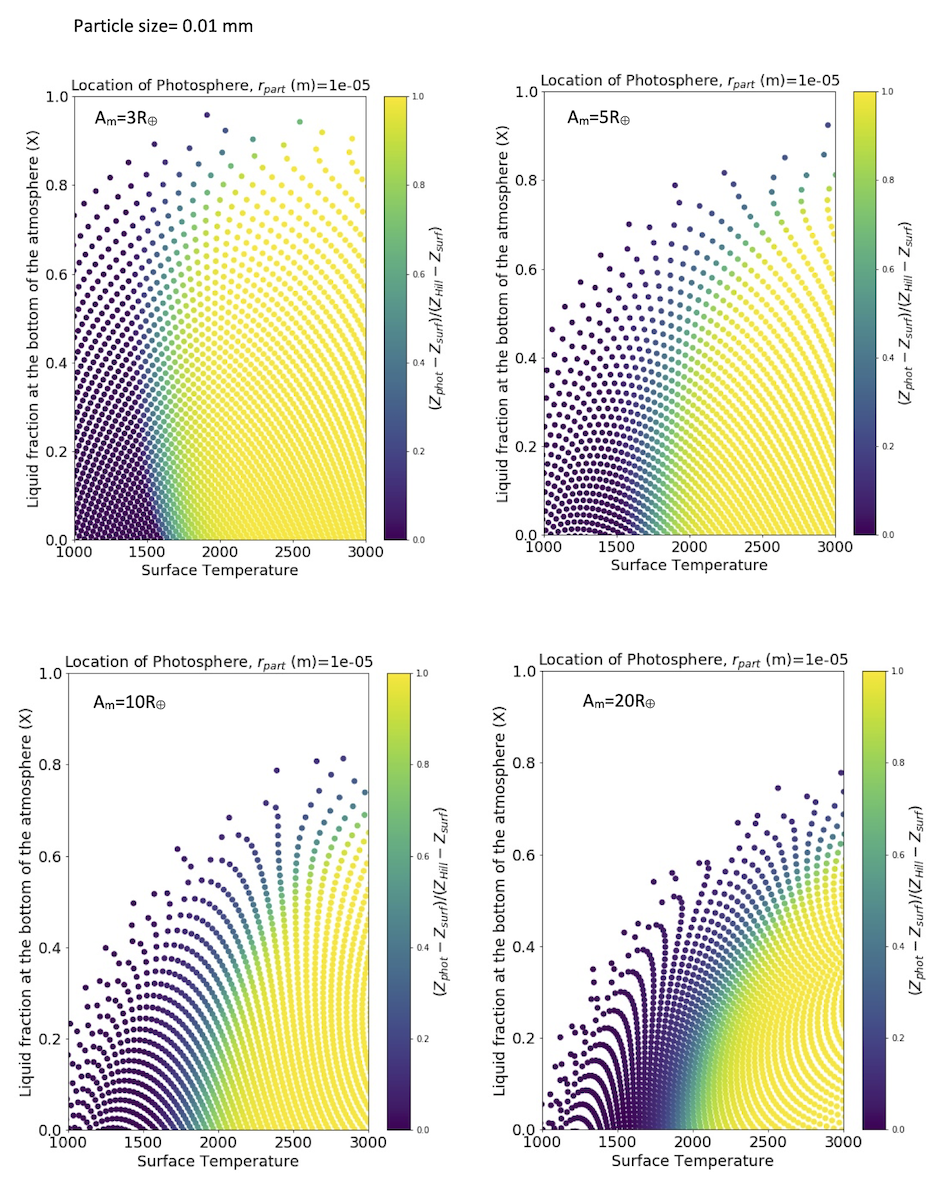}
   \caption{Location of the photosphere (color-coded) as a function of the surface temperature and the liquid mass fraction, $X$, at the surface. The color codes the relative distance of the photosphere compared to the surface or the Hill radius: $(Z_{\rm phot}-R_{\rm m})/(R_{\rm h}-R_{\rm m})$. Yellow color means that the photosphere is at the Hill radius, black color means that the photosphere is at the ground-level (i.e the atmosphere is transparent). Droplet radius is $10^{-5}$ m. The 4 panels correspond to different Earth-Moon distances $A_{\rm m}$. White zone correspond to regions where no escape occurs.}
    \label{Figure_photosphere_rad=1e-5m}

\end{center}
\end{figure}

\begin{figure}[!t]
\begin{center}
   \includegraphics[width=.8\linewidth]{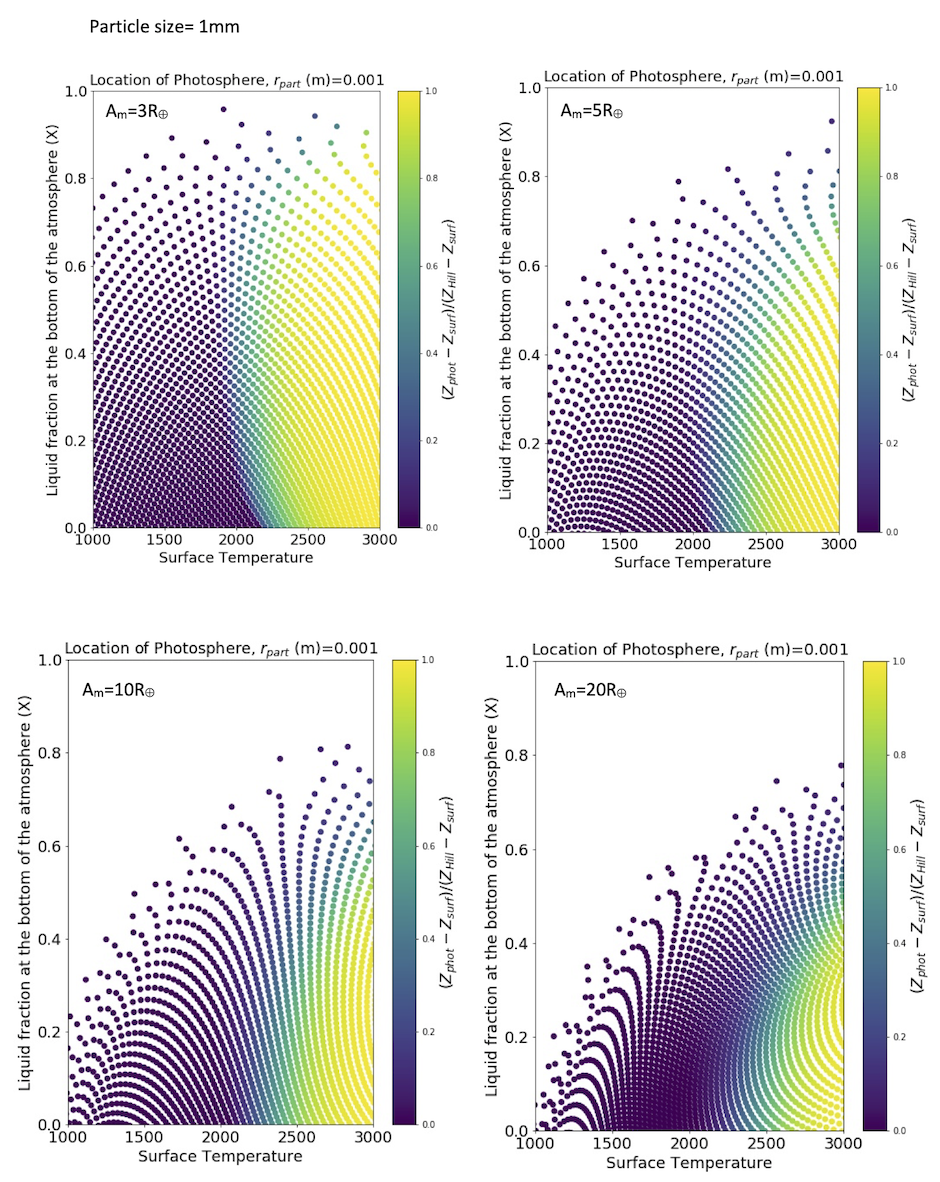}
   \caption{Location of the photosphere (color-coded) as a function of the surface temperature and the liquid mass fraction, X, at the surface. Same legend as above. The droplet radius is $10^{-3}$ m.}
    \label{Figure_photosphere_rad=1e-3m}
\end{center}
\end{figure}

We have computed the atmospheric optical depth $\tau(r)$ assuming that liquid droplets are the only source of opacity. The droplet radius, $r_{\rm d}$ is a fixed parameter and the droplets density is $\rho_{\rm d}$ . With these assumptions the optical depth, integrated from the Hill Sphere down to some distance from the Moon's center, r, is :

\begin{equation}
    \tau(r)=\frac{3}{4\pi\rho_{\rm d} r_{\rm d}}\int_{r}^{R_g} \rho(r)X(r)  dr
\end{equation}

where $R_{\rm h}$ is the  Hill sphere location, $\rho$ is the density of the liquid-gas mixture and  $X$ is the droplet mass fraction. We define the photosphere $Z_{\rm phot}$ as the location where $\tau(Z_{\rm phot})=1$. We take $\rho_{\rm d}=3200 {\rm kg/m}^3$. Results are presented for different Earth-Moon distances and for different droplet radii : 0.01 mm (Figure \ref{Figure_photosphere_rad=1e-5m}) and 1 mm (Figure \ref{Figure_photosphere_rad=1e-3m}). As expected, for larger particle size the photosphere is closer to the ground (i.e. the atmosphere is more transparent), and for a larger Earth-Moon distance also. Interestingly the temperature is the main parameter controlling the location of the photosphere. We find that the transition from an opaque atmosphere (with $Z_{\rm phot} \simeq R_{\rm h}$) to a transparent atmosphere ( $Z_{\rm phot} \simeq R_{\rm m}$) occurs for temperature between 1800 K and 2400 K for an Earth-Moon distance = $5R_{\oplus}$.

\newpage

\section{Composition of the vapor atmosphere}
\label{SOM_THERMO}
The saturated vapor composition of an atmosphere at equilibrium with a magma with BSE composition was recently published  \citep{Visscher_Fegley_2013}. However this work does not extend below 2000 K, a temperature regime of interest to explain the Moon composition. Using up to date thermodynamic data(\cite{Sossi_Fegley_2018}), we have computed the vapor composition for $T<2000$ K.  The equilibrium partial pressure may be calculated for any given vaporisation reaction:

\begin{equation}
\left( M^{x+n}  O_ \frac{x+n}{2} \right) _{l}=\left( M^{x} O_\frac{x}{2} \right) _{g}+\frac{n}{4}O_2
\label{Eq_reaction_1}
\end{equation}

where $l$ denotes the liquid phase, $g$ the gas phase, $x$ the oxidation state of the metal in the gas phase and $n$ the coefficient of stoichiometry (an integer value). We note $K$ the equilibrium constant of this reaction. From Equation \ref{Eq_reaction_1} the equilibrium partial vapor pressure of species $M^x O_{x/2}$ (designated as species $i$ for short) is obtained as follows :

\begin{equation}
p \left( M^x O_{\frac{x}{2}} \right)=\frac{K X(i) \gamma(i)}{f(O_2)^{n/4}}
\end{equation}

where $X(i)$ is the mole fraction of species $i$ (see table below), $\gamma(i)$ its activity coefficient and $f(O_2)$ the oxygen fugacity.

Here we estimate the activity coefficients of Na, K, Cu and Zn based on thermodynamic measurements from laboratory experiments on basaltic melts (Na, De Maria et al. 1971, Mathieu et al. 2008 and Sossi et al. 2019; K, De Maria et al. 1971, Hastie et al. 1983 and Sossi et al. 2019; Cu, Sossi et al. 2019, and Zn, Reyes and Gaskell, 1983 and Sossi et al. 2019). The ratio $\gamma(Na)$/$\gamma(K)$ is set to remain constant as a function of temperature (cf. Charles, 1967) where $\gamma(Na)$ is taken to be $10^{-3}$ at 1700 K and $\gamma(K) = 5 \times 10^{-4}$ at the same temperature. Uncertainties in these values are as large as a factor of 5, see also Jiang et al. (2019). For example, Mathieu et al. 2008 provide a value of $\gamma(Na)$ = $10^{-3}$ at 1623 K, Sossi et al. (2019) report the same value at 1673 K, whereas the Knudsen Effusion Mass Spectrometry measurements of De Maria et al. (1971) suggest a $\gamma(Na)$ of $5 \times 10^{-3}$ at 1500 K. Copper may exist in its 2+ and 1+ oxidation states in silicate melts (Holzheid and Lodders, 2001), however at the oxygen fugacities produced by evaporation of the lunar silicate (~FMQ), it occurs uniquely as $Cu^+$. We assume that the major element oxides ($MgO$, $FeO$ and $SiO_2$) dissolve ideally into silicate melts (e.g. O'Neill and Eggins, 2002; Beckett, 2002).

Here, we take the model of Visscher and Fegley (2013) as an input parameter to calculate $f(O_2)$. This model compares well with measurements of oxygen fugacities above vaporising silicates above 1400 K (see Fig. 4 of ref.  \cite{Sossi_Fegley_2018}). The values of $n$, the stoichiometric coefficient of the vaporisation reaction, is an integer value equal to 1 (Na, K) and 2 (Mg, Fe, Si, Zn). As such, the condensed species considered are: $NaO_{0.5}$, $KO_{0.5}$, $CuO_{0.5}$ $MgO$, $FeO$, $SiO_2$, and $ZnO$, all (l) and the gas species are: $Na^0$, $K^0$, $Cu^0$, $Mg^0$, $Fe^0$, $SiO$, and $Zn^0$. We consider only these eight elements as they have the most abundant species in the vapour <2000 K \citep{Visscher_Fegley_2013}.

Following our experimental data of BSE vaporisation we use the values for $f(O_2)$ reported in Table \ref{table_fO2}.

\medskip
\begin{table}
\begin{tabular} {|l|l|l|l|l|l|l|}
  \hline
  \textbf{Temperature (K)}  & \textbf{1500} & \textbf{1600} & \textbf{1700} & \textbf{1800} & \textbf{1900}  & \textbf{2000} \\
  \hline
  \textbf{$f(O_2)$ (bar)}    & $2.04 \times 10^{-7}$ & $ 5.79 \times 10^{-7} $& $1.53 \times 10^{-6}$ & $3.82 \times 10^{-6}$ & $9.03 \times 10^{-6}$ & $2.04 \times 10^{-5}$ \\
  \hline

\end{tabular}
\caption{The $f(O_2)$ of an atmosphere in equilibrium with a magma ocean of BSE composition. Oxygen fugacities are close to the FMQ buffer, as attested to by measurements above natural basalts (De Maria et al. 1971) and olivine (Costa et al. 2017), and as modelled by Visscher and Fegley (2013). Here we use the values modelled by Visscher and Fegley (2013) for a BSE composition.}
\label{table_fO2}

\end{table}

\medskip

Here are the thermodynamic parameters for the different species of interest. The molar abundances :

\medskip
\begin{table}
\begin{tabular} {|l|l|}
  \hline
  \textbf{Species} & \textbf{X (Mole Fraction)} \\
   \hline
   Na & $5.93 \times 10^{-3}$ \\
   K  & $3.5 \times 10^{-4}$   \\
   Cu & $2.6 \times 10^{-5}$   \\
   Zn & $4.95 \times 10^{-5}$ \\
   Mg & $0.2217$ \\
   Fe & $0.063$ \\
   Si & $0.2122$ \\
   \hline
\end{tabular}
\caption{Composition of BSE used in this paper.}
\label{table_BSE_compo}
\end{table} 

In order to predict the temperature dependence of the activity coefficients of Na, K and Zn in silicate melts, we draw upon pre-existing experimental data. Charles (1967) determined the activity coefficients of $Na_2O$ and $K_2O$ in alkali oxide - $SiO_2$ binary systems by application of the cryoscopic equation to pre-existing vaporisation data, thereby determining heats of solution of oxide components in the silicate melt. The two quantities are related by the van't Hoff equation:

\begin{equation}
\frac{dln{\gamma(i)}}{d(1/T)} = \frac{\Delta H_{sol}}{R}
\end{equation}

We take values of $\Delta H_{sol}$ from Charles, 1967 for $NaO_{0.5}$ (-112 kJ/mol) and fix $\gamma(KO_{0.5})$ as $0.5 \times \gamma(NaO_{0.5})$, while we use the values determined by Reyes and Gaskell (1983), -135 kJ/mol and Sossi et al. (2019), -143 kJ/mol for ZnO, with 121 kJ/mol for $CuO_{0.5}$ from Sossi et al. (2019). All quantities are extracted presuming infinite dilution of the trace component in the silicate melt.

The estimated activity coefficients are therefore the following :

\begin{table}
\begin{tabular} {|l|l|l|l|l|l|l|}
  \hline
  \textbf{Temperature (K)}  & \textbf{1500} & \textbf{1600} & \textbf{1700} & \textbf{1800} & \textbf{1900}  & \textbf{2000} \\
  \hline
  $\gamma$ for $NaO_{0.5}$  &  0.0004 & 0.0008 & 0.001  & 0.002 &  0.003  & 0.004   \\
  $\gamma$ for $KO_{0.5}$  &  0.0002 & 0.0004 & 0.0005 & 0.001 &  0.0015 & 0.002 \\
  $\gamma$ for $CuO_{0.5}$  &  37.6 & 20.5 & 12 & 7.5 &  4.9 & 3.3 \\
  $\gamma$ for $ZnO$  &  0.18   & 0.42  & 0.56    & 0.65  &  0.71    & 0.75  \\

  \hline

\end{tabular}
\caption{Activity coefficients used in this paper.}
\end{table}

\medskip

The equilibrium constant may be calculated using thermodynamic data (e.g. Chase 1998) using the expression:
\begin{equation}
\Delta G=-R T ln(K)
\end{equation}
in which
\begin{equation}
\Delta G= \Delta H-T \Delta S
\end{equation}
We use the following enthalpies and entropies  :

\begin{itemize}
\item $\Delta G(Na)= 260720-T \times 118.09$ J/mol (Lamoreaux and Hildenbrand, 1984)
\item $\Delta G(K)= 210056-T \times 101.71$ J/mol(Lamoreaux and Hildenbrand, 1984)
\item $\Delta G(Zn)= 405910-T \times 174.4$ J/mol (Lamoreaux et al., 1987)
\item $\Delta G(Cu)= 377700-T \times 135.0$ J/mol (O'Neill and Pownceby, 1993)
\item $\Delta G(Mg)= 662155-T \times 184.01$ J/mol (Chase, 1998)
\item $\Delta G(Fe)= 636430-T \times 178.46$ J/mol (Chase, 1998)
\item $\Delta G(Si)= 778440-T \times 242.29$ J/mol (Chase, 1998)

\end{itemize}

The resulting partial pressure are plotted in Figure \ref{Figure_partial_pressure}.

 \section{Caveats and approximations}
 \label{SOM_approximations}
 This paper aims to show that degassing during the lunar assembling is potentially an important process, and may explain the depletion of lunar material in moderately volatile elements. Of course, to make the computation tractable, several approximations must be done, and before making a fully self-consistent model we aim here to show that first-orders calculation show that tidally driven evaporation is a viable process. Our simplifications are summarized below:

 \begin{itemize}
    
    \item Radiative heating and cooling : The present work does not consider radiative heating and cooling processes that may act during the gas expansion. Only adiabatic cooling is considered. This would require to couple the dynamical model to radiative transfer code, that has never been developed for an atmosphere expanding above a magma ocean. Computation of the radiative diffusion timescale following Tremblin et al. (equation 17) shows that in the optically thick case the radiative diffusion timescale is  shorter than the expansion timescale (a few 10s compared to a few $10^{4}$s for droplets in the range of micrometer size), assuming the source of opacity  are the droplets in the flow. This implies that the thermal radiation may potentially modify the temperature during the expansion if important temperature gradient is present.
    In the absence of a full radiative transfer model, we compare two length scales : the temperature scale height ($L_{\rm temp}$):
    \begin{equation}
        L_{\rm temp}=\frac{T}{\partial T/\partial r}
    \end{equation}

    and the photon mean free path ($L_{\rm rad})$:
    \begin{equation}
        L_{\rm rad}=\frac{1}{\kappa \rho}
    \end{equation}
    where $\kappa$ is the local opacity (computed with assuming that the opacity source is the presence or droplets with a fixed radius). If $L_{\rm temp} >> L_{\rm rad}$, this means that a photon experiences several scattering events before reaching a region with significantly different temperature. The assumption of local thermal equilibrium (LTE) is therefore valid, and we do not need to consider the energy loss due to radiation. In the opposite case, the photons can escape without being absorbed and radiative cooling effect should not be neglected.
   We plot in Figure \ref{Figure_Ltemp_Lrad} the ratio between the two length scales for different atmospheric profiles, $L_{\rm temp}$ and $L_{\rm rad}$ being evaluated at the sonic point. We find that at high temperature $L_{\rm temp} > L_{\rm rad}$
    and at low temperature $L_{\rm temp} < L_{\rm rad}$, the transition between the two regimes situating at $T=1400$ K for 1 micron radius droplets and $T=1700$ K for 1 mm radius droplets. We conclude that above such temperature, the present work neglecting radiative cooling effects stays valid, whereas at lower temperature complete radiative modeling should be considered.

\begin{figure}
\begin{center}
   \includegraphics[width=.9\linewidth]{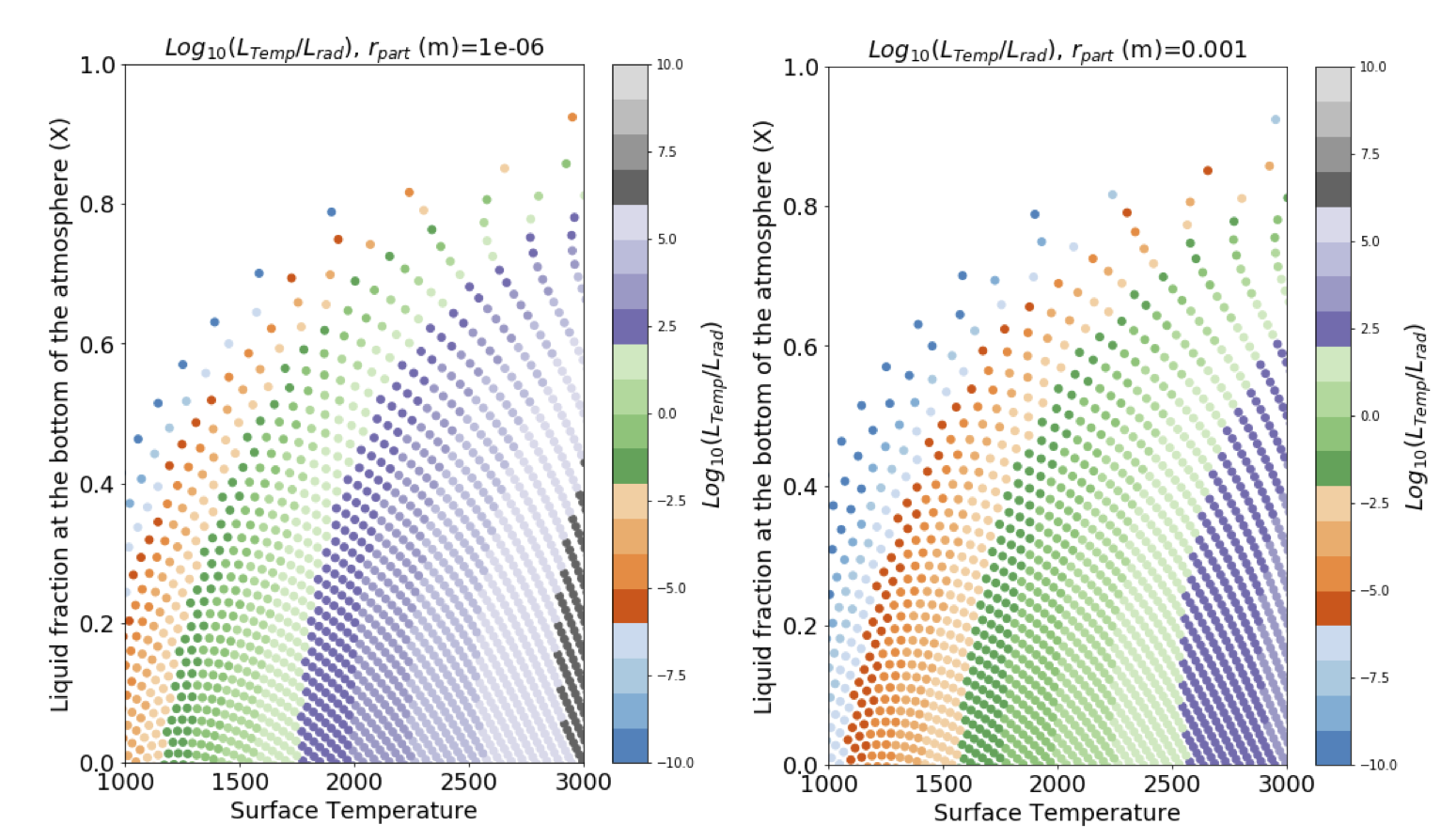}
   \caption{$Log_{10}(L_{\rm temp}/L_{\rm rad})$ evaluated at the sonic point for a Moon at 5 $R_\oplus$ as a function of the surface temperature and the liquid mass fraction (X). Left : for liquid droplets size=1 micron; Right:for liquid droplets size=1mm.}
    \label{Figure_Ltemp_Lrad}
\end{center}
\end{figure}

    Taking into account radiative cooling and heating from the Moon surface would require to develop a radiative transfer model, well beyond the scope of the paper that intends to identify a new escaping process and to point out its potential importance. However, for the simple case of a collisionless gas, with no pressure and at thermal equilibrium with the hot radiating Moon's surface,  we show in Figure \ref{Figure_T_escape}.a that the gas temperature becomes always larger than the necessary temperature for escaping for a Moon within 5$R_\oplus$ of the Earth center'.

 This shows that escape is an unavoidable process for a fully grown Moon. Escape becomes even more intense for a Moon's building block, with a fraction of a lunar mass (Figure \ref{Figure_T_escape}.b, and core text).

 \item 1D calculation : In this paper we have done the calculation in 1D, along the line joining the center of the Earth and the center of the Moon. So, strictly speaking, the computed flow would be only valid along this line, and it cannot be assumed that the flow is emitted at $4 \pi$ steradians on the surface. It is why, for simplicity purpose we have assume that the flow is arbitrarily emitted on a solid angle of 1 steradian.
 The full calculation would need a full 3D treatment, as the gas will probably circulate around the Moon before escaping. But such model is still beyond the state of the art, but could be done in a close future.

\section{Lost mass fractions}
\label{SOM_escaped_mass}
Let $M^M$ and $M^E$ the Moon and Earth Mass respectively today.
Let $\mu_i^E$ the mass fraction of an element $i$ in the BSE, $\mu_i^M$ the mass fraction of an element in the Moon today,  and $\mu_i^v$the mass fraction of an element  $i$ in the vapor at some temperature. We call $M^L$ the total mass lost during degassing of the Moon. $M^L=M^L_v+M^L_l$ where $M^L_v$ is the mass lost in the form of gas, and $M^L_l$ is the mass lost in the form of droplets dragged by the gas. We call $M_i^M$ the total mass of an element in the Moon today and $M_i^E$ the total mass of an element in the BSE. We assume that the Moon is degassed at constant temperature for simplicity so that  $\mu_i^v$  is a constant for every species.
The Initial mass of the Moon is $M^M+M^L$ and final mass is $M^M$.
If we assume that the Moon material was initially BSE composition we have :

\begin{equation}
    M_i^M=\mu_I^E (M^M+M^L) -\mu_i^v M^L_v-\mu_i^E M^L_l
\end{equation}
The first term is the initial mass of element $i$ the second term is the mass lost in vapor form, and the third term is the mass lost in liquid form (assumed to have BSE composition). Since $M^L_l=x M^L$ and $M^L_v=(1-x) M^L$ ; we get
\begin{equation}
    M_i^M=\mu_I^E (M^M+M^L_v) -\mu_i^v M^L_v
\end{equation}

Dividing by $M^M$ , the Moon mass, we obtain
\begin{equation}
    \mu_i^M=\mu_I^E \frac{M^M+M^L_v}{M^M}-\mu_i^v \frac{M^L_v }{M^M} \Rightarrow
\end{equation}
\begin{equation}
    \mu_i^M=\mu_I^E+\frac{M^L_v}{M^M}(\mu_i^E-\mu_i^v)
    \label{Equ_MuiM}
\end{equation}

In the above equation the effect of degassing on the mass fraction in the Moon ($\mu_i^M$) is obvious. No degassing means $M^L_v=0$ so $\mu_i^M$ stays equal to the BSE value $\mu_I^E$. if $\mu_i^v > \mu_i^E$ (which is the case for volatile species in general) then degassing lowers the value of $\mu_i^M$.

 From Equation \ref{Equ_MuiM} we derive the mass fraction of an element remaining in the Moon, compared to its mass fraction in the BSE:

 \begin{equation}
    \frac{\mu_i^M}{\mu_i^E}=1+\frac{M^L_v}{M^M}(1-\frac{\mu_i^v}{\mu_i^E})
    \label{Equ_ratio}
\end{equation}

 From the above equation we see that fractionation, compared to the BSE, is only possible through vapor loss. If $\mu_i^v < \mu_i^E$ then this element will be enriched in the Moon, conversely elements with $\mu_i^v > \mu_i^E$ will be depleted in the Moon.

\section{Enhancement of Volatile Loss due to XUV Heating from the Proto-Earth}
\label{XUVearth}


While our model treats tidally-driven escape heated uniquely from the bottom by the magma ocean, additional heating from the top by XUV irradiation as at exoplanets \citep{erkaev07} and even Kuiper Belt Objects \citep{johnson2015} gives further credence to the Moon's devolatilization by atmospheric escape. By employing the atmospheric evolution model \texttt{DISHOOM} \citep{Gebek2020} we investigate mass loss due to surface and upper atmospheric heating. We find that while the tenuous Jeans escape rates due to surface heating are quite small, catastrophic volatile loss ($< 100$ Myr) must still occur due to a remarkably bright proto-earth\citep{saxena2017} heating the upper atmosphere via XUV energy-limited escape. Figure \ref{Figure_EL_escape} shows three scenarios of three different estimates of the proto-Earth temperature with distance. We adopt a heating efficiency of 0.35 similar to Io's volcanic atmosphere\citep{lellouch92}.  A conservative assumption, neglecting atmospheric sputtering, that the only proto-earth irradiation driving escape is due to the X-ray and UV component ($\sim 3.6 \times 10^{-6} L_{\rm bol}$) still yields a total mass loss of $\sim 0.1 \%$ of a lunar mass, in 1 Myr. The mass loss rates on the order of $\sim 10^{6}$ kg/s are then quite similar to an early magma-ocean planet at 1 AU orbiting an M-dwarf\citep{bower2019} or a close-in exomoon\citep{oza2019b}. Considering the timescales and proximities studied in this work, a catastrophic volatile loss scenario for the Moon seems inevitable.

\begin{figure}
\begin{center}
  \label{Figure_EL_escape}
  \includegraphics[width=.7\linewidth]{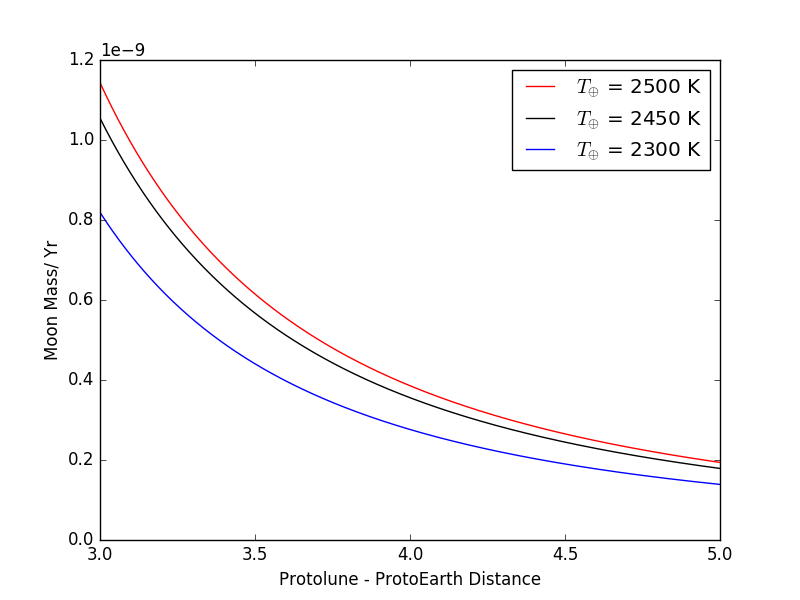}
  \caption{Energy-limited escape from the proto-Moon due to irradiation of a molten proto-Earth at three different effective temperatures $T_{\oplus}$: 2500 K (red), 2450 K (black), 2300 K (blue).
  }

\end{center}
\end{figure}

  \end{itemize}

\end{appendix}
\end{document}